\documentclass[aps,pra,groupedaddress,nofootinbib,notitlepage,showpacs,floatfix,twocolumn,longbibliography]{revtex4-1}
\usepackage{graphicx}
\usepackage{epstopdf}
\usepackage{amsmath}
\usepackage{amssymb}
\usepackage{hyperref}
\usepackage{bm}
\usepackage{subfigure}

\hypersetup{
    colorlinks=true,       
    linkcolor=cyan,          
    citecolor=magenta,        
    filecolor=magenta,      
    urlcolor=cyan,           
    runcolor=cyan
}
\usepackage[pdftex]{color}

\newcommand{\beq}{\begin{equation}}
\newcommand{\eeq}{\end{equation}}
\newcommand{\beqnn}{\begin{equation*}}
\newcommand{\eeqnn}{\end{equation*}}
\newcommand{\bea}{\begin{eqnarray}}
\newcommand{\eea}{\end{eqnarray}}
\newcommand{\beann}{\begin{eqnarray*}}
\newcommand{\eeann}{\end{eqnarray*}}
\newcommand{\bes} {\begin{subequations}}
\newcommand{\ees} {\end{subequations}}

\newcommand{\ket}[1]{ | #1\rangle}
\newcommand{\bra}[1]{\langle #1 | }

\newcommand{\eps}{\varepsilon}

\newcommand{\ignore}[1]{}
\newcommand{\GS}{\mathrm{GS}}

\begin{document}
\title{Quantum annealing correction for random Ising problems}
\author{Kristen L. Pudenz,$^{1,2,3}$ Tameem Albash,$^{2,3,4}$ \& Daniel A. Lidar$^{1,2,3,4,5}$}
\affiliation{$^{(1)}$Department of Electrical Engineering, $^{(2)}$Center for Quantum Information Science \& Technology, $^{(3)}$Information Sciences Institute, $^{(4)}$Department of Physics and Astronomy, $^{(5)}$Department of Chemistry\\
University of Southern California, Los Angeles, California 90089, USA.}

\begin{abstract}
We demonstrate that the performance of a quantum annealer on hard random Ising optimization problems can be substantially improved using quantum annealing correction (QAC).  Our error correction strategy is tailored to the D-Wave Two device.  We find that QAC provides a statistically significant enhancement in the performance of the device over a classical repetition code, improving as a function of problem size as well as hardness. Moreover, QAC provides a mechanism for overcoming the precision limit of the device, in addition to correcting calibration errors.  Performance is robust even to missing qubits.  We present evidence for a constructive role played by quantum effects in our experiments by contrasting the experimental results with the predictions of a classical model of the device.  Our work demonstrates the importance of error correction in appropriately determining the performance of quantum annealers.
\end{abstract}

\maketitle

\section{Introduction}
It is widely accepted that no form of quantum information processing can be scalable without some form of quantum error prevention, suppression, or correction \cite{Lidar-Brun:book}. This applies in particular to quantum annealing \cite{finnila_quantum_1994,kadowaki_quantum_1998,Santoro,morita:125210,RevModPhys.80.1061,Bapst2013} and the closely related quantum adiabatic algorithm \cite{farhi_quantum_2001}, strategies designed to take advantage of quantum mechanics in solving  classical optimization problems, such as finding the ground state of a disordered Ising Hamiltonian, a well-known NP-hard problem \cite{barahona_computational_1982}. Interest in quantum annealing has piqued in recent years since commercial processors comprising hundreds of programmable superconducting flux qubits have become available to the research community \cite{Dwave,Harris:2010kx}, and a lively debate has erupted concerning their quantumness \cite{q-sig,q108,Smolin,comment-SS,SSSV,q-sig2,SSSV-comment,DWave-entanglement} and the possibility of observing a quantum speedup \cite{speedup,2014Katzgraber,Venturelli:2014nx}, for which there exists theoretical evidence via specific examples \cite{Santoro,morita:125210,Somma:2012kx}. 

While error mitigation strategies for quantum annealing and more generally, adiabatic quantum computing have been proposed \cite{Mizel:01,jordan2006error,PhysRevLett.100.160506,PhysRevA.86.042333,Young:13,Sarovar:2013kx,Young:2013fk,Ganti:13,Mizel:2014sp,Bookatz:2014uq} and implemented \cite{PAL:13}, much less is known compared to the relatively mature state of quantum error correction in the circuit model \cite{Lidar-Brun:book,Gaitan:book}. In particular, an accuracy threshold theorem \cite{Knill:98,preskill:12,Gottesman:2013ud} for fault-tolerant quantum annealing remains elusive, in spite of some degree of inherent robustness of adiabatic quantum computation to thermal excitations and control errors \cite{childs_robustness_2001,PhysRevLett.95.250503,PhysRevA.79.022107}.  
Notwithstanding, we recently proposed a practical error suppression and correction strategy for quantum annealing and implemented it using a D-Wave Two (DW2) quantum annealing processor \cite{Bunyk:2014hb} on a toy problem of antiferromagnetic chains \cite{PAL:13}. 
We demonstrated that this quantum annealing correction (QAC) scheme provided a substantial fidelity enhancement in the presence of thermal excitation and control errors. 

Here, we experimentally study the performance of QAC on random Ising problems with quenched disorder using a DW2 processor, on up to $112$ logical qubits comprising $4$ physical qubits each. In contrast to anti-ferromagnetic chains, these are hard optimization problems, of the type studied in recent benchmarking work probing for a quantum speedup \cite{q108,speedup} against simulated annealing \cite{kirkpatrick_optimization_1983}. In this manner we hope to demonstrate the importance of including error correction in future quantum annealing devices, and in particular the utility of QAC in improving the performance of the current D-Wave devices.  
However, rather than attempting to demonstrate a speedup, which seems likely to be precluded in our setting for the reasons discussed in Ref.~\cite{2014Katzgraber}, we focus on establishing a performance improvement, i.e., an enhancement in the success probability of finding the Ising spin glass ground state, when using QAC.
Moreover, we demonstrate that QAC is also effective at extending the precision range of the D-Wave device, thus in effect overcoming control errors. The price to be paid for these improvements is a reduction in the number of qubits and the degree of the qubit connectivity graph due to the use of an encoding, but such tradeoffs seem inevitable if the goal is to reach scalability of quantum information processing.

A key question raised by the performance gains of the QAC strategy is to what extent it is a form of \emph{quantum} error correction. To address this we compare our experimental results for the ground state probability to those we compute numerically using a classical model of interacting spins. This SSSV model \cite{SSSV} has been very successful in reproducing the success probabilities of random Ising instances reported in Ref.~\cite{q108}, and it has played a central role in the quantumness discussion concerning the D-Wave devices \cite{SSSV,q-sig2,SSSV-comment}. We demonstrate that for random Ising problems subject to QAC encoding, there is a strong discrepancy between the SSSV model and the experimental results. This conclusion is robust to varying the parameters of the SSSV model, and it provides indirect evidence that quantum effects play an important role in the success of the QAC strategy in our experiments.\\

\section{Results}
\subsection{Quantum annealing and the D-Wave Two processor}
These topics have been described in detail in a number of publications (see, e.g., Refs.~\cite{q108,speedup,Bunyk:2014hb}), so here we give just the details needed for our work. 

Quantum annealing (QA) is a method for finding the ground state of an Ising spin Hamiltonian 
\beq
H_{\mathrm{Ising}} =  \sum_{i=1}^N h_i \sigma_i^z + \sum_{i<j}^N J_{ij} \sigma_i^z \sigma_j^z \ .
\eeq
The local fields $h_i$ and couplings $J_{ij}$ are given and the problem is to find the spin configuration $\{\sigma_i^z\}_{i=1}^N$ that minimizes $H_{\mathrm{Ising}}$, where each spin variable $\sigma_i^z \in \pm 1$. In QA this is done by adiabatic evolution from the ground state of an initial transverse field, i.e., the time-dependent Hamiltonian is 
\beq
H(t) = A(t) H_X + B(t) H_{\mathrm{Ising}}\ ,
\label{eq:H(t)}
\eeq 
where 
$H_X = \sum_{i=1}^N \sigma_i^x$ and the $\sigma_i$'s are now the standard Pauli spin-$1/2$ matrices acting on the $i$th qubit. The function $A(t)$ decreases monotonically to zero, while $B(t)$ increases monotonically from zero, with $t\in [0,t_f]$. For a closed system the adiabatic theorem guarantees that if the initial state is the ground state then the final state will be arbitrarily close to the ground state provided $t_f$ is large enough compared to the minimum energy gap of $H(t)$ and provided the functions $A(t)$ and $B(t)$ are sufficiently smooth \cite{Jansen:07,lidar:102106}.  

The DW2 processor is a physical realization of the quantum annealing algorithm.  The system is initialized in the thermal Gibbs state of $A(0) H_X$, which has almost its entire weight on the ground state since $kT \ll A(0)$. The idealized conditions of the adiabatic theorem can of course not be realized in a physical device such as the DW2, which operates at a finite temperature and suffers from programming control errors on the $h_i$ and $J_{ij}$ terms. In such an open system, thermal processes can depopulate the ground state, reducing the success probability of the algorithm (though thermal relaxation can sometimes be beneficial \cite{TAQC,DWave-16q}). Control errors can unintentionally cause the annealer to evolve according to the wrong Hamiltonian.\\

\subsection{Quantum Annealing Correction} 
\begin{figure}[t]
\begin{center}
\subfigure[\ ]{\includegraphics[width=0.3\columnwidth]{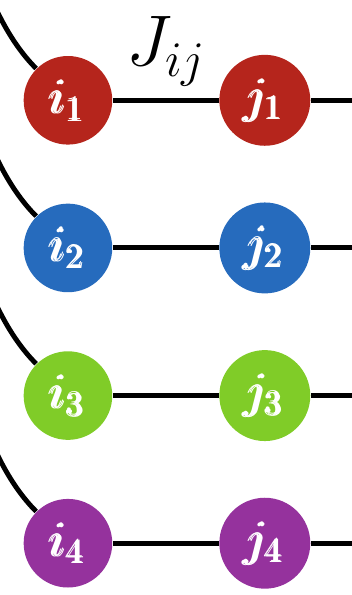}\label{fig:1a}} \hspace{1cm}
\subfigure[\ ]{\includegraphics[width=0.3\columnwidth]{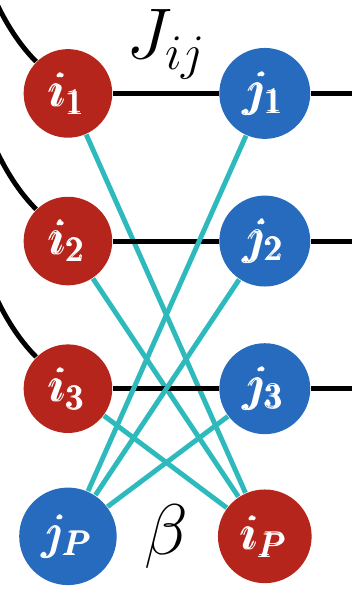}\label{fig:1b}} 
\end{center}
\caption{{(Color online) Embedding of the C and QAC strategies on the D-Wave processor.} (a) C strategy, (b) QAC strategy. The basic unit cell of the D-Wave processor includes 8 qubits (shown as circles), with the qubits on the left side coupling to adjacent unit cells up and down, and those on the right side coupling left and right.  Panel (a) shows how 4 parallel copies of the problem for the C strategy are embedded, each with the same problem couplings $J_{ij}$ (black lines).  Panel (b) shows how two logical qubits (red and blue) for the QAC strategy are embedded within a unit cell; the problem qubits $i_{1, 2, 3}$ are coupled using the same problem couplings $J_{ij}$ (black lines), and the penalty qubit $i_P$ couples ferromagnetically to the problem qubits with magnitude $\beta$ (light blue lines).}
\label{fig:code_embedding}
\end{figure}

What can be done to mitigate the detrimental effect of thermal excitation and control errors? While a variety of theoretical proposals exist \cite{Mizel:01,jordan2006error,PhysRevLett.100.160506,PhysRevA.86.042333,Young:13,Sarovar:2013kx,Young:2013fk,Ganti:13,Mizel:2014sp,Bookatz:2014uq} we require one that is implementable using the DW2 device. Toward that end, we employ two strategies that were proposed and studied in Ref.~\cite{PAL:13}. The first is a purely classical (C) repetition strategy, whereby we evolve $K$ independent copies of the problem, i.e., 
\begin{equation}
H_{\mathrm{C}} = 
\sum_{i=1}^{\overline{N}} \sum_{k=1}^{K}  h_{i}  \sigma_{i_k}^z  + 
\sum_{i<j}^{\overline{N}} \sum_{k=1}^{K}  J_{ij} { \sigma_{i_k}^z \sigma_{j_k}^z} .
\label{eq:C-strategy}
\end{equation}
where $K \overline{N} = N$.  This strategy runs the quantum annealing algorithm $K\geq 2$ times in parallel to increase its chances of finding the ground state. An example of how the C strategy is embedded on the DW2 device is shown in Fig.~\ref{fig:1a}, where $K=4$.

\begin{figure*}[t]
\includegraphics[width=\textwidth]{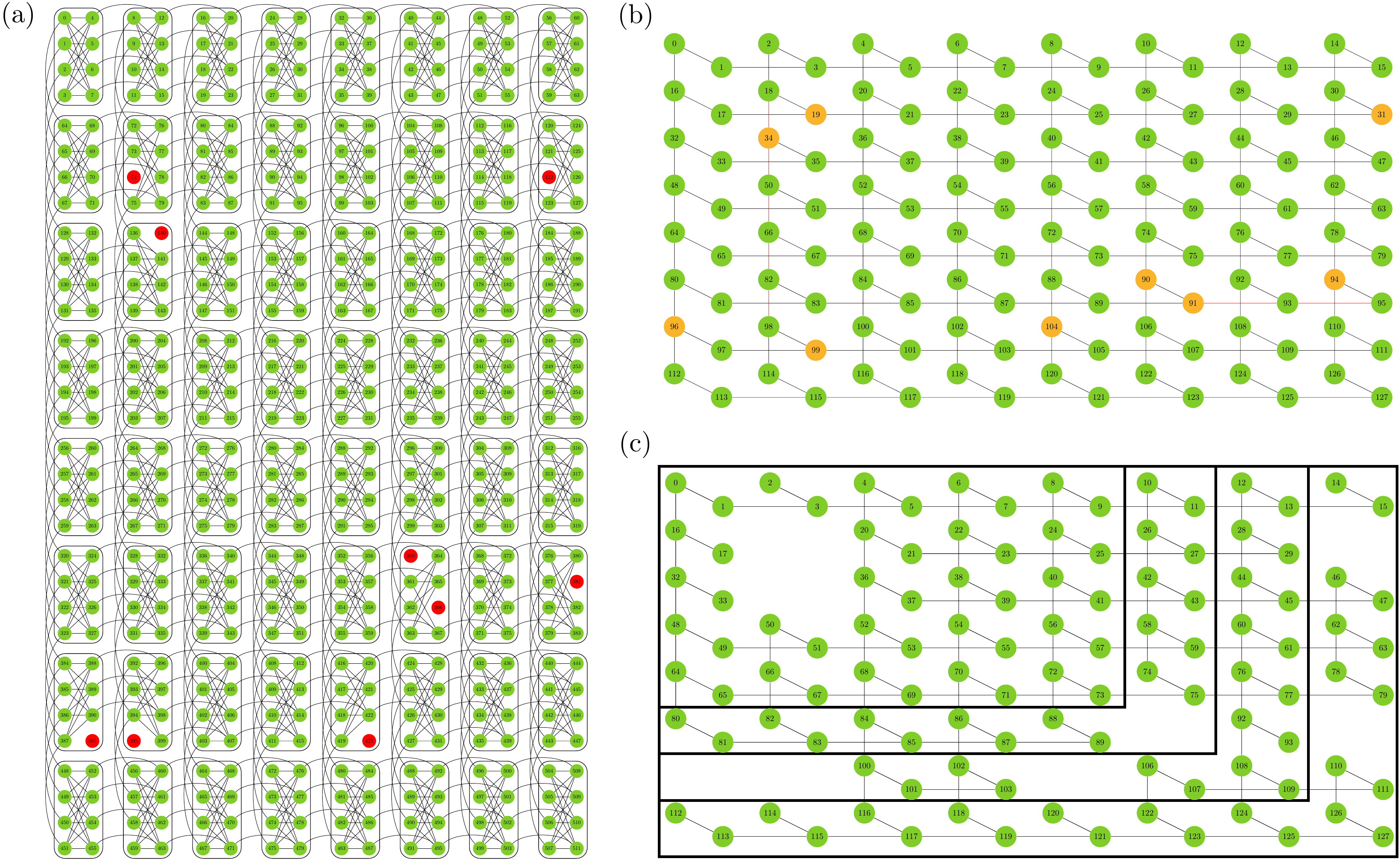}
\caption{(Color online) {Physical and logical qubit connectivity graphs.}  (a) The physical connectivity graph consists of $8\times8$ unit cells of eight qubits (denoted by circles), connected by programmable inductive couplers (lines). The $503$ green (red) circles denote functional (inactive) qubits. In the ideal case, where all qubits are functional and all couplers are present, one obtains the non-planar, degree-$6$ ``Chimera" connectivity graph. (b) The complete degree-$3$ logical connectivity graph, with perfect (imperfect) logical qubits and their couplers shown in green and black (orange and red) respectively. (c) The actual logical connectivity graph.  Imperfect qubits and their couplings are not shown as these were not used. We generated random problems over each of the regions shown in the rectangles of increasing sizes $\overline{N}\in\{46,66,86,112\}$ in (c). These problem sizes were chosen because they consist of square blocks of unit cells, and the treewidth of a planar square lattice grows as a function of the smallest dimension of any rectangular region chosen \cite{Robertson198449}.}
\label{fig:graphs}
\end{figure*}

\begin{figure}[t]
\includegraphics[width=\columnwidth]{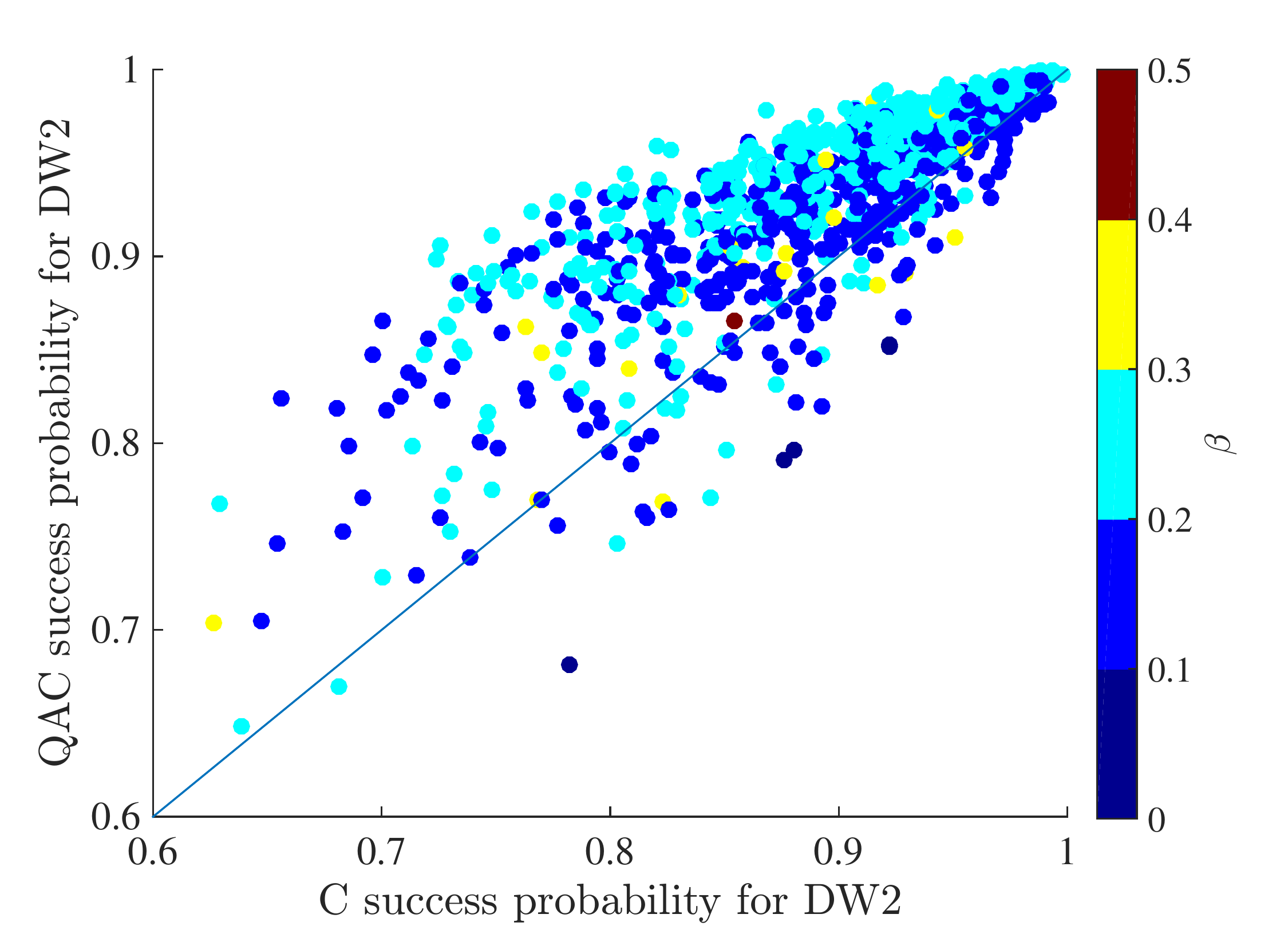}\label{fig:scatter112-DW2}
\caption{(Color online) {Correlation plot of the success probability for the QAC and C strategies for DW2.} The case of $\overline{N} = 112$ logical qubits is shown. Above the diagonal line QAC performs better, while below it C performs better. The color scale indicates the optimal $\beta$ for each instance.  The QAC strategy uses both problem group decoding and logical group decoding, while the C strategy only uses problem group decoding, as discussed in Appendix \ref{app:Methods}.
QAC outperforms C  for the overwhelming majority of instances.  The few instances where C outperforms QAC are those with low-lying undecodable energy states due to logical qubits with $<3$ (and weak) problem couplings to neighbors, arising from holes in the logical connectivity graph (see Appendix \ref{app:error_types}). 
}
\label{fig:scatter112}
\end{figure}

The second strategy, QAC, uses the same physical resources as the C strategy, but adds two important aspects to suppress and correct bit-flip errors: (i) we encode our problem using a $K-1$ qubit repetition code, and (ii) we supplement $H_{\mathrm{Ising}}$ with energy penalty terms in the form of the stabilizer generators of the repetition code. The former allows us to correct bit-flip errors via decoding and to boost the energy scale, while the latter allows us to suppress thermal excitations.  The resulting final Hamiltonian takes the form
\begin{align}
{H}_{\mathrm{QAC}} &=  \alpha \overline{H}_{\mathrm{Ising}} - \beta H_\mathrm{P}\ , \quad \alpha >0, \ \beta \geq 0 \\
\overline{H}_{\mathrm{Ising}} &=  \sum_{i=1}^{\overline{N}} h_i\overline{\sigma_i^z} + \sum_{i<j}^{\overline{N}}  J_{ij} \overline{ \sigma_i^z \sigma_j^z}\ , \quad  H_\mathrm{P} = \sum_{i=1}^{\overline{N}} \overline{\sigma^z_i}   \sigma^z_{i_\mathrm{P}}  \notag \ ,
\end{align}
where $\overline{\sigma^z_i} =\sum_{k=1}^{K-1}  \sigma^z_{i_{k}},\  \overline{\sigma_i^z \sigma_j^z} =  \sum_{k=1}^{K-1}\sigma^z_{i_k}\sigma^z_{j_k}$ are (scaled) logical $\sigma_i^z$ and $\sigma_i^z \sigma_j^z$ operators respectively.  The QAC strategy treats $K-1$ of the $K$ qubits comprising the $i$th logical qubit as ``problem'' qubits which encode the original problem into a repetition code, while the remaining ``penalty'' qubit $i_\mathrm{P}$ is used to implement the energy penalty. This is illustrated in Fig.~\ref{fig:1b} for the DW2.  The parameter $\alpha$ is the problem scale factor, through which we can control the effective noise level on the logical Ising Hamiltonian $\overline{H}_{\mathrm{Ising}}$. The penalty term $H_{\mathrm{P}}$ enforces a ferromagnetic coupling between the problem qubits and their penalty qubit, thus aligning them in agreement and forcing errors that do not commute with $\sigma^z$ to pay an energy penalty. This compensates for the fact that the repetition code can only be used to detect and correct bit-flip errors. However, note that if the dominant dephasing is in the instantaneous energy eigenbasis rather than the computational basis then the adiabatic algorithm is not adversely affected by phase errors \cite{childs_robustness_2001,PhysRevA.79.022107,ABLZ:12-SI}. The parameter $\beta$ is the penalty scale factor, which we optimize to maximize the success probability for each problem instance, balancing $H_{\mathrm{P}}$ against $\overline{H}_{\mathrm{Ising}}$ \cite{PAL:13}. We note that we cannot at the same time encode $H_X$ since this would require $N$-body interactions, and that $\alpha,\beta \leq 1$ on the DW2, thus precluding energy penalty strategies of the type suggested in Refs.~\cite{jordan2006error,Bookatz:2014uq}.  

If the copies are statistically independent in the C strategy and each succeeds with probability $p$, then the probability that at least one copy will succeed is $1-(1-p)^K$, which is greater than $p$ if $0<p<1$ and $K\geq 2$, so C will improve performance compared to a single copy of the same problem instance. In Ref.~\cite{PAL:13} we established that the QAC strategy outperforms the C strategy for sufficiently long antiferromagnetic chains. \ignore{Will QAC continue to outperform C for sufficiently large and hard optimization problems? We take up this question next.} The pertinent question, then, is whether QAC will continue to outperform C for sufficiently large and hard optimization problems.\\

\begin{figure*}[t]
\subfigure[\ ]{\includegraphics[width=0.49\textwidth]{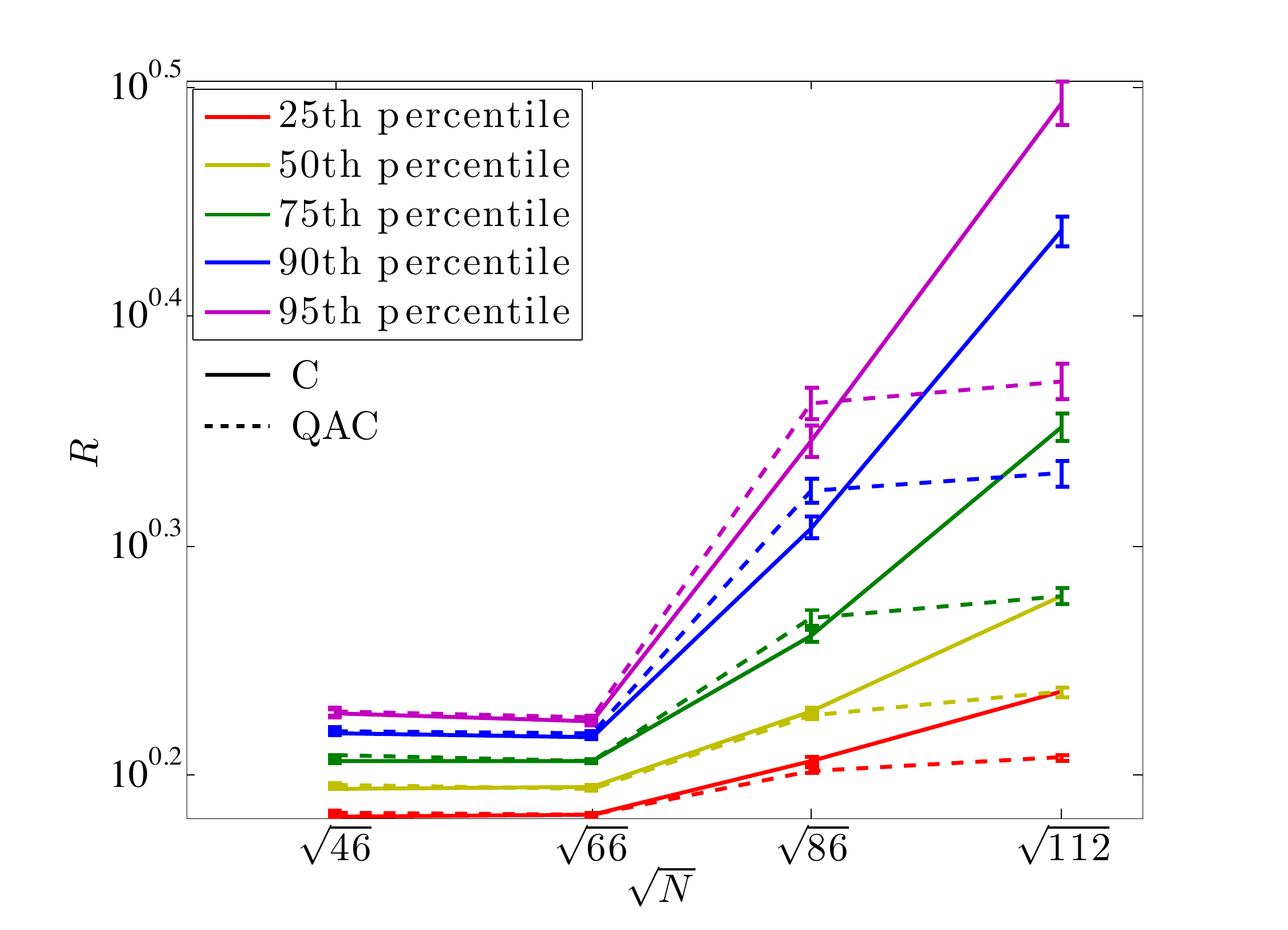} \label{fig:scaling_a1}}
\subfigure[\ ]{\includegraphics[width=0.49\textwidth]{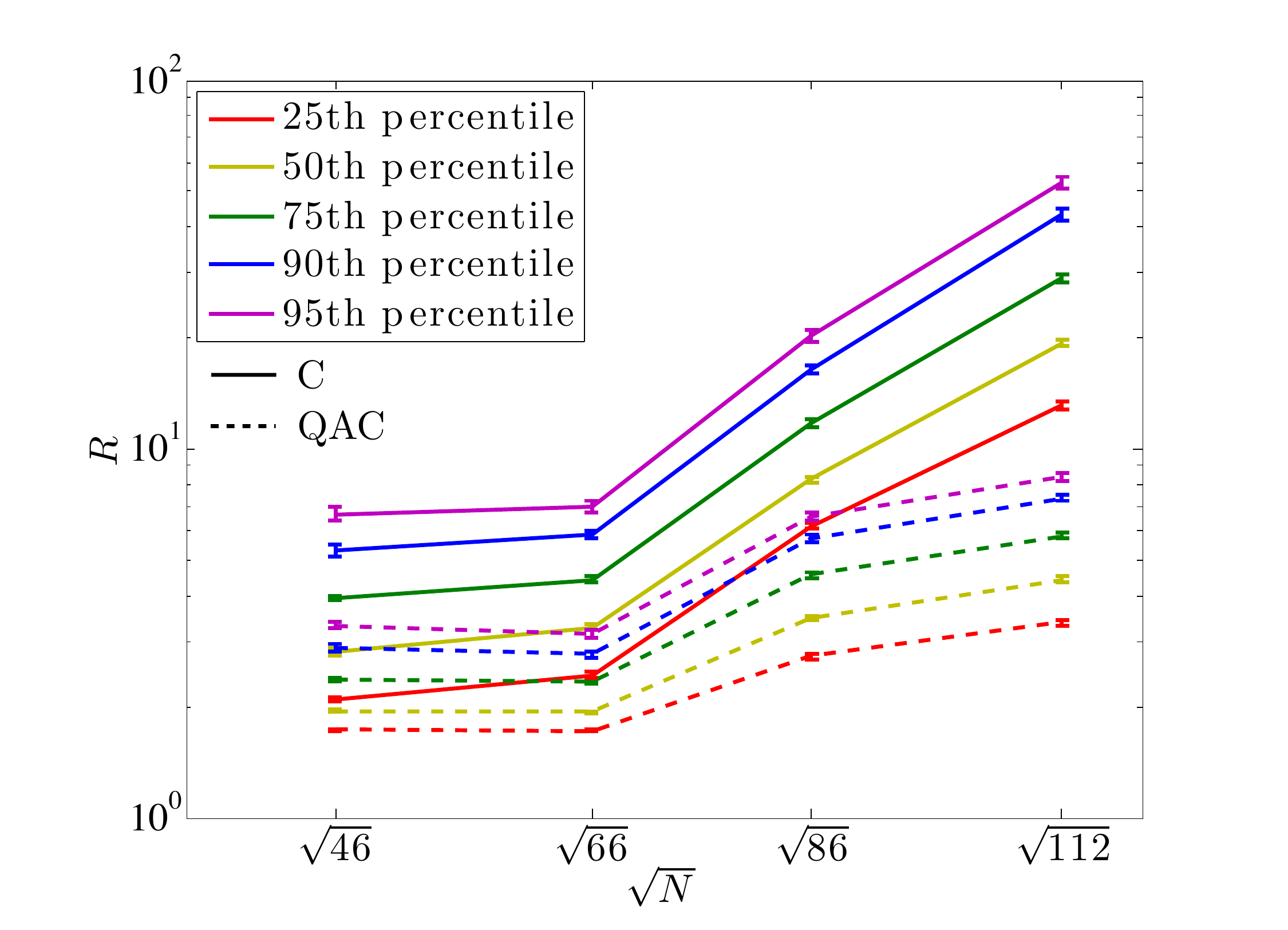} \label{fig:scaling_a05}}
\subfigure[\ ]{\includegraphics[width=0.49\textwidth]{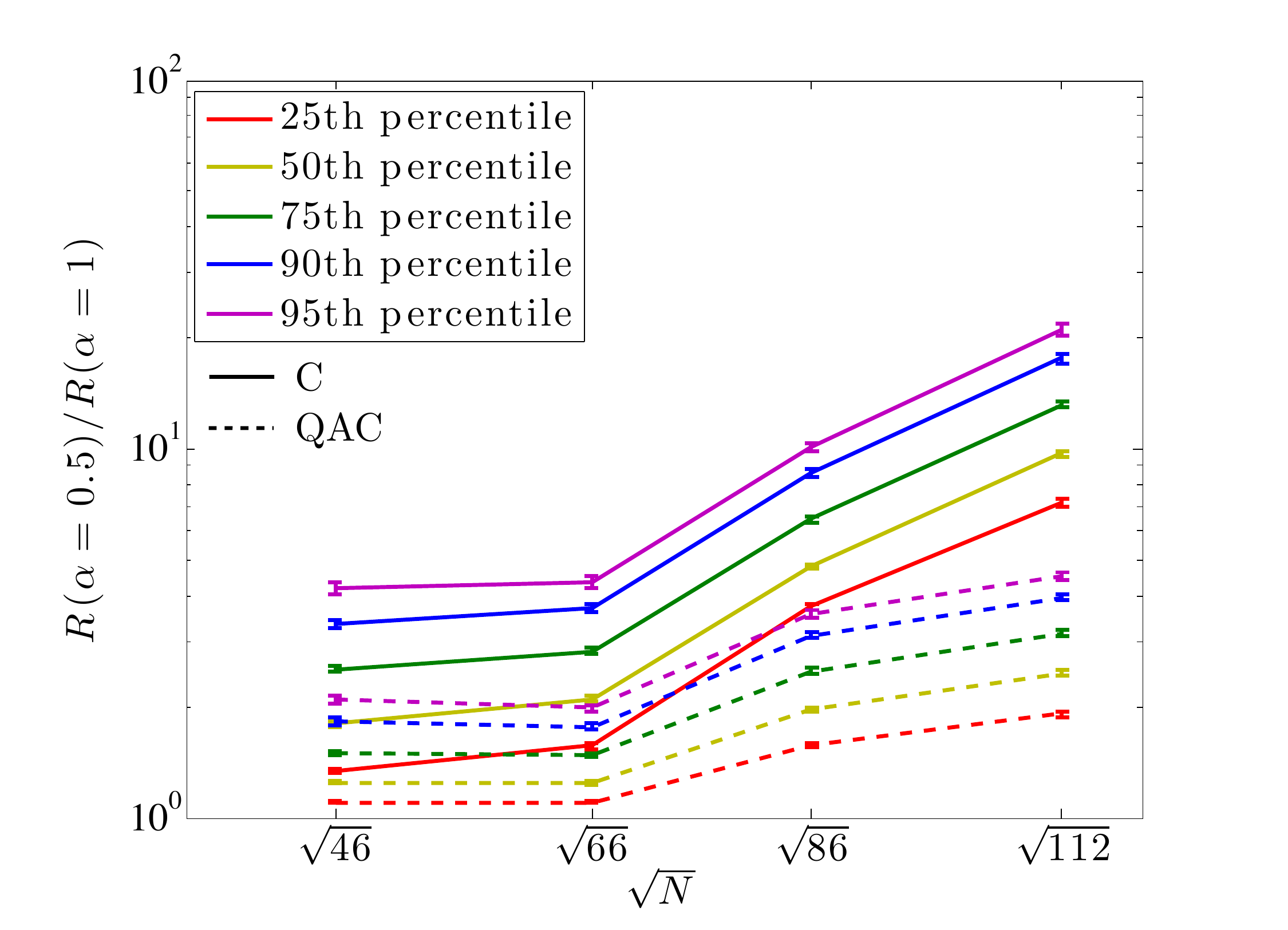}
\label{fig:ratio_scaling}}
\subfigure[\ ]{\includegraphics[width=0.49\textwidth]{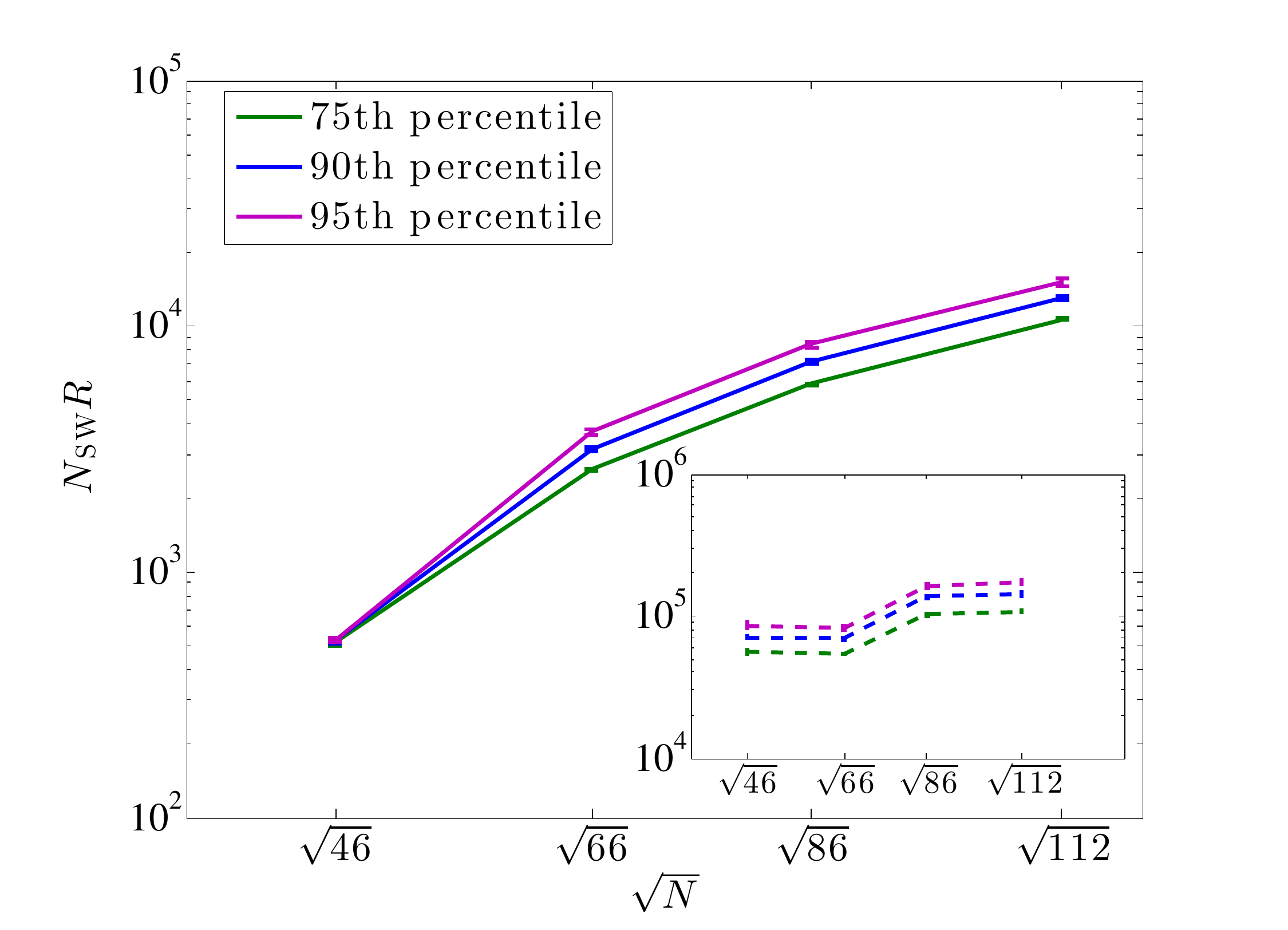}
\label{fig:SAScaling}}
\caption{(Color online) {Problem size dependence of the expected number of annealing runs $R$ for the QAC and C strategies.} (a)  For $\alpha=1$. (b) For $\alpha=0.5$. Various percentiles of problem hardness are included, as shown in the legend. The QAC strategy uses both problem group decoding and logical group decoding, while the C strategy only uses problem group decoding, as discussed in Appendix \ref{app:Methods}.  We observe better performance for the QAC strategy at all percentiles for the largest problem size. (c) Scaling of the ratio of $R$ values for $\alpha=0.5$ and $\alpha = 1$.  The performance of the QAC strategy decreases by a factor of less than $3$ at the hardest percentile and largest problem size, whereas the C strategy's performance decreases by a factor of close to $20$. (d) Scaling of QAC for the SSSV model without coupling noise.  Solid lines use an optimized number of sweeps $N_{\mathrm{SW}}$ for each problem instance, i.e., $N_{\mathrm{SW}}$ is chosen to minimize the expected number of annealing runs $N_{\mathrm{SW}} R$, while the dashed lines in the inset show the scaling for a fixed $N_{\mathrm{SW}} = 10^5$.}
\label{fig:scaling}
\end{figure*}

\subsection{Success probability for random Ising instances} 
Earlier work studied the performance of the D-Wave processors on random Ising problem instances of increasing size $N$ without error correction \cite{speedup,q108} by embedding these problems on the physical ``Chimera" connectivity graph shown in Fig.~\ref{fig:graphs}(a). We now present the results of our study of similar problem instances, using the C and QAC strategies. When the Chimera graph is contracted by replacing each set of $4$ physical qubits by the corresponding logical qubit we obtain the \emph{logical} connectivity graph shown in Fig.~\ref{fig:graphs}(b).
For each problem size $\overline{N}$ we generated $1000$ instances in which all local fields $h_i=0$ and the couplings $J_{ij}$ were specified by drawing uniformly at random from the set $\pm\{\frac{1}{6},\dots,\frac{5}{6},1\}$, slightly above the $\Delta J_{ij} \sim \frac{1}{7}$ precision limit of the DW2 \cite{Trevor}. These instances were embedded on the actual logical connectivity graph shown in Fig.~\ref{fig:graphs}(c), comprising only perfect logical qubits, defined as consisting of $4$ physical qubits (in contrast, an imperfect logical qubit consists of only the $3$ problem qubits, without the penalty qubit).
We ran each instance $1000$ times using different gauges to reduce systematic errors (see Appendix \ref{app:Methods}) and counted the number of times a state with the correct logical ground state energy was found, and we defined this as the success probability for that instance (see Appendix \ref{app:Methods} for more details). Note that we distinguish between the logical and physical ground states: the former is the ground state corresponding to the Ising instance defined in terms of the $\overline{N}$ logical qubits, while the latter is the ground state over the $N$ physical qubits. We also ran each instance with $\beta \in \{0.1,0.2,0.3,0.4,0.5\}$ and defined the optimal $\beta$ as the value $\beta_\textrm{opt}$ that maximized the success probability after gauge-averaging (see Appendix \ref{app:Methods} for more details).

Since $9$ of the physical qubits are missing in the DW2 hardware graph [see Fig.~\ref{fig:graphs}(a)],  the maximum problem size we can study while implementing the full $4$-copy C strategy is $\overline{N} = 112$. This maximal encoded graph has holes in the regular structure due to the missing physical qubits that reduce the number of couplings for some of the QAC logical qubits [see Fig.~\ref{fig:graphs}(c)].  
Figure~\ref{fig:scatter112} shows the performance of QAC \textit{vs} C in this case, where all $\overline{N}=112$ perfect logical qubits were used. Answering our earlier question, we find that QAC is a better strategy choice than C for the overwhelming majority of problem instances. 
We discuss the reason for the appearance of a small number of instances in which C outperforms QAC in Appendix \ref{app:Methods}. 

\begin{figure*}[t]
\subfigure[\ ]{\includegraphics[width=0.32\textwidth]{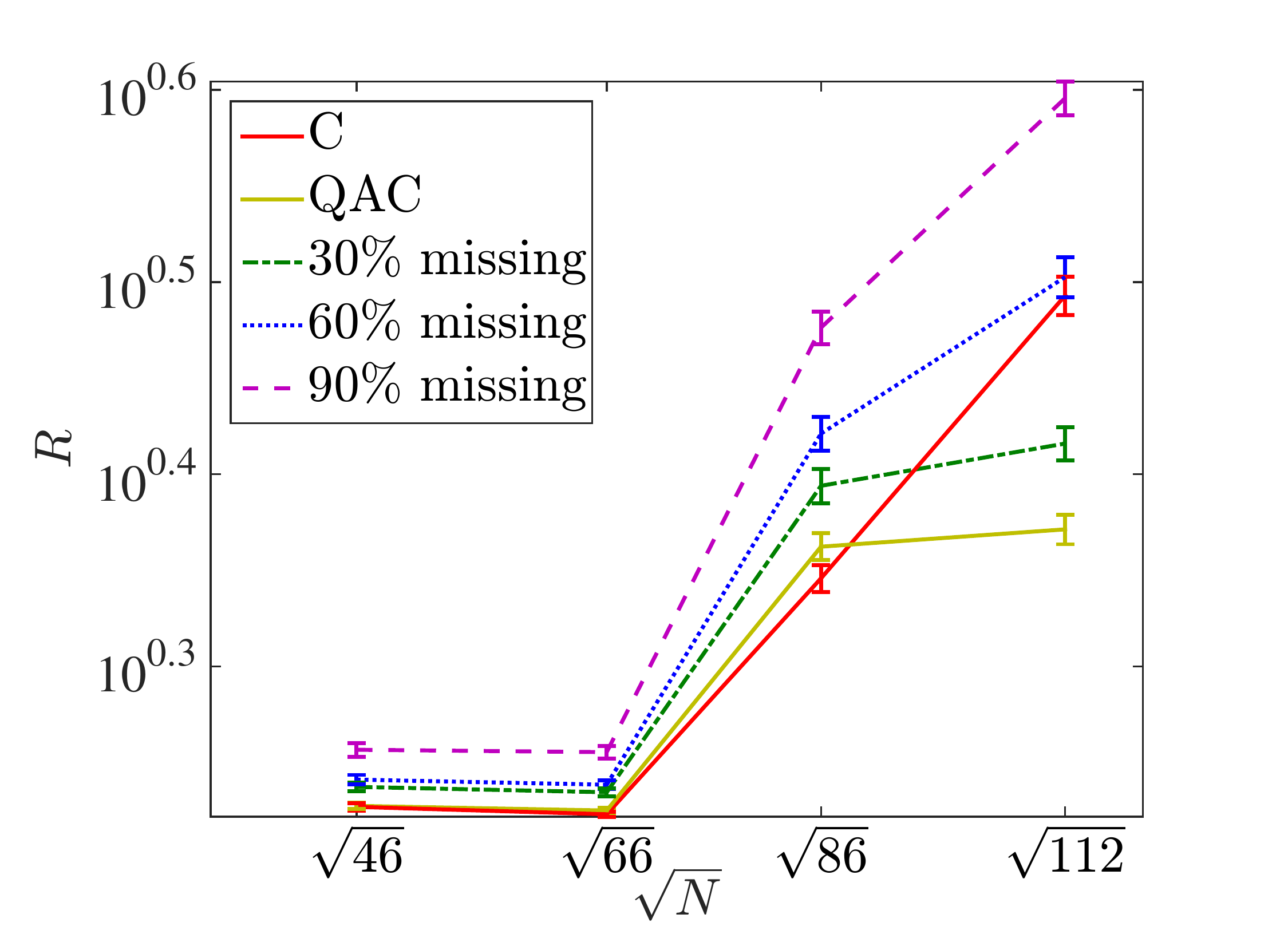} \label{fig:robust_origopt}}
\subfigure[\ ]{\includegraphics[width=0.32\textwidth]{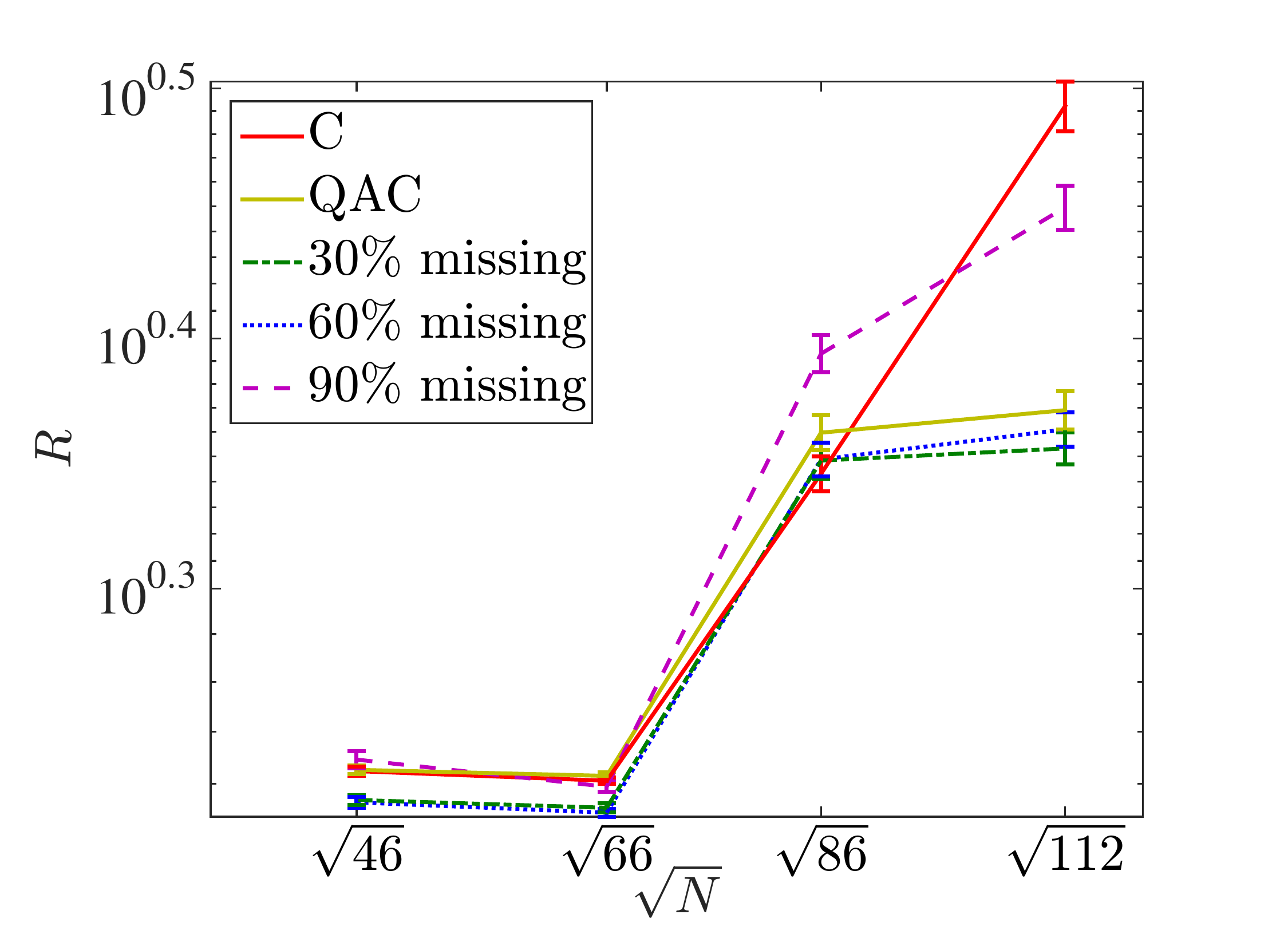} \label{fig:robust_reopt}}
\subfigure[\ ]{\includegraphics[width=0.32\textwidth]{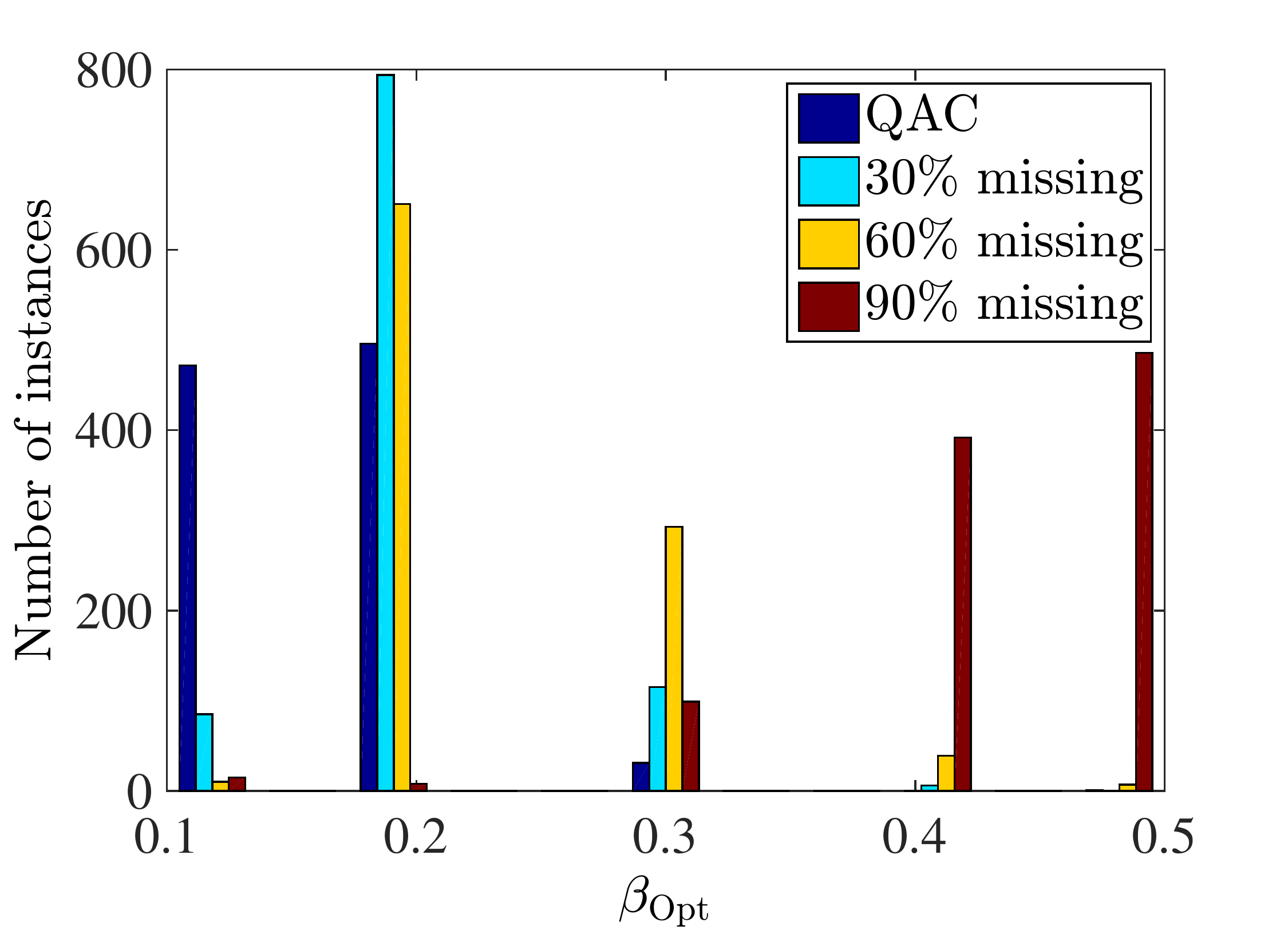} \label{fig:robust_beta2}}
\caption{(Color online) {Robustness of QAC to missing penalty qubits.} Effect of missing penalty qubits for $\alpha=1$ on the $95$th percentile for (a) $\beta_\textrm{opt}$ held constant at its original QAC value, and (b) $\beta$ optimized for the number of missing qubits. The C strategy (solid red) and QAC strategy (solid yellow) lines are the same data that were displayed in Fig.~\ref{fig:scaling_a1}. The new dash-dotted green, dotted blue, and dashed purple series show the effects of randomly removing $30\%$, $60\%$, and $90\%$ of the penalty qubits, respectively. For (a), QAC with $30\%$ loss continues to outperform C, but not at the higher percentages.  For (b), performance at $30\%$ and $60\%$ loss tracks the original QAC closely, suggesting the code is highly resilient to penalty qubit loss if the penalty magnitude is adjusted accordingly. (c) The histogram displays the optimized $\beta$ values from the trial set $\{0.1,0.2,0.3,0.4,0.5\}$, for $1000$ instances. The true optimal $\beta$ for the perfect QAC scheme lies between $0.1$ and $0.2$, but for $30\%$ of penalty qubits missing the optimum shifts to $0.2$, for $60\%$ missing between $0.2$ and $0.3$, and for $90\%$ missing to the maximum allowed value.}
\label{fig:robustness}
\end{figure*}

\subsection{Dependence on problem size}
 In order to quantify performance with respect to the hardness of the Ising instances for DW2, we consider the dependence on $\overline{N}$ of the expected number of annealing runs necessary to observe a success (a logical ground state) at least once with 99\% probability \cite{q108,speedup}:
\beq \label{eqt:T2S}
R = \frac{\log(1 - 0.99)}{\log(1 - p_\mathrm{S})} \ ,
\eeq
where $p_\mathrm{S}$ is the probability of success.  For the C strategy, $p_{\mathrm{S}}$ corresponds to observing a logical ground state in at least one of the four copies (this is almost the same as running a single copy four times, as discussed in Appendix \ref{sec:CorrelationTest}), which we refer to as problem group decoding.  For the QAC strategy, $p_{\mathrm{S}}$ corresponds to observing a logical ground state after decoding using two complementary decoding strategies, problem group decoding over the three problem qubits and logical group decoding.  Details of the decoding strategies is discussed in Appendix \ref{app:Methods}.

 $R$ is a proxy for the time-to-solution $t_a R$, where $t_a$ is the annealing time, set to $20 \mu s$ (the minimum possible with the DW2) in all our experiments. This annealing time is certainly suboptimal, i.e., too long to enable the extraction of meaningful scaling behavior, as discussed in detail in Ref.~\cite{speedup}.  
Therefore $R$ should not be interpreted as reflecting the true value of the time-to-solution for the different strategies. Instead, we focus on the increasing separation between the values of this quantity for QAC and C as a function of problem size to establish the relative performance of the different strategies. 

Consider first the results for $\alpha=1$, in Fig.~\ref{fig:scaling_a1}. Increasing the problem size corresponds to harder problem instances and requires more runs $R$, with a steeper rise occurring for the higher percentiles. While for the three smallest problem sizes $R$ is similar for the C and QAC strategies, when the problem size becomes sufficiently large there is a statistically significant separation for every percentile between the two strategies. The beneficial effect of QAC becomes much more prominent at $\alpha = 0.5$, as seen in Fig.~\ref{fig:scaling_a05}, where the separation between QAC and C is apparent already at the smallest problem size. By reducing the problem energy scale $\alpha$ we have increased the effect of thermal noise, and by halving the values of the couplings we operate the DW2 device in a regime well below the aforementioned $\Delta J_{ij} \sim \frac{1}{7}$ precision limit.  The susceptibility to programming errors is therefore higher, yet QAC continues to work. 
This is further visualized in Fig.~\ref{fig:ratio_scaling}, where we plot the ratio of $R$ values for $\alpha=0.5$ and $\alpha=1$. The ratio rises much more rapidly as a function of problem size for the C case than for the  QAC case, demonstrating the relative (as well as absolute) stability of the latter to both thermal and control errors. \\

\subsection{Robustness of QAC to Qubit Loss} 
A good code should be robust, and to this end we next study the effect of imperfect logical qubits, by systematically removing penalty qubits from perfect logical qubits.  We repeated our QAC experiments on the same $1000$ random problem instances of each size, but we removed $30\%$, $60\%$, and $90\%$ of the penalty qubits from each instance at random, while keeping all problem couplings intact.  We consider two cases: keeping $\beta_\textrm{opt}$ fixed at its QAC value, as well as re-optimizing $\beta$ for each new fraction of removed penalty qubits.

In Fig.~\ref{fig:robustness} we show that the separation between the C and QAC strategies at the largest problem size persists even when a significant number of penalty qubits are removed.  If $\beta_\textrm{opt}$ is held fixed [Fig.~\ref{fig:robust_origopt}] QAC improves on the C strategy even when $30\%$ of the penalty qubits are removed.  When we allow $\beta$ to be adjusted, up to $60\%$ of the penalty qubits can be removed and QAC still shows an advantage over the C strategy [Fig.~\ref{fig:robust_reopt}].  To achieve this $\beta_\textrm{opt}$ must be increased as more penalty qubits are removed [Fig.~\ref{fig:robust_beta2}].  With fewer penalty qubits, the remaining penalty qubits must couple more strongly to their respective problem qubits in order to maintain the benefits of the QAC strategy. At the $30\%$ level $\beta_\textrm{opt}$ is sharply peaked in Fig.~\ref{fig:robust_beta2}, suggesting that the optimum is determined by the number of available penalty qubits, not the particular problem instance being solved. This is important because the advantages provided by QAC would be diluted if it were necessary to try several values of $\beta$ to solve a relevant problem. Instead, it suffices to pick a single $\beta$ value to solve a new problem.\\

\section{Discussion}
We have demonstrated a substantial performance enhancement using QAC over the classical C strategy. A natural question is to what extent QAC is a truly quantum strategy. We now discuss this from two different angles: a solvable analytical model, and a comparison to the classical SSSV model \cite{SSSV}.\\

\subsection{Analytically solvable model with an optimal $\beta$ value}
 Consider the Ising Hamiltonian on a ring with local fields:
\beq \label{eqt:chain}
H_{1\mathrm{D}} =   - h \sum_{i=1}^d \sigma_i^z - \beta \sum_{i=1}^d \sigma_i^z \sigma_{i+1}^z \ ,
\eeq
where $\sigma_{d+1}^z \equiv \sigma_1^z$. We can reinterpret this as a single qubit ``problem Hamiltonian" $-h \overline{\sigma^z}$ encoded into a distance $d$ repetition code with (scaled) logical operator $\overline{\sigma^z} = \sum_{i=1}^d \sigma^z_i$.  The  ground and excited states are encoded as $\ket{\overline{0}} = \ket{000 \dots 0}$ and $\ket{\overline{1}} = \ket{111 \dots 1}$ respectively.  The spin-spin couplings $\sigma_i^z \sigma_{i+1}^z$ are the stabilizer generators of this repetition code, acting as penalty terms that energetically penalize bit flips. Replacing $H_{\mathrm{Ising}}$ by $H_{1\mathrm{D}}$ in Eq.~\eqref{eq:H(t)} and including a transverse field $H_X$ with annealing schedules $A(t)$ and $B(t)$ gives rise to a quantum annealing evolution designed
to terminate in the $\ket{\overline{0}}$ state. The resulting solvable model is a special case of the QAC encoding, without logical spin-spin coupling. It is also known as the transverse field Ising model for a chain with periodic boundary conditions, where it is commonly written as $H(t) \equiv -h_X \sum_i \sigma^x_i - h_Z \sum_i \sigma^z_i - J \sum_i \sigma_i^z \sigma_{i+1}^z$ \cite{Sachdev:book}, where $h_X =  A(t)$, $h_Z = h B(t) $, and $J = \beta B(t)$.
We can study this model analytically in various limits. Here we present the main results (details of the calculations can be found in Appendix \ref{app:Analytics}). 

{The ${\beta \gg h}$ case}: intuitively, in this case the penalty term overwhelms the problem Hamiltonian, which should be detrimental. More rigorously, in this case, corresponding to $J \gg h_Z $, it is well established that in the thermodynamic limit (i.e., $d \to \infty$), there is a critical point at $h_X = J$, and the energy gap at the critical point scales as $\Delta \sim (h_Z/J)^{8/15}$ \cite{Zamolodchikov:1989,Henkel:1989}. 
Clearly, in this limit increasing the energy penalty $\beta$ makes the problem harder to solve at fixed $t_f$ since the shrinking gap will increase the degree of non-adiabaticity and increase thermal excitations.  Furthermore, the state $\ket{\overline{1}}$, which represents a logical error, is actually the first excited state of the Ising Hamiltonian.  Any population lost to this state cannot be decoded and recovered. Having $\beta \gg h$ is thus undesirable.  

\begin{figure} [t]
\includegraphics[width=\columnwidth]{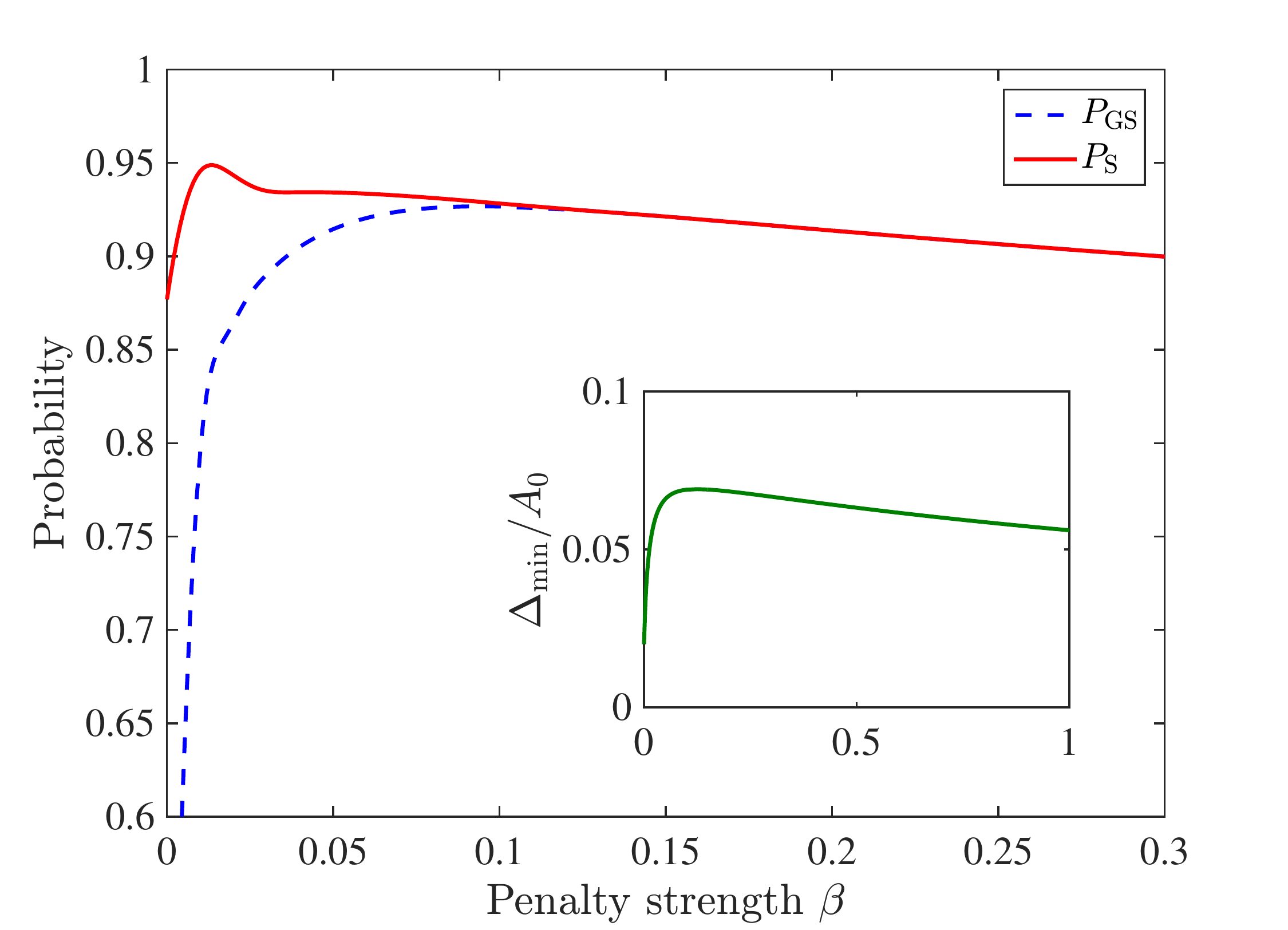}
\caption{(Color online) {Optimal $\beta$ in an open quantum system model.} The main figure shows the final physical ground state probability $P_{\mathrm{GS}}$ and the post-decoding (majority vote with random tie-breaker) success probability $P_{\mathrm{S}}$ of the logical ground state, computed using the adiabatic master equation \cite{ABLZ:12-SI} for quantum annealing using the Hamiltonian~\eqref{eqt:chain} with $d = 4$, with linear annealing schedules $A(t) = A_0 (1-t/t_f)$ and $B(t) = A_0 t/t_f$. Both probabilities exhibit an optimal $\beta$ value.  The inset shows the numerically computed minimum gap between the ground and first excited state as a function of $\beta$.  The peak in $P_{\mathrm{S}}$ and the minimum gap align as expected since a larger minimum gap means fewer transitions out of the ground state.  However, the peak in $P_{\mathrm{GS}}$ occurs earlier, maximizing the population in all decodable states and not just the ground state.  Simulation parameters are: $A_0 = 33.84$ GHz, $t_f = 100$ns, $h = 0.01$, and system bath coupling $g^2 \eta /\hbar^2 = 1.2732 \times 10^{-3}$ (see Eq.~(15) in Ref.~\cite{PAL:13}).}
\label{fig:chains}
\end{figure}

\begin{figure*}
\subfigure[\ ]{\includegraphics[width=0.45\textwidth]{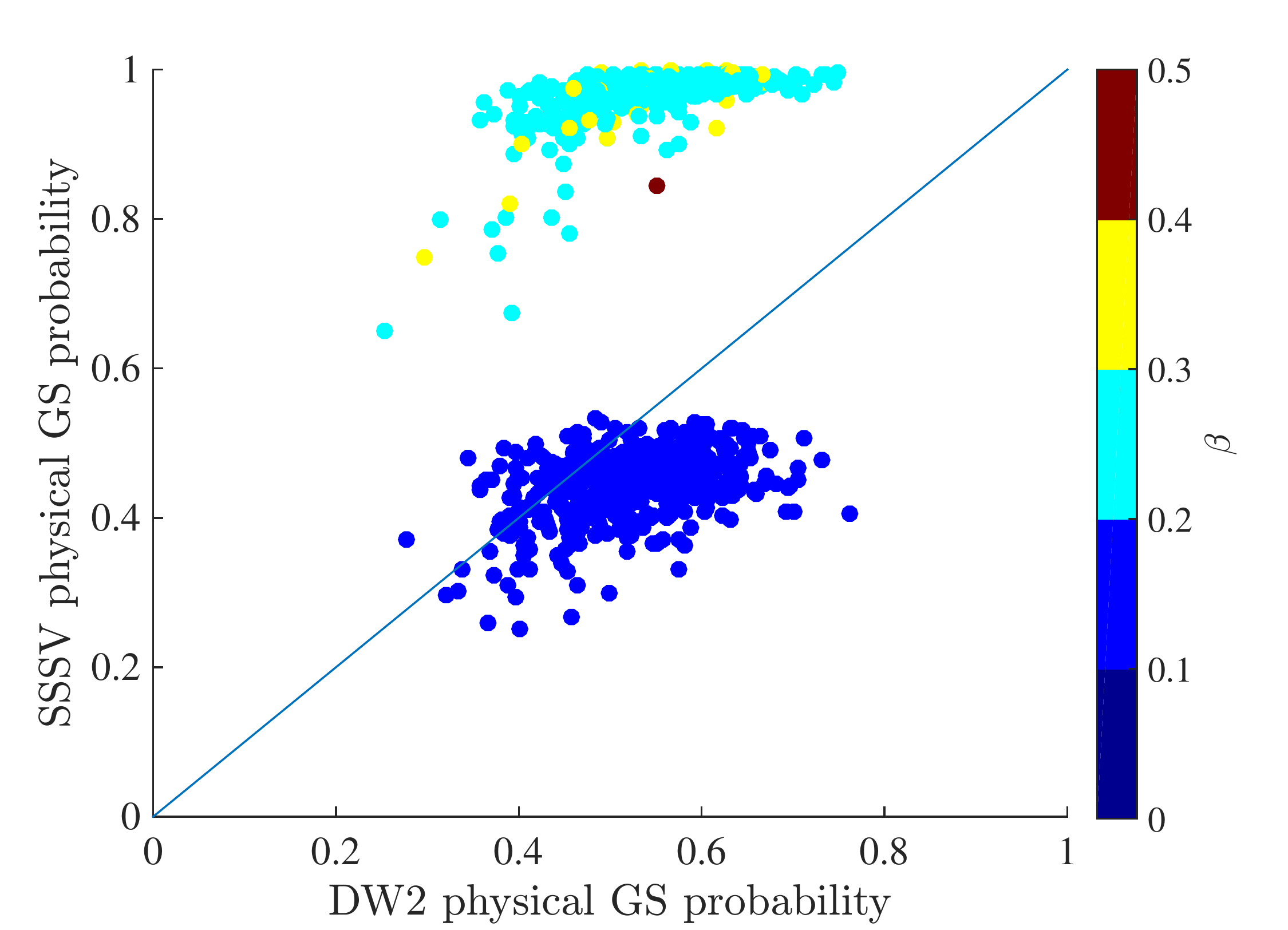}\label{fig:QAC_SSSV1}}
\subfigure[\ ]{\includegraphics[width=0.45\textwidth]{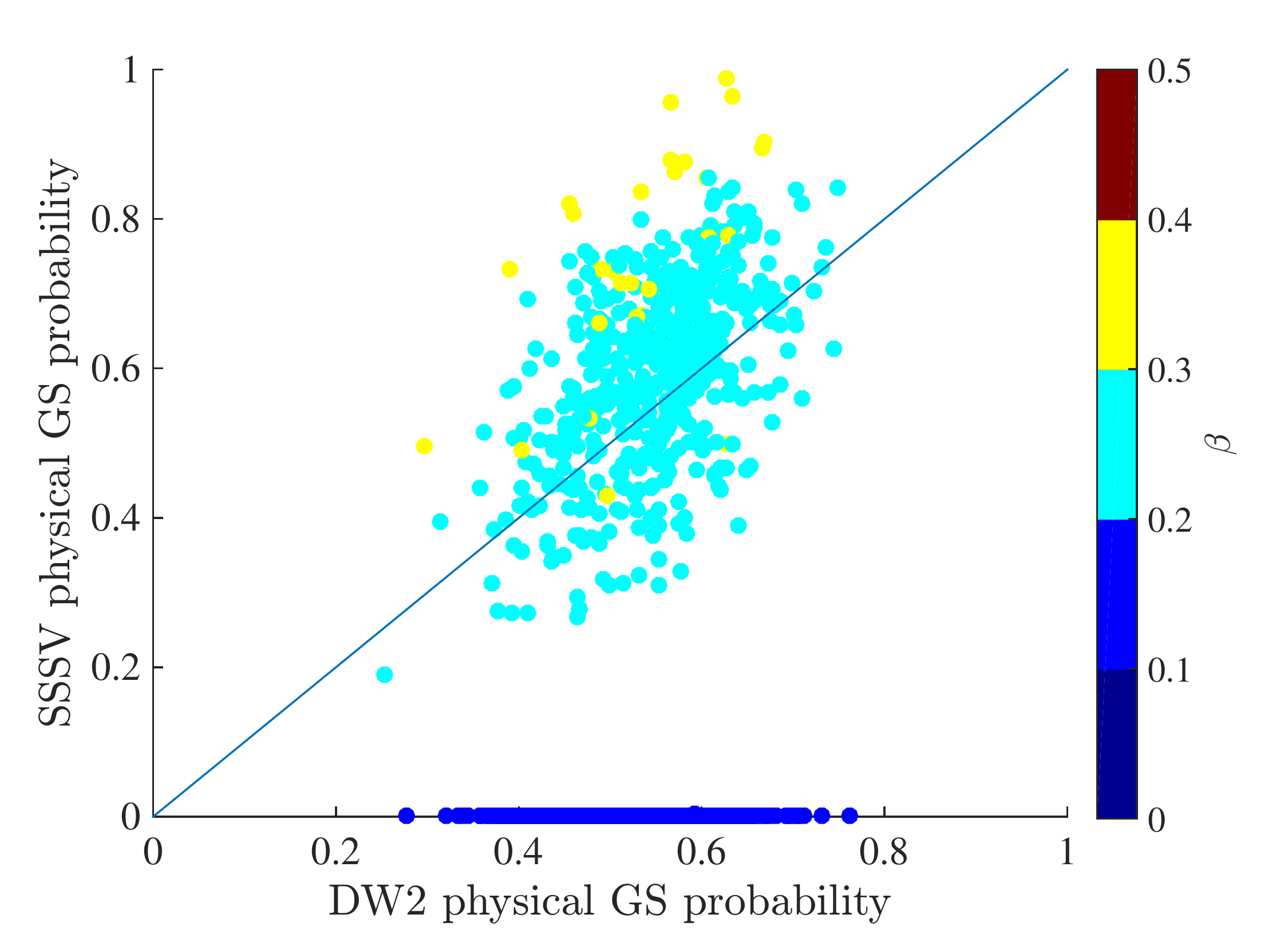}\label{fig:QAC_SSSV2}}
\subfigure[\ ]{\includegraphics[width=0.45\textwidth]{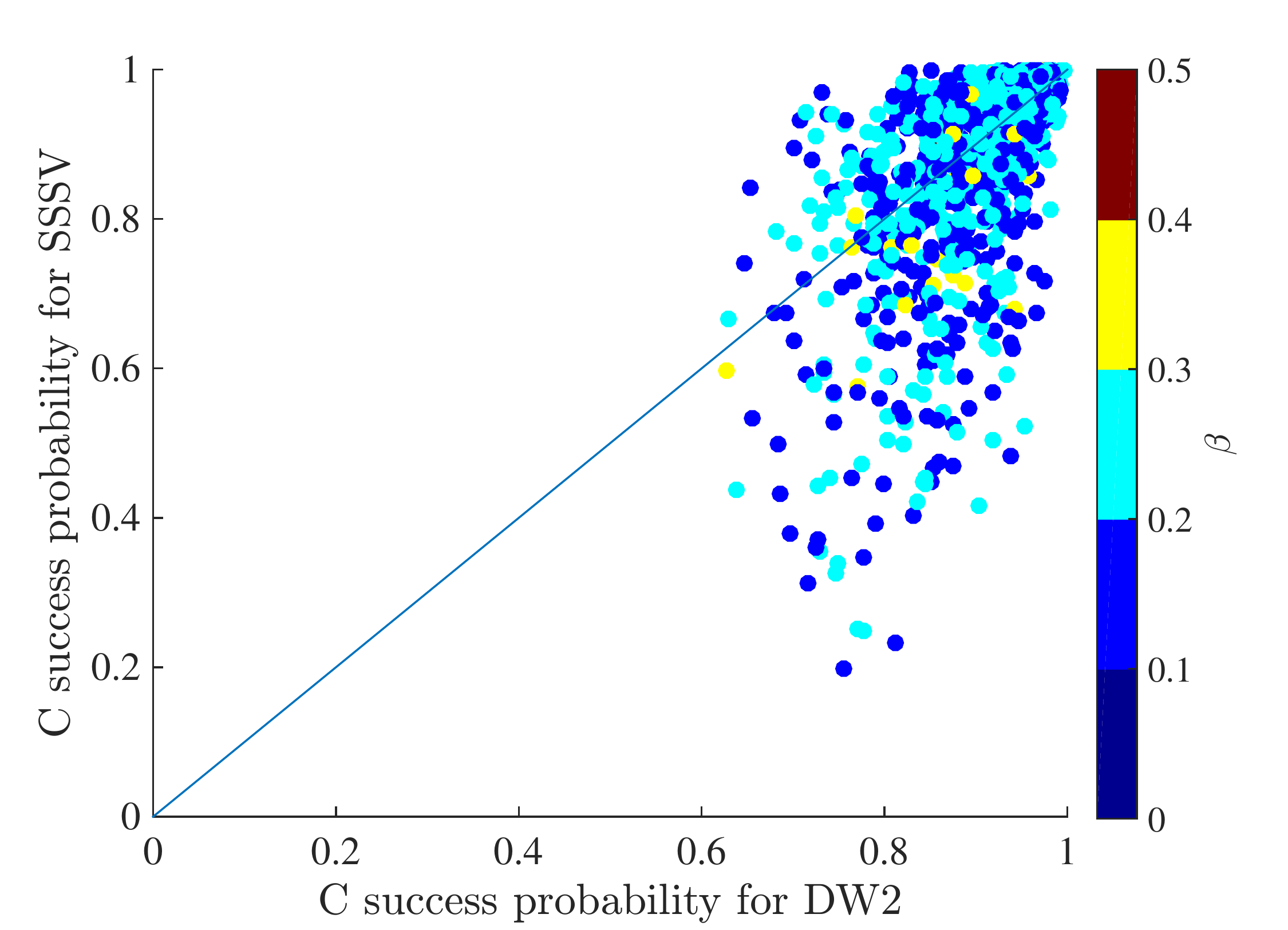}\label{fig:C_SSSV}} 
\subfigure[\ ]{\includegraphics[width=0.45\textwidth]{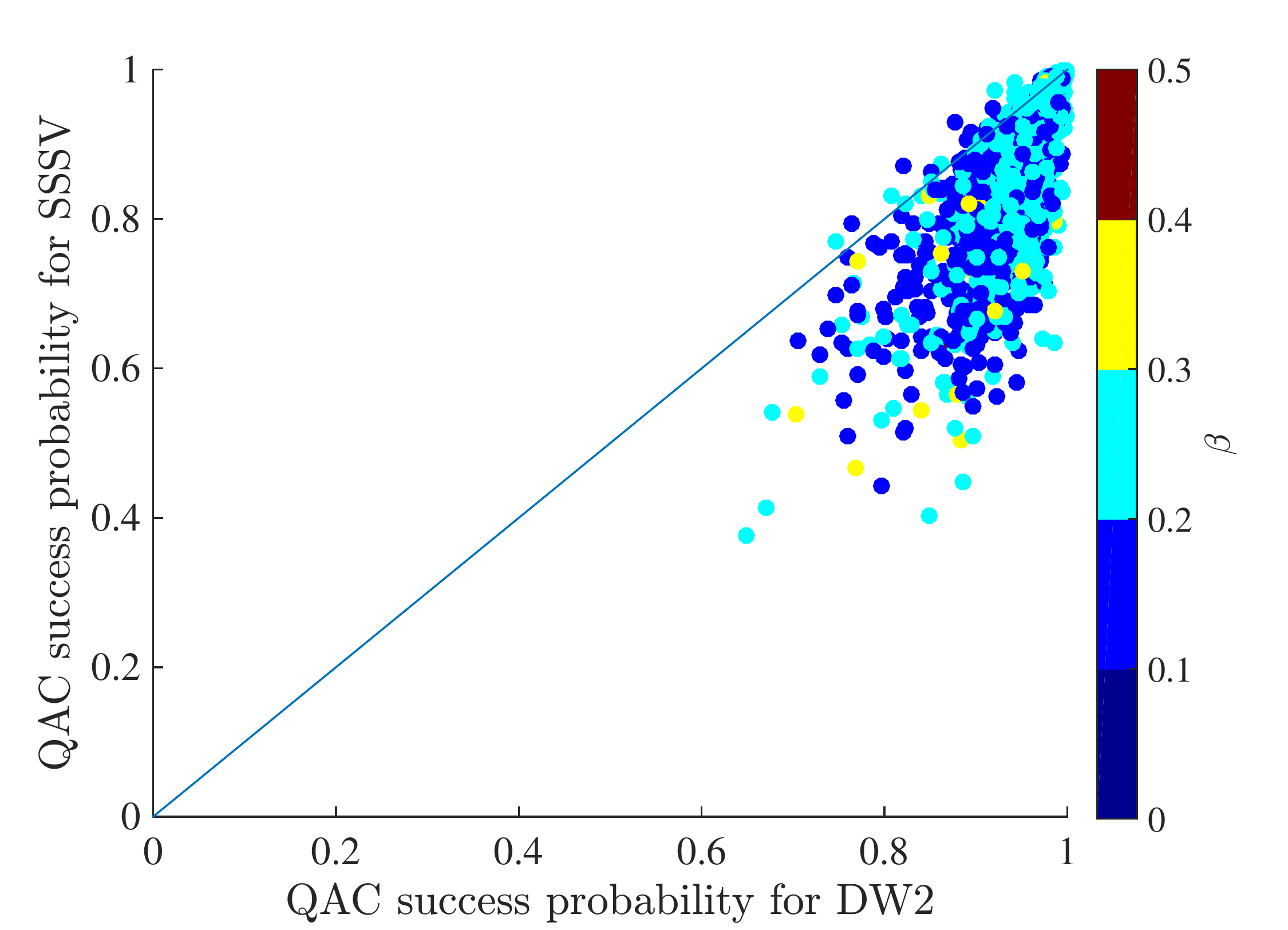}\label{fig:QAC_SSSV}} 
\caption{(Color online) {Quantumness test by comparison to the classical SSSV model.} Shown are the success probability correlations for the physical or logical ground state of the DW2 \textit{vs} SSSV results for $N=448$ physical or $\bar{N} = 112$ logical qubits. (a) QAC, physical, without calibration noise on the couplings ($\sigma = 0$); (b) QAC, physical, with noise ($\sigma = 0.085$); (c) C, logical, with noise ($\sigma = 0.085$); (d) QAC, logical, with noise ($\sigma = 0.085$).  The instances are color-coded according to their $\beta_\textrm{opt}$ as used in the QAC strategy [note that in (c), there is no $\beta$ but we still color-coded the instances by their QAC-optimal $\beta$ values]. For QAC with $\beta_\textrm{opt} = 0$ the $\beta_\textrm{opt}$ value was shifted to $0.1$ to differentiate them from C.  SSSV simulation parameters: temperature $T = 10.56$mK and $1\times 10^5$ Monte Carlo sweeps.  Gaussian noise with standard deviation $\sigma$ was added to the Ising couplings and local fields.}
\label{fig:SSSVCorr}
\end{figure*}

{The ${\beta \ll h}$ case}:   
in the opposite case, where $J \ll h_Z$, the penalty term acts as a perturbation on the problem Hamiltonian.  We show analytically in Appendix \ref{app:Analytics} 
that the introduction of the penalty term increases the gap and shifts it to earlier in the evolution provided $h<\sqrt{2}$.  An increased gap means that the evolution is more adiabatic and thermal excitations are suppressed at fixed $t_f$, suggesting that the introduction of a small $\beta$ improves the adiabaticity of the evolution.  Furthermore, the low energy excited states of the Ising Hamiltonian correspond to a small number of spin flips, which can be corrected via decoding.

There is thus clearly an optimal $\beta>0$ value in the large $d$ limit, a fact that helps to explain the observation of an optimal ${\beta}$ value in our experiments and in Ref.~\cite{PAL:13}. To address this in a more realistic model, we performed numerical simulations for finite $d$ using an adiabatic Markovian master equation of an open quantum system \cite{ABLZ:12-SI}. This master equation has been used extensively in related work where it was shown to be a good model of the D-Wave device \cite{q-sig,q-sig2} (it is briefly reviewed in the Methods section of Ref.~\cite{PAL:13}). 
We use it here to model a qubit encoded into a distance $d = 4$ classical repetition code as given by Eq.~\eqref{eqt:chain}.  
As shown in Fig.~\ref{fig:chains}, both the physical ground state probability $P_{\mathrm{GS}}$ and the logical ground state success probability $P_{\mathrm{S}}$ exhibit a peak as a function of $\beta$.  The position of the $P_{\mathrm{GS}}$ peak coincides with the peak in the minimum energy gap, whereas $P_{\mathrm{S}}$ also depends on the decodability of the excited state spectrum, to which a significant amount of population is lost. The fact that $P_{\mathrm{S}} \geq P_{\mathrm{GS}}$ for all $\beta$ shows that decoding is always beneficial. 

These results of the closed and open system models thus lead us to associate the improved performance of QAC with $\beta>0$ to both the enhancement of the energy gap and the decodability of low-energy excited states.\\

\subsection{QAC for a classical model of the D-Wave device} 
\label{sec:results2}
%
Although we have demonstrated that the QAC strategy provides a significant performance advantage over the C strategy, and that quantum models exhibit an optimal penalty strength just as observed in our experiments, it is difficult to establish to what extent quantum effects are responsible for the overall success of QAC.  To further address this question, we tested the efficacy of the same strategy on a classical model (SSSV) that has been successfully used \cite{SSSV} to reproduce the D-Wave One physical ground state probabilities on random Ising instances reported in Ref.~\cite{q108}.  The SSSV model replaces the qubits by classical planar rotors whose dynamics are governed by Eq.~\eqref{eq:H(t)} with the replacements $\sigma_i^x \mapsto \sin\theta_i$ and $\sigma_i^z \mapsto \cos\theta_i$, with Monte Carlo updates for each angle $\theta_i$. Though there is evidence that the SSSV model does not correctly capture experiments on specially designed Ising instances of up to $20$ qubits \cite{q-sig2}, it remains an excellent classical model for random Ising instances on larger numbers of qubits, setting a high bar for genuine quantum effects. In particular, if the D-Wave device is well described by the SSSV model, then the QAC strategy applied to this model should give the same performance enhancement as we observed experimentally. If, on the other hand, quantum effects play an important role in the performance of the QAC strategy on the D-Wave device, then the SSSV model will not benefit from these effects.

To test this we numerically solved the SSSV model applied to the same set of instances as we used on the DW2, with the same $\beta_\textrm{opt}$ values, and we compared the probability of observing the physical ground state on each (details of the algorithm used here can be found in Ref.~\cite{q-sig2}). As seen in Fig.~\ref{fig:QAC_SSSV1}, we find that  the SSSV physical ground state probabilities separate the instances into two sets according to $\beta_\textrm{opt}$, with a higher probability for those with large $\beta_\textrm{opt}$ values.  This is unlike the DW2 results, where the instances cluster irrespectively of the relevant $\beta_\textrm{opt}$. 

Since it is known that the D-Wave device is susceptible to calibration noise and it is important to account for this when comparing its results to the SSSV model \cite{q-sig2,SSSV-comment}, we checked the robustness of this conclusion by including a variable amount of Gaussian noise with standard deviation $\sigma$ in the range $[0,0.085]$ (the range found to be relevant in Ref.~\cite{q-sig2}) on the couplings $J_{ij}$, and we also included local fields $h_i\sim\mathcal{N}(0,\sigma)$. As shown in Fig.~\ref{fig:QAC_SSSV2}, depending on the amount of calibration noise introduced, we can only correlate the SSSV results with the DW2 results for a subset of the instances.
We find that with no calibration noise, we can correlate the $\beta_\textrm{opt} = 0.1$ instances, while with calibration noise, we can correlate those with $\beta_\textrm{opt} = 0.2$.  Since we know that  calibration noise is present \cite{q-sig2}, the latter case is the more realistic fit.  If we accept this conclusion, then Fig.~\ref{fig:QAC_SSSV2} shows that the SSSV model almost never finds the physical ground state for the $\beta_\textrm{opt} = 0.1$ instances, while the DW2 still has a substantial probability of finding the physical ground state. While increasing the number of sweeps helps to improve the SSSV success probabilities for all quantities presented, this conclusion is robust to varying the temperature and the number of sweeps (Monte Carlo updates per spin) in the range [10mK, 25mK] and [50k, 200k] respectively.

Furthermore, Fig.~\ref{fig:C_SSSV} shows that the separation by $\beta_\textrm{opt}$ does not appear when using the C strategy, suggesting that the separation is an effect of the physical, rather than the logical Ising instance.  Figure~\ref{fig:QAC_SSSV} shows that the separation vanishes when we decode using the QAC strategy, suggesting that the SSSV model suffers from many correctable errors for these instances. 
 While these results of course do not amount to a proof of quantumness, they do support the notion that quantum effects play a relevant role in separating the SSSV model from the DW2 results observed in our QAC experiments.

The SSSV model is useful in another sense. In our earlier discussion of the scaling results shown in Figs.~\ref{fig:scaling_a1} and \ref{fig:scaling_a05} we stressed that they do not exhibit the correct scaling curves for the DW2 device, since the minimal possible annealing time of $20\mu$s is too long and is hence suboptimal \cite{speedup}.  
To illustrate the importance of this point, we show in Fig.~\ref{fig:SAScaling} the scaling of the noise-free SSSV model with and without optimizing the number of annealing sweeps.  The scaling curves differ substantially, and one would be misled about the true scaling by the suboptimal curves. 
Finally, we note that the general shape of the SSSV no-optimization curves is similar to the experimental curves, indicating that the QAC strategy has reduced the effect of control errors on the DW2 device.\\

\section{Conclusions}
 We have demonstrated that QAC substantially enhances the performance of an experimental quantum annealer, boosting its success probabilities on hard random Ising instances well beyond a classical repetition code strategy using equal hardware resources. Moreover, we have demonstrated that quantum effects appear to play an operational role in the success of the QAC strategy.
These results demonstrate that the encouraging conclusions concerning the beneficial role of QAC based on antiferromagnetic chains reported in Ref.~\cite{PAL:13} extend to hard computational problems as well, with increasing benefit as problem instances grow in size and hardness.

While extrapolation of our DW2 scaling results to larger problem sizes would be inappropriate due to the issue of suboptimal annealing times, the improvement in performance relative to a simple classical repetition strategy validates the importance of QAC, especially for benchmarking studies. Future studies will explore harder problem instances, both larger and with a non-zero spin glass phase critical temperature \cite{2014Katzgraber}, where the optimal annealing time will be greater than the minimum currently allowed by the D-Wave device, allowing us to extract its true scaling under QAC. Our work reinforces the importance of the inclusion of error correction in quantum annealing, with the ultimate goal of demonstrating a quantum speedup.\\

\acknowledgements
This research was supported by the Lockheed Martin Corporation, by ARO-MURI Grant No. W911NF-11-1-0268, and by ARO-QA Grant No. W911NF-12-1-0523. This research used resources of the Oak Ridge Leadership Computing Facility at the Oak Ridge National Laboratory, which is supported by the Office of Science of the U.S. Department of Energy under Contract No. DE-AC05-00OR22725.  Computation for the work described in this paper was supported by the University of Southern California's Center for High-Performance Computing (http://hpcc.usc.edu).


\begin{thebibliography}{54}%
\makeatletter
\providecommand \@ifxundefined [1]{%
 \@ifx{#1\undefined}
}%
\providecommand \@ifnum [1]{%
 \ifnum #1\expandafter \@firstoftwo
 \else \expandafter \@secondoftwo
 \fi
}%
\providecommand \@ifx [1]{%
 \ifx #1\expandafter \@firstoftwo
 \else \expandafter \@secondoftwo
 \fi
}%
\providecommand \natexlab [1]{#1}%
\providecommand \enquote  [1]{``#1''}%
\providecommand \bibnamefont  [1]{#1}%
\providecommand \bibfnamefont [1]{#1}%
\providecommand \citenamefont [1]{#1}%
\providecommand \href@noop [0]{\@secondoftwo}%
\providecommand \href [0]{\begingroup \@sanitize@url \@href}%
\providecommand \@href[1]{\@@startlink{#1}\@@href}%
\providecommand \@@href[1]{\endgroup#1\@@endlink}%
\providecommand \@sanitize@url [0]{\catcode `\\12\catcode `\$12\catcode
  `\&12\catcode `\#12\catcode `\^12\catcode `\_12\catcode `\%12\relax}%
\providecommand \@@startlink[1]{}%
\providecommand \@@endlink[0]{}%
\providecommand \url  [0]{\begingroup\@sanitize@url \@url }%
\providecommand \@url [1]{\endgroup\@href {#1}{\urlprefix }}%
\providecommand \urlprefix  [0]{URL }%
\providecommand \Eprint [0]{\href }%
\providecommand \doibase [0]{http://dx.doi.org/}%
\providecommand \selectlanguage [0]{\@gobble}%
\providecommand \bibinfo  [0]{\@secondoftwo}%
\providecommand \bibfield  [0]{\@secondoftwo}%
\providecommand \translation [1]{[#1]}%
\providecommand \BibitemOpen [0]{}%
\providecommand \bibitemStop [0]{}%
\providecommand \bibitemNoStop [0]{.\EOS\space}%
\providecommand \EOS [0]{\spacefactor3000\relax}%
\providecommand \BibitemShut  [1]{\csname bibitem#1\endcsname}%
\let\auto@bib@innerbib\@empty
\bibitem [{\citenamefont {Lidar}\ and\ \citenamefont
  {Brun}(2013)}]{Lidar-Brun:book}%
  \BibitemOpen
  \bibinfo {editor} {\bibfnamefont {D.A.}\ \bibnamefont {Lidar}}\ and\ \bibinfo
  {editor} {\bibfnamefont {T.A.}\ \bibnamefont {Brun}},\ eds.,\ \href
  {http://www.cambridge.org/9780521897877} {\emph {\bibinfo {title} {Quantum
  Error Correction}}}\ (\bibinfo  {publisher} {Cambridge University Press},\
  \bibinfo {address} {{Cambride, UK}},\ \bibinfo {year} {2013})\BibitemShut
  {NoStop}%
\bibitem [{\citenamefont {Finnila}\ \emph {et~al.}(1994)\citenamefont
  {Finnila}, \citenamefont {Gomez}, \citenamefont {Sebenik}, \citenamefont
  {Stenson},\ and\ \citenamefont {Doll}}]{finnila_quantum_1994}%
  \BibitemOpen
  \bibfield  {author} {\bibinfo {author} {\bibfnamefont {A.~B.}\ \bibnamefont
  {Finnila}}, \bibinfo {author} {\bibfnamefont {M.~A.}\ \bibnamefont {Gomez}},
  \bibinfo {author} {\bibfnamefont {C.}~\bibnamefont {Sebenik}}, \bibinfo
  {author} {\bibfnamefont {C.}~\bibnamefont {Stenson}}, \ and\ \bibinfo
  {author} {\bibfnamefont {J.~D.}\ \bibnamefont {Doll}},\ }\bibfield  {title}
  {\enquote {\bibinfo {title} {Quantum annealing: A new method for minimizing
  multidimensional functions},}\ }\href {\doibase
  http://dx.doi.org/10.1016/0009-2614(94)00117-0} {\bibfield  {journal}
  {\bibinfo  {journal} {Chemical Physics Letters}\ }\textbf {\bibinfo {volume}
  {219}},\ \bibinfo {pages} {343--348} (\bibinfo {year} {1994})}\BibitemShut
  {NoStop}%
\bibitem [{\citenamefont {Kadowaki}\ and\ \citenamefont
  {Nishimori}(1998)}]{kadowaki_quantum_1998}%
  \BibitemOpen
  \bibfield  {author} {\bibinfo {author} {\bibfnamefont {Tadashi}\ \bibnamefont
  {Kadowaki}}\ and\ \bibinfo {author} {\bibfnamefont {Hidetoshi}\ \bibnamefont
  {Nishimori}},\ }\bibfield  {title} {\enquote {\bibinfo {title} {Quantum
  annealing in the transverse \uppercase{I}sing model},}\ }\href {\doibase
  10.1103/PhysRevE.58.5355} {\bibfield  {journal} {\bibinfo  {journal} {Phys.
  Rev. E}\ }\textbf {\bibinfo {volume} {58}},\ \bibinfo {pages} {5355}
  (\bibinfo {year} {1998})}\BibitemShut {NoStop}%
\bibitem [{\citenamefont {Santoro}\ \emph {et~al.}(2002)\citenamefont
  {Santoro}, \citenamefont {Marto\v{n}\'{a}k}, \citenamefont {Tosatti},\ and\
  \citenamefont {Car}}]{Santoro}%
  \BibitemOpen
  \bibfield  {author} {\bibinfo {author} {\bibfnamefont {Giuseppe~E.}\
  \bibnamefont {Santoro}}, \bibinfo {author} {\bibfnamefont {Roman}\
  \bibnamefont {Marto\v{n}\'{a}k}}, \bibinfo {author} {\bibfnamefont {Erio}\
  \bibnamefont {Tosatti}}, \ and\ \bibinfo {author} {\bibfnamefont {Roberto}\
  \bibnamefont {Car}},\ }\bibfield  {title} {\enquote {\bibinfo {title} {Theory
  of quantum annealing of an {I}sing spin glass},}\ }\href {\doibase
  10.1126/science.1068774} {\bibfield  {journal} {\bibinfo  {journal}
  {Science}\ }\textbf {\bibinfo {volume} {295}},\ \bibinfo {pages} {2427--2430}
  (\bibinfo {year} {2002})}\BibitemShut {NoStop}%
\bibitem [{\citenamefont {Morita}\ and\ \citenamefont
  {Nishimori}(2008)}]{morita:125210}%
  \BibitemOpen
  \bibfield  {author} {\bibinfo {author} {\bibfnamefont {Satoshi}\ \bibnamefont
  {Morita}}\ and\ \bibinfo {author} {\bibfnamefont {Hidetoshi}\ \bibnamefont
  {Nishimori}},\ }\bibfield  {title} {\enquote {\bibinfo {title} {Mathematical
  foundation of quantum annealing},}\ }\href
  {http://dx.doi.org/10.1063/1.2995837} {\bibfield  {journal} {\bibinfo
  {journal} {J. Math. Phys.}\ }\textbf {\bibinfo {volume} {49}},\ \bibinfo
  {pages} {125210--47} (\bibinfo {year} {2008})}\BibitemShut {NoStop}%
\bibitem [{\citenamefont {Das}\ and\ \citenamefont
  {Chakrabarti}(2008)}]{RevModPhys.80.1061}%
  \BibitemOpen
  \bibfield  {author} {\bibinfo {author} {\bibfnamefont {Arnab}\ \bibnamefont
  {Das}}\ and\ \bibinfo {author} {\bibfnamefont {Bikas~K.}\ \bibnamefont
  {Chakrabarti}},\ }\bibfield  {title} {\enquote {\bibinfo {title}
  {\textit{Colloquium}: Quantum annealing and analog quantum computation},}\
  }\href {\doibase 10.1103/RevModPhys.80.1061} {\bibfield  {journal} {\bibinfo
  {journal} {Rev. Mod. Phys.}\ }\textbf {\bibinfo {volume} {80}},\ \bibinfo
  {pages} {1061--1081} (\bibinfo {year} {2008})}\BibitemShut {NoStop}%
\bibitem [{\citenamefont {Bapst}\ \emph {et~al.}(2013)\citenamefont {Bapst},
  \citenamefont {Foini}, \citenamefont {Krzakala}, \citenamefont {Semerjian},\
  and\ \citenamefont {Zamponi}}]{Bapst2013}%
  \BibitemOpen
  \bibfield  {author} {\bibinfo {author} {\bibfnamefont {V.}~\bibnamefont
  {Bapst}}, \bibinfo {author} {\bibfnamefont {L.}~\bibnamefont {Foini}},
  \bibinfo {author} {\bibfnamefont {F.}~\bibnamefont {Krzakala}}, \bibinfo
  {author} {\bibfnamefont {G.}~\bibnamefont {Semerjian}}, \ and\ \bibinfo
  {author} {\bibfnamefont {F.}~\bibnamefont {Zamponi}},\ }\bibfield  {title}
  {\enquote {\bibinfo {title} {The quantum adiabatic algorithm applied to
  random optimization problems: The quantum spin glass perspective},}\ }\href
  {\doibase 10.1016/j.physrep.2012.10.002} {\bibfield  {journal} {\bibinfo
  {journal} {Physics Reports}\ }\textbf {\bibinfo {volume} {523}},\ \bibinfo
  {pages} {127 -- 205} (\bibinfo {year} {2013})}\BibitemShut {NoStop}%
\bibitem [{\citenamefont {Farhi}\ \emph {et~al.}(2001)\citenamefont {Farhi},
  \citenamefont {Goldstone}, \citenamefont {Gutmann}, \citenamefont {Lapan},
  \citenamefont {Lundgren},\ and\ \citenamefont {Preda}}]{farhi_quantum_2001}%
  \BibitemOpen
  \bibfield  {author} {\bibinfo {author} {\bibfnamefont {Edward}\ \bibnamefont
  {Farhi}}, \bibinfo {author} {\bibfnamefont {Jeffrey}\ \bibnamefont
  {Goldstone}}, \bibinfo {author} {\bibfnamefont {Sam}\ \bibnamefont
  {Gutmann}}, \bibinfo {author} {\bibfnamefont {Joshua}\ \bibnamefont {Lapan}},
  \bibinfo {author} {\bibfnamefont {Andrew}\ \bibnamefont {Lundgren}}, \ and\
  \bibinfo {author} {\bibfnamefont {Daniel}\ \bibnamefont {Preda}},\ }\bibfield
   {title} {\enquote {\bibinfo {title} {A quantum adiabatic evolution algorithm
  applied to random instances of an {NP-Complete} problem},}\ }\href {\doibase
  10.1126/science.1057726} {\bibfield  {journal} {\bibinfo  {journal}
  {Science}\ }\textbf {\bibinfo {volume} {292}},\ \bibinfo {pages} {472--475}
  (\bibinfo {year} {2001})}\BibitemShut {NoStop}%
\bibitem [{\citenamefont {Barahona}(1982)}]{barahona_computational_1982}%
  \BibitemOpen
  \bibfield  {author} {\bibinfo {author} {\bibfnamefont {F.}~\bibnamefont
  {Barahona}},\ }\bibfield  {title} {\enquote {\bibinfo {title} {On the
  computational complexity of \uppercase{I}sing spin glass models},}\
  }\href@noop {} {\bibfield  {journal} {\bibinfo  {journal} {J. Phys. A: Math.
  Gen}\ }\textbf {\bibinfo {volume} {15}},\ \bibinfo {pages} {3241--3253}
  (\bibinfo {year} {1982})}\BibitemShut {NoStop}%
\bibitem [{\citenamefont {Johnson}\ \emph {et~al.}(2011)\citenamefont
  {Johnson}, \citenamefont {Amin}, \citenamefont {Gildert}, \citenamefont
  {Lanting}, \citenamefont {Hamze}, \citenamefont {Dickson}, \citenamefont
  {Harris}, \citenamefont {Berkley}, \citenamefont {Johansson}, \citenamefont
  {Bunyk}, \citenamefont {Chapple}, \citenamefont {Enderud}, \citenamefont
  {Hilton}, \citenamefont {Karimi}, \citenamefont {Ladizinsky}, \citenamefont
  {Ladizinsky}, \citenamefont {Oh}, \citenamefont {Perminov}, \citenamefont
  {Rich}, \citenamefont {Thom}, \citenamefont {Tolkacheva}, \citenamefont
  {Truncik}, \citenamefont {Uchaikin}, \citenamefont {Wang}, \citenamefont
  {Wilson},\ and\ \citenamefont {Rose}}]{Dwave}%
  \BibitemOpen
  \bibfield  {author} {\bibinfo {author} {\bibfnamefont {M.~W.}\ \bibnamefont
  {Johnson}}, \bibinfo {author} {\bibfnamefont {M.~H.~S.}\ \bibnamefont
  {Amin}}, \bibinfo {author} {\bibfnamefont {S.}~\bibnamefont {Gildert}},
  \bibinfo {author} {\bibfnamefont {T.}~\bibnamefont {Lanting}}, \bibinfo
  {author} {\bibfnamefont {F.}~\bibnamefont {Hamze}}, \bibinfo {author}
  {\bibfnamefont {N.}~\bibnamefont {Dickson}}, \bibinfo {author} {\bibfnamefont
  {R.}~\bibnamefont {Harris}}, \bibinfo {author} {\bibfnamefont {A.~J.}\
  \bibnamefont {Berkley}}, \bibinfo {author} {\bibfnamefont {J.}~\bibnamefont
  {Johansson}}, \bibinfo {author} {\bibfnamefont {P.}~\bibnamefont {Bunyk}},
  \bibinfo {author} {\bibfnamefont {E.~M.}\ \bibnamefont {Chapple}}, \bibinfo
  {author} {\bibfnamefont {C.}~\bibnamefont {Enderud}}, \bibinfo {author}
  {\bibfnamefont {J.~P.}\ \bibnamefont {Hilton}}, \bibinfo {author}
  {\bibfnamefont {K.}~\bibnamefont {Karimi}}, \bibinfo {author} {\bibfnamefont
  {E.}~\bibnamefont {Ladizinsky}}, \bibinfo {author} {\bibfnamefont
  {N.}~\bibnamefont {Ladizinsky}}, \bibinfo {author} {\bibfnamefont
  {T.}~\bibnamefont {Oh}}, \bibinfo {author} {\bibfnamefont {I.}~\bibnamefont
  {Perminov}}, \bibinfo {author} {\bibfnamefont {C.}~\bibnamefont {Rich}},
  \bibinfo {author} {\bibfnamefont {M.~C.}\ \bibnamefont {Thom}}, \bibinfo
  {author} {\bibfnamefont {E.}~\bibnamefont {Tolkacheva}}, \bibinfo {author}
  {\bibfnamefont {C.~J.~S.}\ \bibnamefont {Truncik}}, \bibinfo {author}
  {\bibfnamefont {S.}~\bibnamefont {Uchaikin}}, \bibinfo {author}
  {\bibfnamefont {J.}~\bibnamefont {Wang}}, \bibinfo {author} {\bibfnamefont
  {B.}~\bibnamefont {Wilson}}, \ and\ \bibinfo {author} {\bibfnamefont
  {G.}~\bibnamefont {Rose}},\ }\bibfield  {title} {\enquote {\bibinfo {title}
  {Quantum annealing with manufactured spins},}\ }\href {\doibase
  10.1038/nature10012} {\bibfield  {journal} {\bibinfo  {journal} {Nature}\
  }\textbf {\bibinfo {volume} {473}},\ \bibinfo {pages} {194--198} (\bibinfo
  {year} {2011})}\BibitemShut {NoStop}%
\bibitem [{\citenamefont {Harris}\ \emph {et~al.}(2010)\citenamefont {Harris},
  \citenamefont {Johnson}, \citenamefont {Lanting}, \citenamefont {Berkley},
  \citenamefont {Johansson}, \citenamefont {Bunyk}, \citenamefont {Tolkacheva},
  \citenamefont {Ladizinsky}, \citenamefont {Ladizinsky}, \citenamefont {Oh},
  \citenamefont {Cioata}, \citenamefont {Perminov}, \citenamefont {Spear},
  \citenamefont {Enderud}, \citenamefont {Rich}, \citenamefont {Uchaikin},
  \citenamefont {Thom}, \citenamefont {Chapple}, \citenamefont {Wang},
  \citenamefont {Wilson}, \citenamefont {Amin}, \citenamefont {Dickson},
  \citenamefont {Karimi}, \citenamefont {Macready}, \citenamefont {Truncik},\
  and\ \citenamefont {Rose}}]{Harris:2010kx}%
  \BibitemOpen
  \bibfield  {author} {\bibinfo {author} {\bibfnamefont {R.}~\bibnamefont
  {Harris}}, \bibinfo {author} {\bibfnamefont {M.~W.}\ \bibnamefont {Johnson}},
  \bibinfo {author} {\bibfnamefont {T.}~\bibnamefont {Lanting}}, \bibinfo
  {author} {\bibfnamefont {A.~J.}\ \bibnamefont {Berkley}}, \bibinfo {author}
  {\bibfnamefont {J.}~\bibnamefont {Johansson}}, \bibinfo {author}
  {\bibfnamefont {P.}~\bibnamefont {Bunyk}}, \bibinfo {author} {\bibfnamefont
  {E.}~\bibnamefont {Tolkacheva}}, \bibinfo {author} {\bibfnamefont
  {E.}~\bibnamefont {Ladizinsky}}, \bibinfo {author} {\bibfnamefont
  {N.}~\bibnamefont {Ladizinsky}}, \bibinfo {author} {\bibfnamefont
  {T.}~\bibnamefont {Oh}}, \bibinfo {author} {\bibfnamefont {F.}~\bibnamefont
  {Cioata}}, \bibinfo {author} {\bibfnamefont {I.}~\bibnamefont {Perminov}},
  \bibinfo {author} {\bibfnamefont {P.}~\bibnamefont {Spear}}, \bibinfo
  {author} {\bibfnamefont {C.}~\bibnamefont {Enderud}}, \bibinfo {author}
  {\bibfnamefont {C.}~\bibnamefont {Rich}}, \bibinfo {author} {\bibfnamefont
  {S.}~\bibnamefont {Uchaikin}}, \bibinfo {author} {\bibfnamefont {M.~C.}\
  \bibnamefont {Thom}}, \bibinfo {author} {\bibfnamefont {E.~M.}\ \bibnamefont
  {Chapple}}, \bibinfo {author} {\bibfnamefont {J.}~\bibnamefont {Wang}},
  \bibinfo {author} {\bibfnamefont {B.}~\bibnamefont {Wilson}}, \bibinfo
  {author} {\bibfnamefont {M.~H.~S.}\ \bibnamefont {Amin}}, \bibinfo {author}
  {\bibfnamefont {N.}~\bibnamefont {Dickson}}, \bibinfo {author} {\bibfnamefont
  {K.}~\bibnamefont {Karimi}}, \bibinfo {author} {\bibfnamefont
  {B.}~\bibnamefont {Macready}}, \bibinfo {author} {\bibfnamefont {C.~J.~S.}\
  \bibnamefont {Truncik}}, \ and\ \bibinfo {author} {\bibfnamefont
  {G.}~\bibnamefont {Rose}},\ }\bibfield  {title} {\enquote {\bibinfo {title}
  {Experimental investigation of an eight-qubit unit cell in a superconducting
  optimization processor},}\ }\href {\doibase 10.1103/PhysRevB.82.024511}
  {\bibfield  {journal} {\bibinfo  {journal} {Phys. Rev. B}\ }\textbf {\bibinfo
  {volume} {82}},\ \bibinfo {pages} {024511} (\bibinfo {year}
  {2010})}\BibitemShut {NoStop}%
\bibitem [{\citenamefont {Boixo}\ \emph {et~al.}(2013)\citenamefont {Boixo},
  \citenamefont {Albash}, \citenamefont {Spedalieri}, \citenamefont
  {Chancellor},\ and\ \citenamefont {Lidar}}]{q-sig}%
  \BibitemOpen
  \bibfield  {author} {\bibinfo {author} {\bibfnamefont {Sergio}\ \bibnamefont
  {Boixo}}, \bibinfo {author} {\bibfnamefont {Tameem}\ \bibnamefont {Albash}},
  \bibinfo {author} {\bibfnamefont {Federico~M.}\ \bibnamefont {Spedalieri}},
  \bibinfo {author} {\bibfnamefont {Nicholas}\ \bibnamefont {Chancellor}}, \
  and\ \bibinfo {author} {\bibfnamefont {Daniel~A.}\ \bibnamefont {Lidar}},\
  }\bibfield  {title} {\enquote {\bibinfo {title} {Experimental signature of
  programmable quantum annealing},}\ }\href
  {http://dx.doi.org/10.1038/ncomms3067} {\bibfield  {journal} {\bibinfo
  {journal} {Nat Commun}\ }\textbf {\bibinfo {volume} {4}} (\bibinfo {year}
  {2013})}\BibitemShut {NoStop}%
\bibitem [{\citenamefont {Boixo}\ \emph {et~al.}(2014)\citenamefont {Boixo},
  \citenamefont {Ronnow}, \citenamefont {Isakov}, \citenamefont {Wang},
  \citenamefont {Wecker}, \citenamefont {Lidar}, \citenamefont {Martinis},\
  and\ \citenamefont {Troyer}}]{q108}%
  \BibitemOpen
  \bibfield  {author} {\bibinfo {author} {\bibfnamefont {Sergio}\ \bibnamefont
  {Boixo}}, \bibinfo {author} {\bibfnamefont {Troels~F.}\ \bibnamefont
  {Ronnow}}, \bibinfo {author} {\bibfnamefont {Sergei~V.}\ \bibnamefont
  {Isakov}}, \bibinfo {author} {\bibfnamefont {Zhihui}\ \bibnamefont {Wang}},
  \bibinfo {author} {\bibfnamefont {David}\ \bibnamefont {Wecker}}, \bibinfo
  {author} {\bibfnamefont {Daniel~A.}\ \bibnamefont {Lidar}}, \bibinfo {author}
  {\bibfnamefont {John~M.}\ \bibnamefont {Martinis}}, \ and\ \bibinfo {author}
  {\bibfnamefont {Matthias}\ \bibnamefont {Troyer}},\ }\bibfield  {title}
  {\enquote {\bibinfo {title} {Evidence for quantum annealing with more than
  one hundred qubits},}\ }\href {http://dx.doi.org/10.1038/nphys2900}
  {\bibfield  {journal} {\bibinfo  {journal} {Nat Phys}\ }\textbf {\bibinfo
  {volume} {10}},\ \bibinfo {pages} {218--224} (\bibinfo {year}
  {2014})}\BibitemShut {NoStop}%
\bibitem [{\citenamefont {Smolin}\ and\ \citenamefont {Smith}(2013)}]{Smolin}%
  \BibitemOpen
  \bibfield  {author} {\bibinfo {author} {\bibfnamefont {John~A.}\ \bibnamefont
  {Smolin}}\ and\ \bibinfo {author} {\bibfnamefont {Graeme}\ \bibnamefont
  {Smith}},\ }\bibfield  {title} {\enquote {\bibinfo {title} {Classical
  signature of quantum annealing},}\ }\href {http://arXiv.org/abs/1305.4904}
  {\bibfield  {journal} {\bibinfo  {journal} {arXiv:1305.4904}\ } (\bibinfo
  {year} {2013})}\BibitemShut {NoStop}%
\bibitem [{\citenamefont {Wang}\ \emph {et~al.}(2013)\citenamefont {Wang},
  \citenamefont {R{\o}nnow}, \citenamefont {Boixo}, \citenamefont {Isakov},
  \citenamefont {Wang}, \citenamefont {Wecker}, \citenamefont {Lidar},
  \citenamefont {Martinis},\ and\ \citenamefont {Troyer}}]{comment-SS}%
  \BibitemOpen
  \bibfield  {author} {\bibinfo {author} {\bibfnamefont {Lei}\ \bibnamefont
  {Wang}}, \bibinfo {author} {\bibfnamefont {Troels~F.}\ \bibnamefont
  {R{\o}nnow}}, \bibinfo {author} {\bibfnamefont {Sergio}\ \bibnamefont
  {Boixo}}, \bibinfo {author} {\bibfnamefont {Sergei~V.}\ \bibnamefont
  {Isakov}}, \bibinfo {author} {\bibfnamefont {Zhihui}\ \bibnamefont {Wang}},
  \bibinfo {author} {\bibfnamefont {David}\ \bibnamefont {Wecker}}, \bibinfo
  {author} {\bibfnamefont {Daniel~A.}\ \bibnamefont {Lidar}}, \bibinfo {author}
  {\bibfnamefont {John~M.}\ \bibnamefont {Martinis}}, \ and\ \bibinfo {author}
  {\bibfnamefont {Matthias}\ \bibnamefont {Troyer}},\ }\bibfield  {title}
  {\enquote {\bibinfo {title} {Comment on: `{Classical signature of quantum
  annealing}'},}\ }\href {http://arxiv.org/abs/1305.5837} {\  (\bibinfo {year}
  {2013})},\ \Eprint {http://arxiv.org/abs/arXiv:1305.5837} {arXiv:1305.5837}
  \BibitemShut {NoStop}%
\bibitem [{\citenamefont {Shin}\ \emph
  {et~al.}(2014{\natexlab{a}})\citenamefont {Shin}, \citenamefont {Smith},
  \citenamefont {Smolin},\ and\ \citenamefont {Vazirani}}]{SSSV}%
  \BibitemOpen
  \bibfield  {author} {\bibinfo {author} {\bibfnamefont {Seung~Woo}\
  \bibnamefont {Shin}}, \bibinfo {author} {\bibfnamefont {Graeme}\ \bibnamefont
  {Smith}}, \bibinfo {author} {\bibfnamefont {John~A.}\ \bibnamefont {Smolin}},
  \ and\ \bibinfo {author} {\bibfnamefont {Umesh}\ \bibnamefont {Vazirani}},\
  }\bibfield  {title} {\enquote {\bibinfo {title} {How ``quantum" is the
  {D-Wave} machine?}}\ }\href {http://arXiv.org/abs/1401.7087} {\bibfield
  {journal} {\bibinfo  {journal} {arXiv:1401.7087}\ } (\bibinfo {year}
  {2014}{\natexlab{a}})}\BibitemShut {NoStop}%
\bibitem [{\citenamefont {Vinci}\ \emph {et~al.}(2014)\citenamefont {Vinci},
  \citenamefont {Albash}, \citenamefont {Mishra}, \citenamefont {Warburton},\
  and\ \citenamefont {Lidar}}]{q-sig2}%
  \BibitemOpen
  \bibfield  {author} {\bibinfo {author} {\bibfnamefont {Walter}\ \bibnamefont
  {Vinci}}, \bibinfo {author} {\bibfnamefont {Tameem}\ \bibnamefont {Albash}},
  \bibinfo {author} {\bibfnamefont {Anurag}\ \bibnamefont {Mishra}}, \bibinfo
  {author} {\bibfnamefont {Paul~A.}\ \bibnamefont {Warburton}}, \ and\ \bibinfo
  {author} {\bibfnamefont {Daniel~A.}\ \bibnamefont {Lidar}},\ }\bibfield
  {title} {\enquote {\bibinfo {title} {Distinguishing classical and quantum
  models for the {D-Wave} device},}\ }\href {http://arXiv.org/abs/1403.4228}
  {\bibfield  {journal} {\bibinfo  {journal} {arXiv:1403.4228}\ } (\bibinfo
  {year} {2014})}\BibitemShut {NoStop}%
\bibitem [{\citenamefont {Shin}\ \emph
  {et~al.}(2014{\natexlab{b}})\citenamefont {Shin}, \citenamefont {Smith},
  \citenamefont {Smolin},\ and\ \citenamefont {Vazirani}}]{SSSV-comment}%
  \BibitemOpen
  \bibfield  {author} {\bibinfo {author} {\bibfnamefont {Seung~Woo}\
  \bibnamefont {Shin}}, \bibinfo {author} {\bibfnamefont {Graeme}\ \bibnamefont
  {Smith}}, \bibinfo {author} {\bibfnamefont {John~A.}\ \bibnamefont {Smolin}},
  \ and\ \bibinfo {author} {\bibfnamefont {Umesh}\ \bibnamefont {Vazirani}},\
  }\bibfield  {title} {\enquote {\bibinfo {title} {Comment on "distinguishing
  classical and quantum models for the {D-Wave} device"},}\ }\href
  {http://arXiv.org/abs/1404.6499} {\bibfield  {journal} {\bibinfo  {journal}
  {arXiv:1404.6499}\ } (\bibinfo {year} {2014}{\natexlab{b}})}\BibitemShut
  {NoStop}%
\bibitem [{\citenamefont {Lanting}\ \emph {et~al.}(2014)\citenamefont
  {Lanting}, \citenamefont {Przybysz}, \citenamefont {Smirnov}, \citenamefont
  {Spedalieri}, \citenamefont {Amin}, \citenamefont {Berkley}, \citenamefont
  {Harris}, \citenamefont {Altomare}, \citenamefont {Boixo}, \citenamefont
  {Bunyk}, \citenamefont {Dickson}, \citenamefont {Enderud}, \citenamefont
  {Hilton}, \citenamefont {Hoskinson}, \citenamefont {Johnson}, \citenamefont
  {Ladizinsky}, \citenamefont {Ladizinsky}, \citenamefont {Neufeld},
  \citenamefont {Oh}, \citenamefont {Perminov}, \citenamefont {Rich},
  \citenamefont {Thom}, \citenamefont {Tolkacheva}, \citenamefont {Uchaikin},
  \citenamefont {Wilson},\ and\ \citenamefont {Rose}}]{DWave-entanglement}%
  \BibitemOpen
  \bibfield  {author} {\bibinfo {author} {\bibfnamefont {T.}~\bibnamefont
  {Lanting}}, \bibinfo {author} {\bibfnamefont {A.~J.}\ \bibnamefont
  {Przybysz}}, \bibinfo {author} {\bibfnamefont {A.~Yu.}\ \bibnamefont
  {Smirnov}}, \bibinfo {author} {\bibfnamefont {F.~M.}\ \bibnamefont
  {Spedalieri}}, \bibinfo {author} {\bibfnamefont {M.~H.}\ \bibnamefont
  {Amin}}, \bibinfo {author} {\bibfnamefont {A.~J.}\ \bibnamefont {Berkley}},
  \bibinfo {author} {\bibfnamefont {R.}~\bibnamefont {Harris}}, \bibinfo
  {author} {\bibfnamefont {F.}~\bibnamefont {Altomare}}, \bibinfo {author}
  {\bibfnamefont {S.}~\bibnamefont {Boixo}}, \bibinfo {author} {\bibfnamefont
  {P.}~\bibnamefont {Bunyk}}, \bibinfo {author} {\bibfnamefont
  {N.}~\bibnamefont {Dickson}}, \bibinfo {author} {\bibfnamefont
  {C.}~\bibnamefont {Enderud}}, \bibinfo {author} {\bibfnamefont {J.~P.}\
  \bibnamefont {Hilton}}, \bibinfo {author} {\bibfnamefont {E.}~\bibnamefont
  {Hoskinson}}, \bibinfo {author} {\bibfnamefont {M.~W.}\ \bibnamefont
  {Johnson}}, \bibinfo {author} {\bibfnamefont {E.}~\bibnamefont {Ladizinsky}},
  \bibinfo {author} {\bibfnamefont {N.}~\bibnamefont {Ladizinsky}}, \bibinfo
  {author} {\bibfnamefont {R.}~\bibnamefont {Neufeld}}, \bibinfo {author}
  {\bibfnamefont {T.}~\bibnamefont {Oh}}, \bibinfo {author} {\bibfnamefont
  {I.}~\bibnamefont {Perminov}}, \bibinfo {author} {\bibfnamefont
  {C.}~\bibnamefont {Rich}}, \bibinfo {author} {\bibfnamefont {M.~C.}\
  \bibnamefont {Thom}}, \bibinfo {author} {\bibfnamefont {E.}~\bibnamefont
  {Tolkacheva}}, \bibinfo {author} {\bibfnamefont {S.}~\bibnamefont
  {Uchaikin}}, \bibinfo {author} {\bibfnamefont {A.~B.}\ \bibnamefont
  {Wilson}}, \ and\ \bibinfo {author} {\bibfnamefont {G.}~\bibnamefont
  {Rose}},\ }\bibfield  {title} {\enquote {\bibinfo {title} {Entanglement in a
  quantum annealing processor},}\ }\href
  {http://link.aps.org/doi/10.1103/PhysRevX.4.021041} {\bibfield  {journal}
  {\bibinfo  {journal} {Physical Review X}\ }\textbf {\bibinfo {volume} {4}},\
  \bibinfo {pages} {021041--} (\bibinfo {year} {2014})}\BibitemShut {NoStop}%
\bibitem [{\citenamefont {R{\o}nnow}\ \emph {et~al.}(2014)\citenamefont
  {R{\o}nnow}, \citenamefont {Wang}, \citenamefont {Job}, \citenamefont
  {Boixo}, \citenamefont {Isakov}, \citenamefont {Wecker}, \citenamefont
  {Martinis}, \citenamefont {Lidar},\ and\ \citenamefont {Troyer}}]{speedup}%
  \BibitemOpen
  \bibfield  {author} {\bibinfo {author} {\bibfnamefont {Troels~F.}\
  \bibnamefont {R{\o}nnow}}, \bibinfo {author} {\bibfnamefont {Zhihui}\
  \bibnamefont {Wang}}, \bibinfo {author} {\bibfnamefont {Joshua}\ \bibnamefont
  {Job}}, \bibinfo {author} {\bibfnamefont {Sergio}\ \bibnamefont {Boixo}},
  \bibinfo {author} {\bibfnamefont {Sergei~V.}\ \bibnamefont {Isakov}},
  \bibinfo {author} {\bibfnamefont {David}\ \bibnamefont {Wecker}}, \bibinfo
  {author} {\bibfnamefont {John~M.}\ \bibnamefont {Martinis}}, \bibinfo
  {author} {\bibfnamefont {Daniel~A.}\ \bibnamefont {Lidar}}, \ and\ \bibinfo
  {author} {\bibfnamefont {Matthias}\ \bibnamefont {Troyer}},\ }\bibfield
  {title} {\enquote {\bibinfo {title} {Defining and detecting quantum
  speedup},}\ }\href {http://www.sciencemag.org/content/345/6195/420.abstract
  N2 - The development of small-scale quantum devices raises the question of
  how to fairly assess and detect quantum speedup. Here, we show how to define
  and measure quantum speedup and how to avoid pitfalls that might mask or fake
  such a speedup. We illustrate our discussion with data from tests run on a
  D-Wave Two device with up to 503 qubits. By using random spin glass instances
  as a benchmark, we found no evidence of quantum speedup when the entire data
  set is considered and obtained inconclusive results when comparing subsets of
  instances on an instance-by-instance basis. Our results do not rule out the
  possibility of speedup for other classes of problems and illustrate the
  subtle nature of the quantum speedup question.} {\bibfield  {journal}
  {\bibinfo  {journal} {Science}\ }\textbf {\bibinfo {volume} {345}},\ \bibinfo
  {pages} {420--424} (\bibinfo {year} {2014})}\BibitemShut {NoStop}%
\bibitem [{\citenamefont {Katzgraber}\ \emph {et~al.}(2014)\citenamefont
  {Katzgraber}, \citenamefont {Hamze},\ and\ \citenamefont
  {Andrist}}]{2014Katzgraber}%
  \BibitemOpen
  \bibfield  {author} {\bibinfo {author} {\bibfnamefont {Helmut~G.}\
  \bibnamefont {Katzgraber}}, \bibinfo {author} {\bibfnamefont {Firas}\
  \bibnamefont {Hamze}}, \ and\ \bibinfo {author} {\bibfnamefont {Ruben~S.}\
  \bibnamefont {Andrist}},\ }\bibfield  {title} {\enquote {\bibinfo {title}
  {Glassy chimeras could be blind to quantum speedup: Designing better
  benchmarks for quantum annealing machines},}\ }\href
  {http://link.aps.org/doi/10.1103/PhysRevX.4.021008} {\bibfield  {journal}
  {\bibinfo  {journal} {Physical Review X}\ }\textbf {\bibinfo {volume} {4}},\
  \bibinfo {pages} {021008--} (\bibinfo {year} {2014})}\BibitemShut {NoStop}%
\bibitem [{\citenamefont {Venturelli}\ \emph {et~al.}(2014)\citenamefont
  {Venturelli}, \citenamefont {Mandr{\`a}}, \citenamefont {Knysh},
  \citenamefont {O'Gorman}, \citenamefont {Biswas},\ and\ \citenamefont
  {Smelyanskiy}}]{Venturelli:2014nx}%
  \BibitemOpen
  \bibfield  {author} {\bibinfo {author} {\bibfnamefont {Davide}\ \bibnamefont
  {Venturelli}}, \bibinfo {author} {\bibfnamefont {Salvatore}\ \bibnamefont
  {Mandr{\`a}}}, \bibinfo {author} {\bibfnamefont {Sergey}\ \bibnamefont
  {Knysh}}, \bibinfo {author} {\bibfnamefont {Bryan}\ \bibnamefont {O'Gorman}},
  \bibinfo {author} {\bibfnamefont {Rupak}\ \bibnamefont {Biswas}}, \ and\
  \bibinfo {author} {\bibfnamefont {Vadim}\ \bibnamefont {Smelyanskiy}},\
  }\bibfield  {title} {\enquote {\bibinfo {title} {Quantum optimization of
  fully-connected spin glasses},}\ }\href {http://arXiv.org/abs/1406.7553}
  {\bibfield  {journal} {\bibinfo  {journal} {arXiv:1406.7553}\ } (\bibinfo
  {year} {2014})}\BibitemShut {NoStop}%
\bibitem [{\citenamefont {Somma}\ \emph {et~al.}(2012)\citenamefont {Somma},
  \citenamefont {Nagaj},\ and\ \citenamefont {Kieferov{\'a}}}]{Somma:2012kx}%
  \BibitemOpen
  \bibfield  {author} {\bibinfo {author} {\bibfnamefont {Rolando~D.}\
  \bibnamefont {Somma}}, \bibinfo {author} {\bibfnamefont {Daniel}\
  \bibnamefont {Nagaj}}, \ and\ \bibinfo {author} {\bibfnamefont {M{\'a}ria}\
  \bibnamefont {Kieferov{\'a}}},\ }\bibfield  {title} {\enquote {\bibinfo
  {title} {Quantum speedup by quantum annealing},}\ }\href
  {http://link.aps.org/doi/10.1103/PhysRevLett.109.050501} {\bibfield
  {journal} {\bibinfo  {journal} {Physical Review Letters}\ }\textbf {\bibinfo
  {volume} {109}},\ \bibinfo {pages} {050501--} (\bibinfo {year}
  {2012})}\BibitemShut {NoStop}%
\bibitem [{\citenamefont {Mizel}\ \emph {et~al.}(2001)\citenamefont {Mizel},
  \citenamefont {Mitchell},\ and\ \citenamefont {Cohen}}]{Mizel:01}%
  \BibitemOpen
  \bibfield  {author} {\bibinfo {author} {\bibfnamefont {Ari}\ \bibnamefont
  {Mizel}}, \bibinfo {author} {\bibfnamefont {M.~W.}\ \bibnamefont {Mitchell}},
  \ and\ \bibinfo {author} {\bibfnamefont {Marvin~L.}\ \bibnamefont {Cohen}},\
  }\bibfield  {title} {\enquote {\bibinfo {title} {Energy barrier to
  decoherence},}\ }\href {http://link.aps.org/doi/10.1103/PhysRevA.63.040302}
  {\bibfield  {journal} {\bibinfo  {journal} {Physical Review A}\ }\textbf
  {\bibinfo {volume} {63}},\ \bibinfo {pages} {040302--} (\bibinfo {year}
  {2001})}\BibitemShut {NoStop}%
\bibitem [{\citenamefont {Jordan}\ \emph {et~al.}(2006)\citenamefont {Jordan},
  \citenamefont {Farhi},\ and\ \citenamefont {Shor}}]{jordan2006error}%
  \BibitemOpen
  \bibfield  {author} {\bibinfo {author} {\bibfnamefont {S.~P.}\ \bibnamefont
  {Jordan}}, \bibinfo {author} {\bibfnamefont {E.}~\bibnamefont {Farhi}}, \
  and\ \bibinfo {author} {\bibfnamefont {P.~W.}\ \bibnamefont {Shor}},\
  }\bibfield  {title} {\enquote {\bibinfo {title} {Error-correcting codes for
  adiabatic quantum computation},}\ }\href
  {http://link.aps.org/doi/10.1103/PhysRevA.74.052322} {\bibfield  {journal}
  {\bibinfo  {journal} {{Phys. Rev. A}}\ }\textbf {\bibinfo {volume} {74}},\
  \bibinfo {pages} {052322} (\bibinfo {year} {2006})}\BibitemShut {NoStop}%
\bibitem [{\citenamefont {Lidar}(2008)}]{PhysRevLett.100.160506}%
  \BibitemOpen
  \bibfield  {author} {\bibinfo {author} {\bibfnamefont {D.~A.}\ \bibnamefont
  {Lidar}},\ }\bibfield  {title} {\enquote {\bibinfo {title} {Towards fault
  tolerant adiabatic quantum computation},}\ }\href
  {http://link.aps.org/doi/10.1103/PhysRevLett.100.160506} {\bibfield
  {journal} {\bibinfo  {journal} {{Phys.~Rev.~Lett.}}\ }\textbf {\bibinfo
  {volume} {100}},\ \bibinfo {pages} {160506} (\bibinfo {year}
  {2008})}\BibitemShut {NoStop}%
\bibitem [{\citenamefont {Quiroz}\ and\ \citenamefont
  {Lidar}(2012)}]{PhysRevA.86.042333}%
  \BibitemOpen
  \bibfield  {author} {\bibinfo {author} {\bibfnamefont {G.}~\bibnamefont
  {Quiroz}}\ and\ \bibinfo {author} {\bibfnamefont {D.~A.}\ \bibnamefont
  {Lidar}},\ }\bibfield  {title} {\enquote {\bibinfo {title} {High-fidelity
  adiabatic quantum computation via dynamical decoupling},}\ }\href {\doibase
  10.1103/PhysRevA.86.042333} {\bibfield  {journal} {\bibinfo  {journal} {Phys.
  Rev. A}\ }\textbf {\bibinfo {volume} {86}},\ \bibinfo {pages} {042333}
  (\bibinfo {year} {2012})}\BibitemShut {NoStop}%
\bibitem [{\citenamefont {Young}\ \emph
  {et~al.}(2013{\natexlab{a}})\citenamefont {Young}, \citenamefont {Sarovar},\
  and\ \citenamefont {Blume-Kohout}}]{Young:13}%
  \BibitemOpen
  \bibfield  {author} {\bibinfo {author} {\bibfnamefont {Kevin~C.}\
  \bibnamefont {Young}}, \bibinfo {author} {\bibfnamefont {Mohan}\ \bibnamefont
  {Sarovar}}, \ and\ \bibinfo {author} {\bibfnamefont {Robin}\ \bibnamefont
  {Blume-Kohout}},\ }\bibfield  {title} {\enquote {\bibinfo {title} {Error
  suppression and error correction in adiabatic quantum computation: Techniques
  and challenges},}\ }\href {http://link.aps.org/doi/10.1103/PhysRevX.3.041013}
  {\bibfield  {journal} {\bibinfo  {journal} {Physical Review X}\ }\textbf
  {\bibinfo {volume} {3}},\ \bibinfo {pages} {041013--} (\bibinfo {year}
  {2013}{\natexlab{a}})}\BibitemShut {NoStop}%
\bibitem [{\citenamefont {Sarovar}\ and\ \citenamefont
  {Young}(2013)}]{Sarovar:2013kx}%
  \BibitemOpen
  \bibfield  {author} {\bibinfo {author} {\bibfnamefont {Mohan}\ \bibnamefont
  {Sarovar}}\ and\ \bibinfo {author} {\bibfnamefont {Kevin~C}\ \bibnamefont
  {Young}},\ }\bibfield  {title} {\enquote {\bibinfo {title} {Error suppression
  and error correction in adiabatic quantum computation: non-equilibrium
  dynamics},}\ }\href {http://stacks.iop.org/1367-2630/15/i=12/a=125032}
  {\bibfield  {journal} {\bibinfo  {journal} {New Journal of Physics}\ }\textbf
  {\bibinfo {volume} {15}},\ \bibinfo {pages} {125032} (\bibinfo {year}
  {2013})}\BibitemShut {NoStop}%
\bibitem [{\citenamefont {Young}\ \emph
  {et~al.}(2013{\natexlab{b}})\citenamefont {Young}, \citenamefont
  {Blume-Kohout},\ and\ \citenamefont {Lidar}}]{Young:2013fk}%
  \BibitemOpen
  \bibfield  {author} {\bibinfo {author} {\bibfnamefont {Kevin~C.}\
  \bibnamefont {Young}}, \bibinfo {author} {\bibfnamefont {Robin}\ \bibnamefont
  {Blume-Kohout}}, \ and\ \bibinfo {author} {\bibfnamefont {Daniel~A.}\
  \bibnamefont {Lidar}},\ }\bibfield  {title} {\enquote {\bibinfo {title}
  {Adiabatic quantum optimization with the wrong hamiltonian},}\ }\href
  {http://link.aps.org/doi/10.1103/PhysRevA.88.062314} {\bibfield  {journal}
  {\bibinfo  {journal} {Physical Review A}\ }\textbf {\bibinfo {volume} {88}},\
  \bibinfo {pages} {062314--} (\bibinfo {year}
  {2013}{\natexlab{b}})}\BibitemShut {NoStop}%
\bibitem [{\citenamefont {Ganti}\ \emph {et~al.}(2014)\citenamefont {Ganti},
  \citenamefont {Onunkwo},\ and\ \citenamefont {Young}}]{Ganti:13}%
  \BibitemOpen
  \bibfield  {author} {\bibinfo {author} {\bibfnamefont {Anand}\ \bibnamefont
  {Ganti}}, \bibinfo {author} {\bibfnamefont {Uzoma}\ \bibnamefont {Onunkwo}},
  \ and\ \bibinfo {author} {\bibfnamefont {Kevin}\ \bibnamefont {Young}},\
  }\bibfield  {title} {\enquote {\bibinfo {title} {Family of [[6k,2k,2]] codes
  for practical, scalable adiabatic quantum computation},}\ }\href
  {http://link.aps.org/doi/10.1103/PhysRevA.89.042313} {\bibfield  {journal}
  {\bibinfo  {journal} {Physical Review A}\ }\textbf {\bibinfo {volume} {89}},\
  \bibinfo {pages} {042313--} (\bibinfo {year} {2014})}\BibitemShut {NoStop}%
\bibitem [{\citenamefont {Mizel}(2014)}]{Mizel:2014sp}%
  \BibitemOpen
  \bibfield  {author} {\bibinfo {author} {\bibfnamefont {Ari}\ \bibnamefont
  {Mizel}},\ }\bibfield  {title} {\enquote {\bibinfo {title} {Fault-tolerant,
  universal adiabatic quantum computation},}\ }\href
  {http://arXiv.org/abs/1403.7694} {\bibfield  {journal} {\bibinfo  {journal}
  {arXiv:1403.7694}\ } (\bibinfo {year} {2014})}\BibitemShut {NoStop}%
\bibitem [{\citenamefont {Bookatz}\ \emph {et~al.}(2014)\citenamefont
  {Bookatz}, \citenamefont {Farhi},\ and\ \citenamefont
  {Zhou}}]{Bookatz:2014uq}%
  \BibitemOpen
  \bibfield  {author} {\bibinfo {author} {\bibfnamefont {Adam~D.}\ \bibnamefont
  {Bookatz}}, \bibinfo {author} {\bibfnamefont {Edward}\ \bibnamefont {Farhi}},
  \ and\ \bibinfo {author} {\bibfnamefont {Leo}\ \bibnamefont {Zhou}},\
  }\bibfield  {title} {\enquote {\bibinfo {title} {Error suppression in
  hamiltonian based quantum computation using energy penalties},}\ }\href
  {http://arXiv.org/abs/1407.1485} {\bibfield  {journal} {\bibinfo  {journal}
  {arXiv:1407.1485}\ } (\bibinfo {year} {2014})}\BibitemShut {NoStop}%
\bibitem [{\citenamefont {Pudenz}\ \emph {et~al.}(2014)\citenamefont {Pudenz},
  \citenamefont {Albash},\ and\ \citenamefont {Lidar}}]{PAL:13}%
  \BibitemOpen
  \bibfield  {author} {\bibinfo {author} {\bibfnamefont {Kristen~L}\
  \bibnamefont {Pudenz}}, \bibinfo {author} {\bibfnamefont {Tameem}\
  \bibnamefont {Albash}}, \ and\ \bibinfo {author} {\bibfnamefont {Daniel~A}\
  \bibnamefont {Lidar}},\ }\bibfield  {title} {\enquote {\bibinfo {title}
  {Error-corrected quantum annealing with hundreds of qubits},}\ }\href
  {http://dx.doi.org/10.1038/ncomms4243} {\bibfield  {journal} {\bibinfo
  {journal} {Nat Commun}\ }\textbf {\bibinfo {volume} {5}} (\bibinfo {year}
  {2014})}\BibitemShut {NoStop}%
\bibitem [{\citenamefont {Gaitan}(2008)}]{Gaitan:book}%
  \BibitemOpen
  \bibfield  {author} {\bibinfo {author} {\bibfnamefont {F.}~\bibnamefont
  {Gaitan}},\ }\href {http://books.google.com/books?id=zwvlqspyOK8C} {\emph
  {\bibinfo {title} {Quantum Error Correction and Fault Tolerant Quantum
  Computing}}}\ (\bibinfo  {publisher} {Taylor \& Francis Group},\ \bibinfo
  {address} {Boca Raton},\ \bibinfo {year} {2008})\BibitemShut {NoStop}%
\bibitem [{\citenamefont {Knill}\ \emph {et~al.}(1998)\citenamefont {Knill},
  \citenamefont {Laflamme},\ and\ \citenamefont {Zurek}}]{Knill:98}%
  \BibitemOpen
  \bibfield  {author} {\bibinfo {author} {\bibfnamefont {Emanuel}\ \bibnamefont
  {Knill}}, \bibinfo {author} {\bibfnamefont {Raymond}\ \bibnamefont
  {Laflamme}}, \ and\ \bibinfo {author} {\bibfnamefont {Wojciech~H.}\
  \bibnamefont {Zurek}},\ }\bibfield  {title} {\enquote {\bibinfo {title}
  {Resilient quantum computation},}\ }\href
  {http://www.sciencemag.org/content/279/5349/342.abstract N2 - Practical
  realization of quantum computers will require overcoming decoherence and
  operational errors, which lead to problems that are more severe than in
  classical computation. It is shown that arbitrarily accurate quantum
  computation is possible provided that the error per operation is below a
  threshold value.} {\bibfield  {journal} {\bibinfo  {journal} {Science}\
  }\textbf {\bibinfo {volume} {279}},\ \bibinfo {pages} {342--345} (\bibinfo
  {year} {1998})}\BibitemShut {NoStop}%
\bibitem [{\citenamefont {{J. Preskill}}(2013)}]{preskill:12}%
  \BibitemOpen
  \bibfield  {author} {\bibinfo {author} {\bibnamefont {{J. Preskill}}},\
  }\bibfield  {title} {\enquote {\bibinfo {title} {Sufficient condition on
  noise correlations for scalable quantum computing},}\ }\href
  {http://arXiv.org/abs/1207.6131} {\bibfield  {journal} {\bibinfo  {journal}
  {{Quant. Inf. Comput.}}\ }\textbf {\bibinfo {volume} {13}},\ \bibinfo {pages}
  {181} (\bibinfo {year} {2013})}\BibitemShut {NoStop}%
\bibitem [{\citenamefont {Gottesman}(2013)}]{Gottesman:2013ud}%
  \BibitemOpen
  \bibfield  {author} {\bibinfo {author} {\bibfnamefont {Daniel}\ \bibnamefont
  {Gottesman}},\ }\bibfield  {title} {\enquote {\bibinfo {title}
  {Fault-tolerant quantum computation with constant overhead},}\ }\href
  {http://arXiv.org/abs/1310.2984} {\bibfield  {journal} {\bibinfo  {journal}
  {arXiv:1310.2984}\ } (\bibinfo {year} {2013})}\BibitemShut {NoStop}%
\bibitem [{\citenamefont {Childs}\ \emph {et~al.}(2001)\citenamefont {Childs},
  \citenamefont {Farhi},\ and\ \citenamefont
  {Preskill}}]{childs_robustness_2001}%
  \BibitemOpen
  \bibfield  {author} {\bibinfo {author} {\bibfnamefont {Andrew~M.}\
  \bibnamefont {Childs}}, \bibinfo {author} {\bibfnamefont {Edward}\
  \bibnamefont {Farhi}}, \ and\ \bibinfo {author} {\bibfnamefont {John}\
  \bibnamefont {Preskill}},\ }\bibfield  {title} {\enquote {\bibinfo {title}
  {Robustness of adiabatic quantum computation},}\ }\href {\doibase
  10.1103/PhysRevA.65.012322} {\bibfield  {journal} {\bibinfo  {journal} {Phys.
  Rev. A}\ }\textbf {\bibinfo {volume} {65}},\ \bibinfo {pages} {012322}
  (\bibinfo {year} {2001})}\BibitemShut {NoStop}%
\bibitem [{\citenamefont {Sarandy}\ and\ \citenamefont
  {Lidar}(2005)}]{PhysRevLett.95.250503}%
  \BibitemOpen
  \bibfield  {author} {\bibinfo {author} {\bibfnamefont {M.~S.}\ \bibnamefont
  {Sarandy}}\ and\ \bibinfo {author} {\bibfnamefont {D.~A.}\ \bibnamefont
  {Lidar}},\ }\bibfield  {title} {\enquote {\bibinfo {title} {Adiabatic quantum
  computation in open systems},}\ }\href
  {http://link.aps.org/doi/10.1103/PhysRevLett.95.250503} {\bibfield  {journal}
  {\bibinfo  {journal} {Physical Review Letters}\ }\textbf {\bibinfo {volume}
  {95}},\ \bibinfo {pages} {250503--} (\bibinfo {year} {2005})}\BibitemShut
  {NoStop}%
\bibitem [{\citenamefont {Amin}\ \emph {et~al.}(2009)\citenamefont {Amin},
  \citenamefont {Averin},\ and\ \citenamefont
  {Nesteroff}}]{PhysRevA.79.022107}%
  \BibitemOpen
  \bibfield  {author} {\bibinfo {author} {\bibfnamefont {M.~H.~S.}\
  \bibnamefont {Amin}}, \bibinfo {author} {\bibfnamefont {Dmitri~V.}\
  \bibnamefont {Averin}}, \ and\ \bibinfo {author} {\bibfnamefont {James~A.}\
  \bibnamefont {Nesteroff}},\ }\bibfield  {title} {\enquote {\bibinfo {title}
  {Decoherence in adiabatic quantum computation},}\ }\href {\doibase
  10.1103/PhysRevA.79.022107} {\bibfield  {journal} {\bibinfo  {journal} {Phys.
  Rev. A}\ }\textbf {\bibinfo {volume} {79}},\ \bibinfo {pages} {022107}
  (\bibinfo {year} {2009})}\BibitemShut {NoStop}%
\bibitem [{\citenamefont {Bunyk}\ \emph {et~al.}(2014)\citenamefont {Bunyk},
  \citenamefont {Hoskinson}, \citenamefont {Johnson}, \citenamefont
  {Tolkacheva}, \citenamefont {Altomare}, \citenamefont {Berkley},
  \citenamefont {Harris}, \citenamefont {Hilton}, \citenamefont {Lanting},\
  and\ \citenamefont {Whittaker}}]{Bunyk:2014hb}%
  \BibitemOpen
  \bibfield  {author} {\bibinfo {author} {\bibfnamefont {P.~I.}\ \bibnamefont
  {Bunyk}}, \bibinfo {author} {\bibfnamefont {E.}~\bibnamefont {Hoskinson}},
  \bibinfo {author} {\bibfnamefont {M.~W.}\ \bibnamefont {Johnson}}, \bibinfo
  {author} {\bibfnamefont {E.}~\bibnamefont {Tolkacheva}}, \bibinfo {author}
  {\bibfnamefont {F.}~\bibnamefont {Altomare}}, \bibinfo {author}
  {\bibfnamefont {A.~J.}\ \bibnamefont {Berkley}}, \bibinfo {author}
  {\bibfnamefont {R.}~\bibnamefont {Harris}}, \bibinfo {author} {\bibfnamefont
  {J.~P.}\ \bibnamefont {Hilton}}, \bibinfo {author} {\bibfnamefont
  {T.}~\bibnamefont {Lanting}}, \ and\ \bibinfo {author} {\bibfnamefont
  {J.}~\bibnamefont {Whittaker}},\ }\bibfield  {title} {\enquote {\bibinfo
  {title} {Architectural considerations in the design of a superconducting
  quantum annealing processor},}\ }\href {http://arXiv.org/abs/1401.5504}
  {\bibfield  {journal} {\bibinfo  {journal} {arXiv:1401.5504}\ } (\bibinfo
  {year} {2014})}\BibitemShut {NoStop}%
\bibitem [{\citenamefont {Kirkpatrick}\ \emph {et~al.}(1983)\citenamefont
  {Kirkpatrick}, \citenamefont {Gelatt},\ and\ \citenamefont
  {Vecchi}}]{kirkpatrick_optimization_1983}%
  \BibitemOpen
  \bibfield  {author} {\bibinfo {author} {\bibfnamefont {S.}~\bibnamefont
  {Kirkpatrick}}, \bibinfo {author} {\bibfnamefont {C.~D.}\ \bibnamefont
  {Gelatt}}, \ and\ \bibinfo {author} {\bibfnamefont {M.~P.}\ \bibnamefont
  {Vecchi}},\ }\bibfield  {title} {\enquote {\bibinfo {title} {Optimization by
  simulated annealing},}\ }\href {\doibase 10.1126/science.220.4598.671}
  {\bibfield  {journal} {\bibinfo  {journal} {Science}\ }\textbf {\bibinfo
  {volume} {220}},\ \bibinfo {pages} {671--680} (\bibinfo {year}
  {1983})}\BibitemShut {NoStop}%
\bibitem [{\citenamefont {Jansen}\ \emph {et~al.}(2007)\citenamefont {Jansen},
  \citenamefont {Ruskai},\ and\ \citenamefont {Seiler}}]{Jansen:07}%
  \BibitemOpen
  \bibfield  {author} {\bibinfo {author} {\bibfnamefont {Sabine}\ \bibnamefont
  {Jansen}}, \bibinfo {author} {\bibfnamefont {Mary-Beth}\ \bibnamefont
  {Ruskai}}, \ and\ \bibinfo {author} {\bibfnamefont {Ruedi}\ \bibnamefont
  {Seiler}},\ }\bibfield  {title} {\enquote {\bibinfo {title} {Bounds for the
  adiabatic approximation with applications to quantum computation},}\ }\href
  {http://scitation.aip.org/content/aip/journal/jmp/48/10/10.1063/1.2798382}
  {\bibfield  {journal} {\bibinfo  {journal} {Journal of Mathematical Physics}\
  }\textbf {\bibinfo {volume} {48}},\ \bibinfo {pages} {--} (\bibinfo {year}
  {2007})}\BibitemShut {NoStop}%
\bibitem [{\citenamefont {Lidar}\ \emph {et~al.}(2009)\citenamefont {Lidar},
  \citenamefont {Rezakhani},\ and\ \citenamefont {Hamma}}]{lidar:102106}%
  \BibitemOpen
  \bibfield  {author} {\bibinfo {author} {\bibfnamefont {Daniel~A.}\
  \bibnamefont {Lidar}}, \bibinfo {author} {\bibfnamefont {Ali~T.}\
  \bibnamefont {Rezakhani}}, \ and\ \bibinfo {author} {\bibfnamefont
  {Alioscia}\ \bibnamefont {Hamma}},\ }\bibfield  {title} {\enquote {\bibinfo
  {title} {Adiabatic approximation with exponential accuracy for many-body
  systems and quantum computation},}\ }\href
  {http://scitation.aip.org/content/aip/journal/jmp/50/10/10.1063/1.3236685}
  {\bibfield  {journal} {\bibinfo  {journal} {Journal of Mathematical Physics}\
  }\textbf {\bibinfo {volume} {50}},\ \bibinfo {pages} {--} (\bibinfo {year}
  {2009})}\BibitemShut {NoStop}%
\bibitem [{\citenamefont {Amin}\ \emph {et~al.}(2008)\citenamefont {Amin},
  \citenamefont {Love},\ and\ \citenamefont {Truncik}}]{TAQC}%
  \BibitemOpen
  \bibfield  {author} {\bibinfo {author} {\bibfnamefont {M.~H.~S.}\
  \bibnamefont {Amin}}, \bibinfo {author} {\bibfnamefont {Peter~J.}\
  \bibnamefont {Love}}, \ and\ \bibinfo {author} {\bibfnamefont {C.~J.~S.}\
  \bibnamefont {Truncik}},\ }\bibfield  {title} {\enquote {\bibinfo {title}
  {Thermally assisted adiabatic quantum computation},}\ }\href {\doibase
  10.1103/PhysRevLett.100.060503} {\bibfield  {journal} {\bibinfo  {journal}
  {Phys. Rev. Lett.}\ }\textbf {\bibinfo {volume} {100}},\ \bibinfo {pages}
  {060503} (\bibinfo {year} {2008})}\BibitemShut {NoStop}%
\bibitem [{\citenamefont {Dickson}\ \emph {et~al.}(2013)\citenamefont
  {Dickson}, \citenamefont {Johnson}, \citenamefont {Amin}, \citenamefont
  {Harris}, \citenamefont {Altomare}, \citenamefont {Berkley}, \citenamefont
  {Bunyk}, \citenamefont {Cai}, \citenamefont {Chapple}, \citenamefont
  {Chavez}, \citenamefont {Cioata}, \citenamefont {Cirip}, \citenamefont
  {deBuen}, \citenamefont {Drew-Brook}, \citenamefont {Enderud}, \citenamefont
  {Gildert}, \citenamefont {Hamze}, \citenamefont {Hilton}, \citenamefont
  {Hoskinson}, \citenamefont {Karimi}, \citenamefont {Ladizinsky},
  \citenamefont {Ladizinsky}, \citenamefont {Lanting}, \citenamefont {Mahon},
  \citenamefont {Neufeld}, \citenamefont {Oh}, \citenamefont {Perminov},
  \citenamefont {Petroff}, \citenamefont {Przybysz}, \citenamefont {Rich},
  \citenamefont {Spear}, \citenamefont {Tcaciuc}, \citenamefont {Thom},
  \citenamefont {Tolkacheva}, \citenamefont {Uchaikin}, \citenamefont {Wang},
  \citenamefont {Wilson}, \citenamefont {Merali},\ and\ \citenamefont
  {Rose}}]{DWave-16q}%
  \BibitemOpen
  \bibfield  {author} {\bibinfo {author} {\bibfnamefont {N.~G.}\ \bibnamefont
  {Dickson}}, \bibinfo {author} {\bibfnamefont {M.~W.}\ \bibnamefont
  {Johnson}}, \bibinfo {author} {\bibfnamefont {M.~H.}\ \bibnamefont {Amin}},
  \bibinfo {author} {\bibfnamefont {R.}~\bibnamefont {Harris}}, \bibinfo
  {author} {\bibfnamefont {F.}~\bibnamefont {Altomare}}, \bibinfo {author}
  {\bibfnamefont {A.~J.}\ \bibnamefont {Berkley}}, \bibinfo {author}
  {\bibfnamefont {P.}~\bibnamefont {Bunyk}}, \bibinfo {author} {\bibfnamefont
  {J.}~\bibnamefont {Cai}}, \bibinfo {author} {\bibfnamefont {E.~M.}\
  \bibnamefont {Chapple}}, \bibinfo {author} {\bibfnamefont {P.}~\bibnamefont
  {Chavez}}, \bibinfo {author} {\bibfnamefont {F.}~\bibnamefont {Cioata}},
  \bibinfo {author} {\bibfnamefont {T.}~\bibnamefont {Cirip}}, \bibinfo
  {author} {\bibfnamefont {P.}~\bibnamefont {deBuen}}, \bibinfo {author}
  {\bibfnamefont {M.}~\bibnamefont {Drew-Brook}}, \bibinfo {author}
  {\bibfnamefont {C.}~\bibnamefont {Enderud}}, \bibinfo {author} {\bibfnamefont
  {S.}~\bibnamefont {Gildert}}, \bibinfo {author} {\bibfnamefont
  {F.}~\bibnamefont {Hamze}}, \bibinfo {author} {\bibfnamefont {J.~P.}\
  \bibnamefont {Hilton}}, \bibinfo {author} {\bibfnamefont {E.}~\bibnamefont
  {Hoskinson}}, \bibinfo {author} {\bibfnamefont {K.}~\bibnamefont {Karimi}},
  \bibinfo {author} {\bibfnamefont {E.}~\bibnamefont {Ladizinsky}}, \bibinfo
  {author} {\bibfnamefont {N.}~\bibnamefont {Ladizinsky}}, \bibinfo {author}
  {\bibfnamefont {T.}~\bibnamefont {Lanting}}, \bibinfo {author} {\bibfnamefont
  {T.}~\bibnamefont {Mahon}}, \bibinfo {author} {\bibfnamefont
  {R.}~\bibnamefont {Neufeld}}, \bibinfo {author} {\bibfnamefont
  {T.}~\bibnamefont {Oh}}, \bibinfo {author} {\bibfnamefont {I.}~\bibnamefont
  {Perminov}}, \bibinfo {author} {\bibfnamefont {C.}~\bibnamefont {Petroff}},
  \bibinfo {author} {\bibfnamefont {A.}~\bibnamefont {Przybysz}}, \bibinfo
  {author} {\bibfnamefont {C.}~\bibnamefont {Rich}}, \bibinfo {author}
  {\bibfnamefont {P.}~\bibnamefont {Spear}}, \bibinfo {author} {\bibfnamefont
  {A.}~\bibnamefont {Tcaciuc}}, \bibinfo {author} {\bibfnamefont {M.~C.}\
  \bibnamefont {Thom}}, \bibinfo {author} {\bibfnamefont {E.}~\bibnamefont
  {Tolkacheva}}, \bibinfo {author} {\bibfnamefont {S.}~\bibnamefont
  {Uchaikin}}, \bibinfo {author} {\bibfnamefont {J.}~\bibnamefont {Wang}},
  \bibinfo {author} {\bibfnamefont {A.~B.}\ \bibnamefont {Wilson}}, \bibinfo
  {author} {\bibfnamefont {Z.}~\bibnamefont {Merali}}, \ and\ \bibinfo {author}
  {\bibfnamefont {G.}~\bibnamefont {Rose}},\ }\bibfield  {title} {\enquote
  {\bibinfo {title} {Thermally assisted quantum annealing of a 16-qubit
  problem},}\ }\href {\doibase 10.1038/ncomms2920} {\bibfield  {journal}
  {\bibinfo  {journal} {Nat. Commun.}\ }\textbf {\bibinfo {volume} {4}},\
  \bibinfo {pages} {1903} (\bibinfo {year} {2013})}\BibitemShut {NoStop}%
\bibitem [{\citenamefont {Robertson}\ and\ \citenamefont
  {Seymour}(1984)}]{Robertson198449}%
  \BibitemOpen
  \bibfield  {author} {\bibinfo {author} {\bibfnamefont {Neil}\ \bibnamefont
  {Robertson}}\ and\ \bibinfo {author} {\bibfnamefont {P.D}\ \bibnamefont
  {Seymour}},\ }\bibfield  {title} {\enquote {\bibinfo {title} {Graph minors.
  iii. planar tree-width},}\ }\href {\doibase
  http://dx.doi.org/10.1016/0095-8956(84)90013-3} {\bibfield  {journal}
  {\bibinfo  {journal} {Journal of Combinatorial Theory, Series B}\ }\textbf
  {\bibinfo {volume} {36}},\ \bibinfo {pages} {49 -- 64} (\bibinfo {year}
  {1984})}\BibitemShut {NoStop}%
\bibitem [{\citenamefont {Albash}\ \emph {et~al.}(2012)\citenamefont {Albash},
  \citenamefont {Boixo}, \citenamefont {Lidar},\ and\ \citenamefont
  {Zanardi}}]{ABLZ:12-SI}%
  \BibitemOpen
  \bibfield  {author} {\bibinfo {author} {\bibfnamefont {Tameem}\ \bibnamefont
  {Albash}}, \bibinfo {author} {\bibfnamefont {Sergio}\ \bibnamefont {Boixo}},
  \bibinfo {author} {\bibfnamefont {Daniel~A}\ \bibnamefont {Lidar}}, \ and\
  \bibinfo {author} {\bibfnamefont {Paolo}\ \bibnamefont {Zanardi}},\
  }\bibfield  {title} {\enquote {\bibinfo {title} {Quantum adiabatic markovian
  master equations},}\ }\href {\doibase 10.1088/1367-2630/14/12/123016}
  {\bibfield  {journal} {\bibinfo  {journal} {New J. of Phys.}\ }\textbf
  {\bibinfo {volume} {14}},\ \bibinfo {pages} {123016} (\bibinfo {year}
  {2012})}\BibitemShut {NoStop}%
\bibitem [{\citenamefont {Lanting}(2013)}]{Trevor}%
  \BibitemOpen
  \bibfield  {author} {\bibinfo {author} {\bibfnamefont {Trevor}\ \bibnamefont
  {Lanting}},\ }\href@noop {} {}\bibinfo {howpublished} {D-Wave Inc., private
  communications} (\bibinfo {year} {2013})\BibitemShut {NoStop}%
\bibitem [{\citenamefont {Sachdev}(2001)}]{Sachdev:book}%
  \BibitemOpen
  \bibfield  {author} {\bibinfo {author} {\bibfnamefont {S.}~\bibnamefont
  {Sachdev}},\ }\href@noop {} {\emph {\bibinfo {title} {Quantum Phase
  Transitions}}}\ (\bibinfo  {publisher} {Cambridge University Press},\
  \bibinfo {address} {Cambridge, UK},\ \bibinfo {year} {2001})\BibitemShut
  {NoStop}%
\bibitem [{\citenamefont {Zamolodchikov}(1989)}]{Zamolodchikov:1989}%
  \BibitemOpen
  \bibfield  {author} {\bibinfo {author} {\bibfnamefont {A.~B.}\ \bibnamefont
  {Zamolodchikov}},\ }\bibfield  {title} {\enquote {\bibinfo {title} {Integrals
  of motion and s-matrix of the (scaled) {$T = T_c$} ising model with magnetic
  field},}\ }\href {\doibase 10.1142/S0217751X8900176X} {\bibfield  {journal}
  {\bibinfo  {journal} {International Journal of Modern Physics A}\ }\textbf
  {\bibinfo {volume} {04}},\ \bibinfo {pages} {4235--4248} (\bibinfo {year}
  {1989})}\BibitemShut {NoStop}%
\bibitem [{\citenamefont {Henkel}\ and\ \citenamefont
  {Saleur}(1989)}]{Henkel:1989}%
  \BibitemOpen
  \bibfield  {author} {\bibinfo {author} {\bibfnamefont {M}~\bibnamefont
  {Henkel}}\ and\ \bibinfo {author} {\bibfnamefont {H}~\bibnamefont {Saleur}},\
  }\bibfield  {title} {\enquote {\bibinfo {title} {The two-dimensional ising
  model in the magnetic field: a numerical check of zamolodchikov's
  conjecture},}\ }\href {http://stacks.iop.org/0305-4470/22/i=11/a=011}
  {\bibfield  {journal} {\bibinfo  {journal} {Journal of Physics A:
  Mathematical and General}\ }\textbf {\bibinfo {volume} {22}},\ \bibinfo
  {pages} {L513} (\bibinfo {year} {1989})}\BibitemShut {NoStop}%
\bibitem [{\citenamefont {Noschese}\ \emph {et~al.}(2013)\citenamefont
  {Noschese}, \citenamefont {Pasquini},\ and\ \citenamefont
  {Reichel}}]{NLA:NLA1811}%
  \BibitemOpen
  \bibfield  {author} {\bibinfo {author} {\bibfnamefont {Silvia}\ \bibnamefont
  {Noschese}}, \bibinfo {author} {\bibfnamefont {Lionello}\ \bibnamefont
  {Pasquini}}, \ and\ \bibinfo {author} {\bibfnamefont {Lothar}\ \bibnamefont
  {Reichel}},\ }\bibfield  {title} {\enquote {\bibinfo {title} {Tridiagonal
  toeplitz matrices: properties and novel applications},}\ }\href {\doibase
  10.1002/nla.1811} {\bibfield  {journal} {\bibinfo  {journal} {Numerical
  Linear Algebra with Applications}\ }\textbf {\bibinfo {volume} {20}},\
  \bibinfo {pages} {302--326} (\bibinfo {year} {2013})}\BibitemShut {NoStop}%
\end{thebibliography}
%

\appendix

\section{Methods} \label{app:Methods}
\subsection{Experiment details} 
\begin{figure}[t]
{\includegraphics[width=\columnwidth]{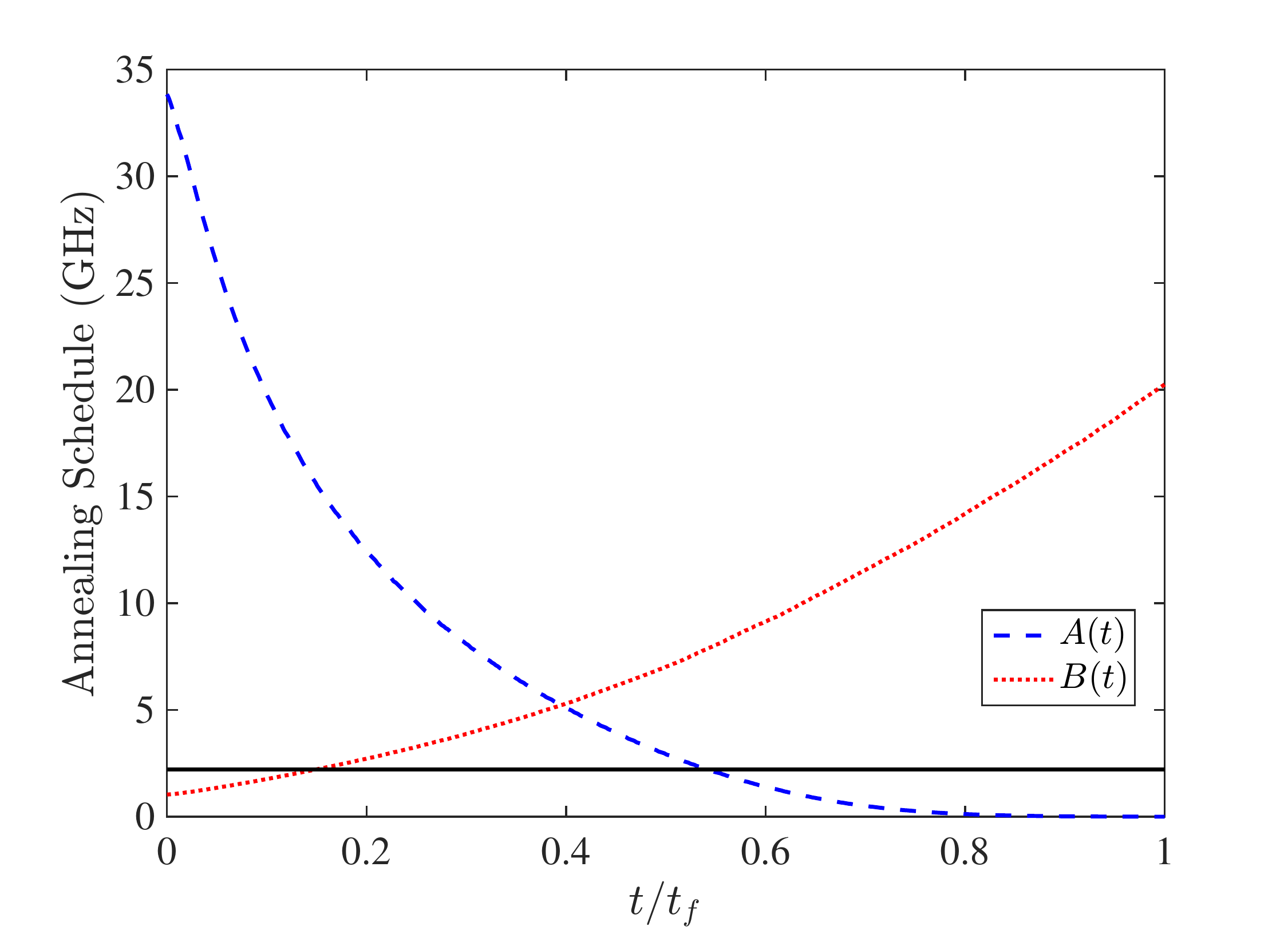}}
\caption{(Color online) {DW2 annealing schedule.} The functions $A$ and $B$ are the ones appearing in Eq.~\eqref{eq:H(t)}. The solid horizontal black line is the operating temperature ($17$mK).}
\label{fig:anneal-schedule-DW2}
\end{figure}

Our experiments were performed on the DW2 ``Vesuvius" processor at the Information Sciences Institute of the University of Southern California. The device has been described in detail elsewhere \cite{Harris:2010kx,Bunyk:2014hb}.  
The D-Wave processors are organized into unit cells consisting of eight  superconducting flux qubits arranged in a complete, balanced bipartite graph, with each side of the graph connecting to a neighboring unit cell, as seen in Fig.~\ref{fig:graphs}(a). The annealing schedule is shown in Fig.~\ref{fig:anneal-schedule-DW2}.

We ran a single copy of each problem instance $1000$ times, and we repeated this for each gauge. A gauge is a transformation of the couplings and fields that leaves the Ising spectrum invariant: pick $a_i = \pm 1$ at random for each spin variable $i$ and map $J_{ij} \mapsto a_i a_j J_{ij}, h_i \mapsto a_i h_i$ along with $\sigma^z_i \mapsto a_i \sigma_i^z$ \cite{q108}. For each problem instance we selected the lowest energy found among all runs and gauges, and we declared this to be the ground state energy. We are confident that every problem instance was solved correctly at least once since we never observed a lower energy using either the C or QAC strategies. We then implemented the C and QAC strategies as described in the main text.

\begin{figure} [t]
\includegraphics[width=\columnwidth]{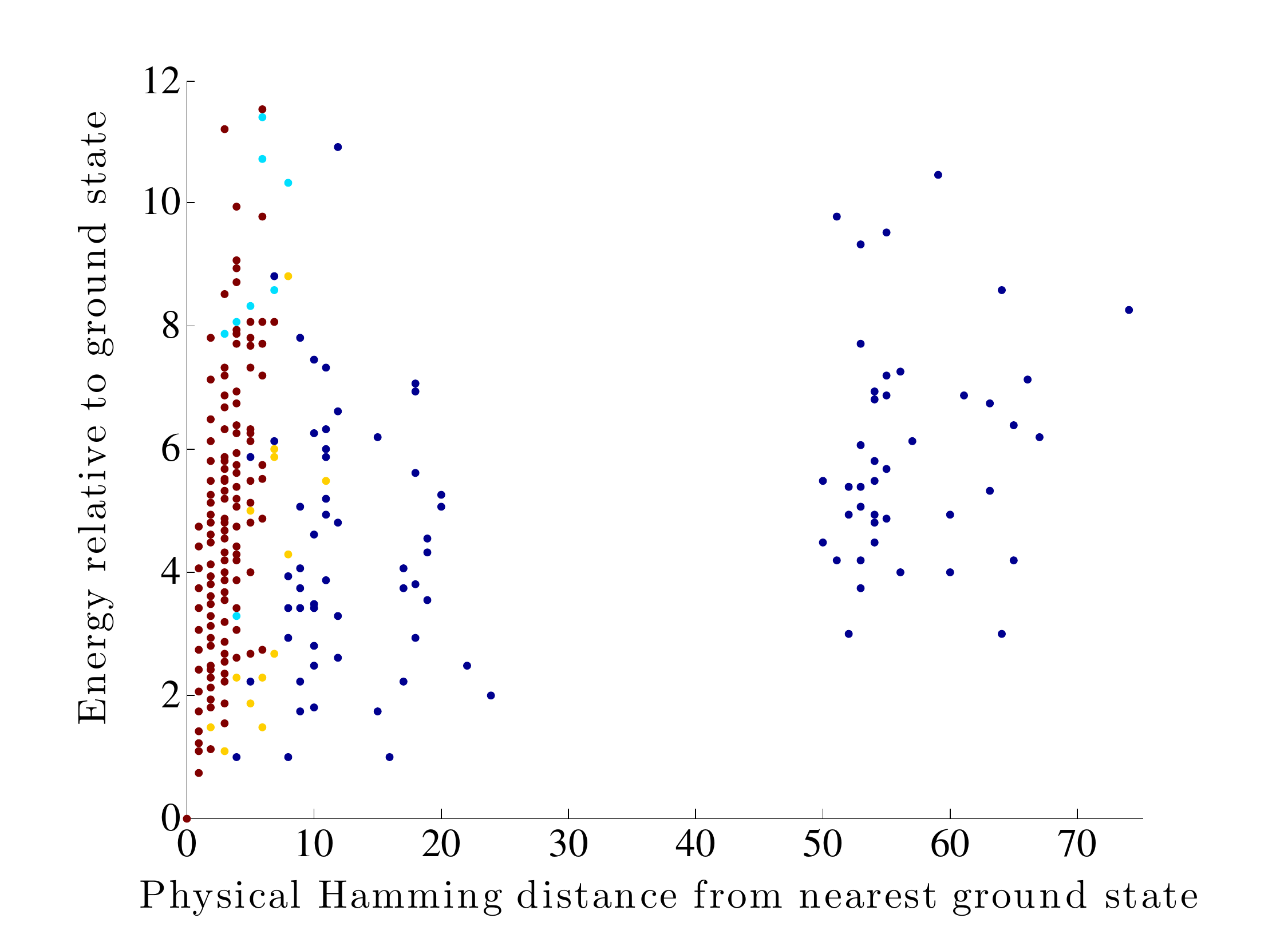}
\caption{(Color online) {Decodability of observed states.} Shown are the energy and Hamming distance relative to the ground state of all observed states in $1000$ annealing cycles of one problem instance with $\overline{N}=112$ that lies near the $95$th percentile in terms of the time to solution.
States colored red are decodable by both logical group and problem group decoding, and tend to be low in Hamming distance. States colored light blue (yellow) are decodable by logical (problem) group only, and occupy higher Hamming distances, with logical group decodable states generally higher in energy than problem group decodable states. Dark blue states are undecodable, and cluster in groups which represent sets of logical qubits flipping together (see Appendix \ref{app:error_types}).}
\label{fig:E_vs_hamming}
\end{figure}

\subsection{Data analysis methodology}
The following method was used to generate the number of annealing runs ($R$) data with their associated error bars, shown in Figs.~\ref{fig:scaling} and \ref{fig:robustness}.  Each instance $i$ for $\overline{N}=112$ $(86,66,46)$ was run for $G = 16$ $(8)$ gauges, and each gauge was run $M=1000$ times.  For each gauge $g$, the success probability $p_{i,g}$ for QAC or C between runs may be correlated, so a binning test was performed to determine the uncertainty in $p_{i,g}$ due to the possible correlations as well as the finite number of samples.  The length $M$ sequence of successes/failures was binned into $B$ subsequences of length $L=M/B$.  The success probability $p_{i,g,l}$ of each sequence $l$ was calculated, and the error associated with this binning was determined by:
\beq
\Delta p_{i,g}(L) = \frac{1}{\sqrt{L}} \sqrt{ \frac{\sum_{l=1}^L (p_{i,g,l} - p_{i,g})^2}{L-1}}
\eeq
If $\Delta p_{i,g}(L)$ converges as $B$ is increased, then the converged value is taken to be the error $\Delta p_{i,g}$ associated with the success probability $p_{i,g}$.  If a gauge did not yield a converged value for the error, then this gauge was discarded.  

Next, we determined the gauge-averaged number of repetitions.  We performed $1000$ bootstraps over the gauges, where for each gauge in each bootstrap, we calculated the number of repetitions using $\mathcal{N}(p_{i,g},\Delta p_{i,g})$ instead of simply $p_{i,g}$.  
For each bootstrap we calculated the mean number of repetitions, then took the mean over the bootstraps to determine the gauge-averaged number of repetitions $R_i$, with the standard deviation over the bootstraps giving the error $\Delta R_i$.

Finally for Figs.~\ref{fig:scaling} and \ref{fig:robustness}, we performed a bootstrap over the instances at a given $\overline{N}$, where for each instance in the bootstrap we used $\mathcal{N}(R_i, \Delta R_i)$.  For each bootstrap we calculated the different percentiles for the number of repetitions.  Then the mean of the percentiles over the bootstraps is the number of repetitions at size $\overline{N}$ with the error bar given by the standard deviation of the percentiles over the bootstraps.

\subsection{Decoding Strategies}
To recover the correct solution to the encoded problem as part of the QAC strategy, we used two post-readout classical decoding methods. The methods are complementary and each has complexity linear in problem size.
The first method, \emph{logical group decoding}, is the standard scheme for decoding a repetition code. It consists of taking a majority vote over the three problem qubits within each logical qubit, which yields a single value for each logical qubit in the original problem we seek to solve.
The second method, \emph{problem group decoding}, is equivalent to the interpretation of the results of the C method but applied only to the three copies of the problem embodied by the problem qubits, disregarding the penalty qubits.  Further details of the decoding methods are discussed in Appendix \ref{app:error_types}.

To illustrate the role of decoding, Fig.~\ref{fig:E_vs_hamming} shows all the states that were observed in $1000$ annealing cycles of a particular problem instance. We observe that  successful decoding depends strongly on the errors having low Hamming weight, but that a significant tolerance to excitations is permissible. This illustrates that under QAC the standard adiabatic criterion of remaining in the ground state is relaxed and replaced by maintaining correctability.

\begin{figure} [h]
\includegraphics[width=\columnwidth]{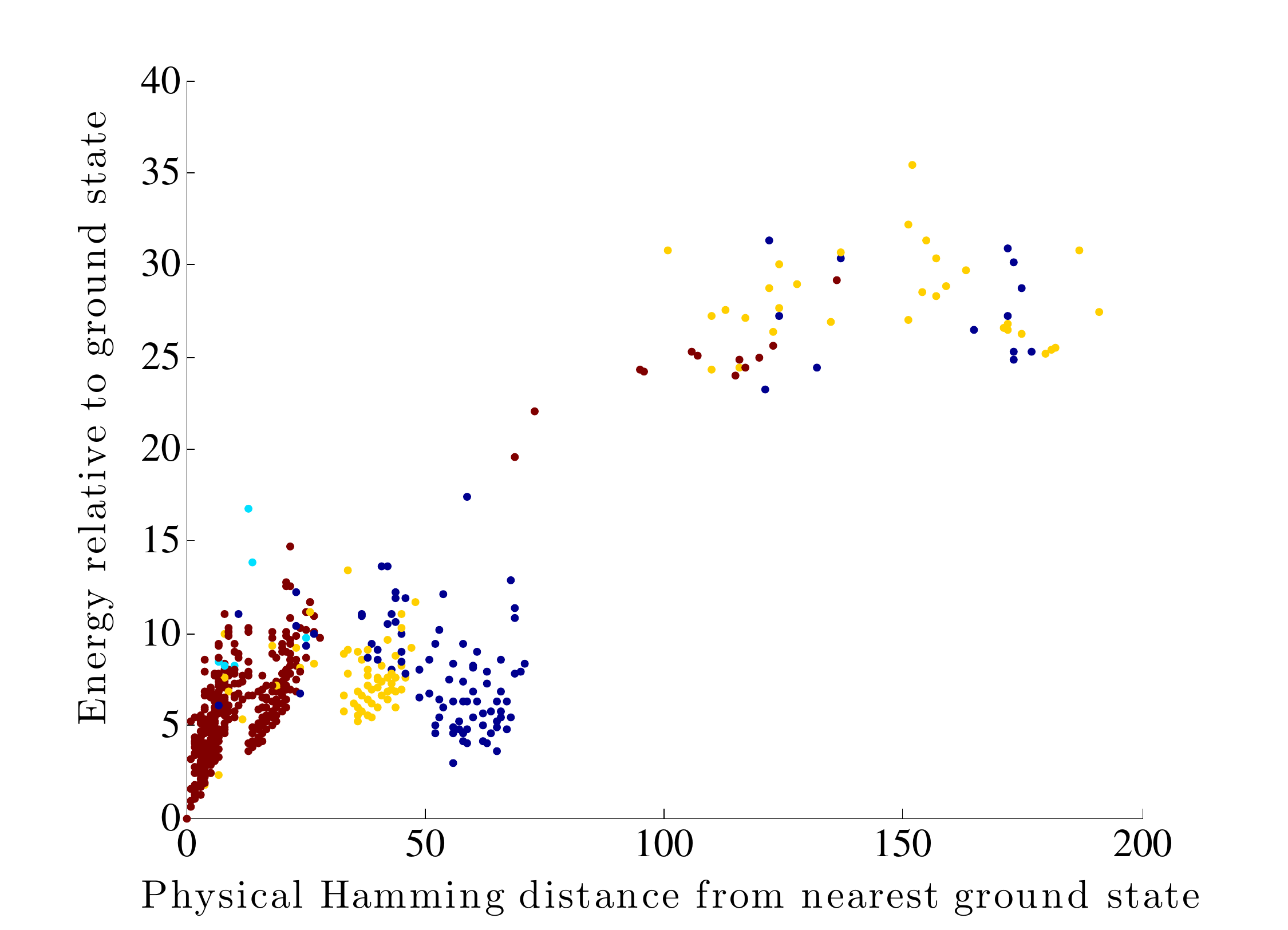}
\caption{{Decodability of observed states for a problem instance in which C outperforms QAC}. States are colored by decodability as in Fig.~\ref{fig:E_vs_hamming}.  Note the large cluster of undecodable (dark blue) states at relatively low energy but at high Hamming distance from the ground state. Here $\beta_{\textrm{opt}}=0.1$.}
\label{fig:E_vs_hamming2}
\end{figure}

\begin{figure*}[b]
\subfigure[\ ]{\includegraphics[width=0.48\textwidth]{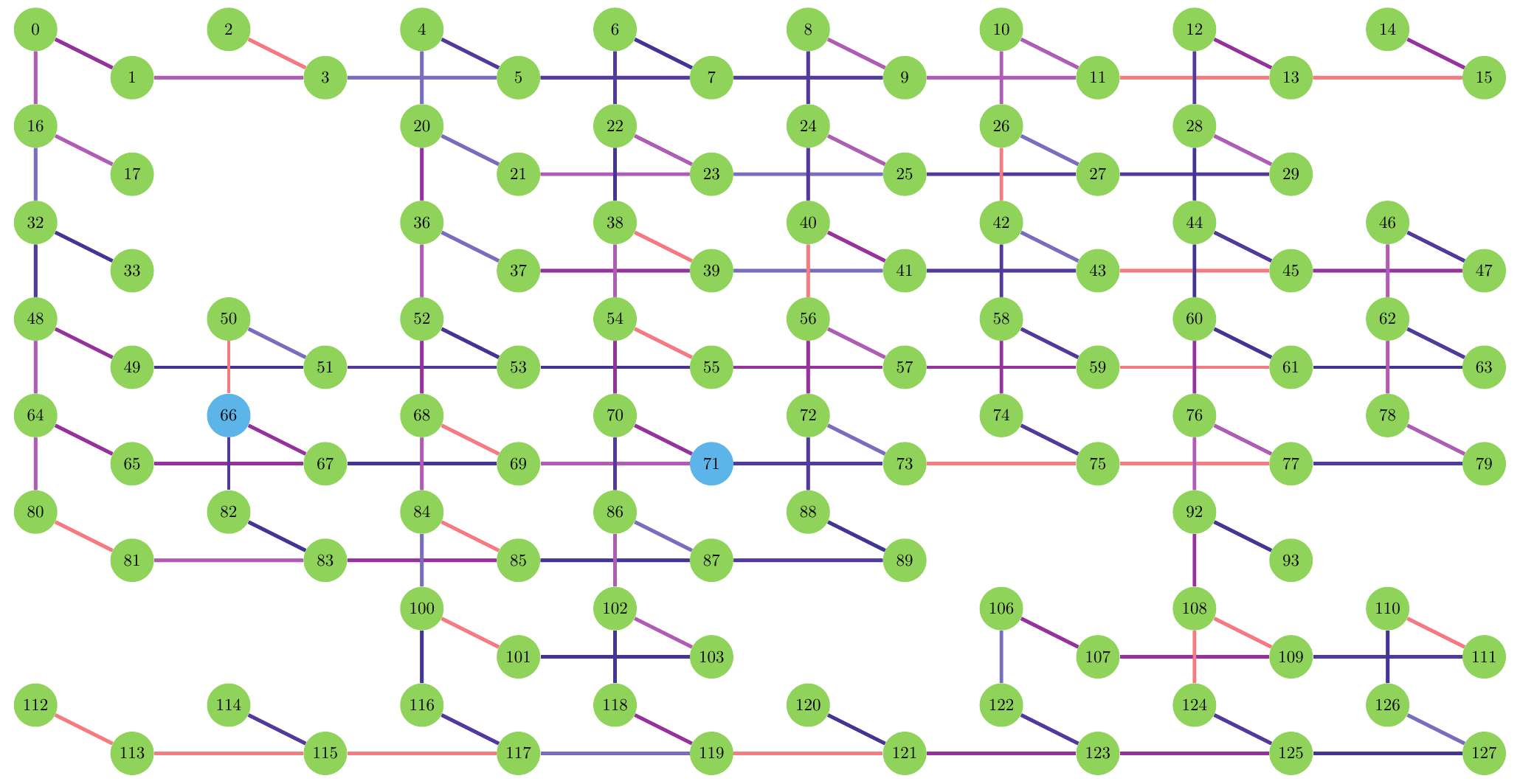}\label{fig:overlap_decodable}}
\subfigure[\ ]{\includegraphics[width=0.48\textwidth]{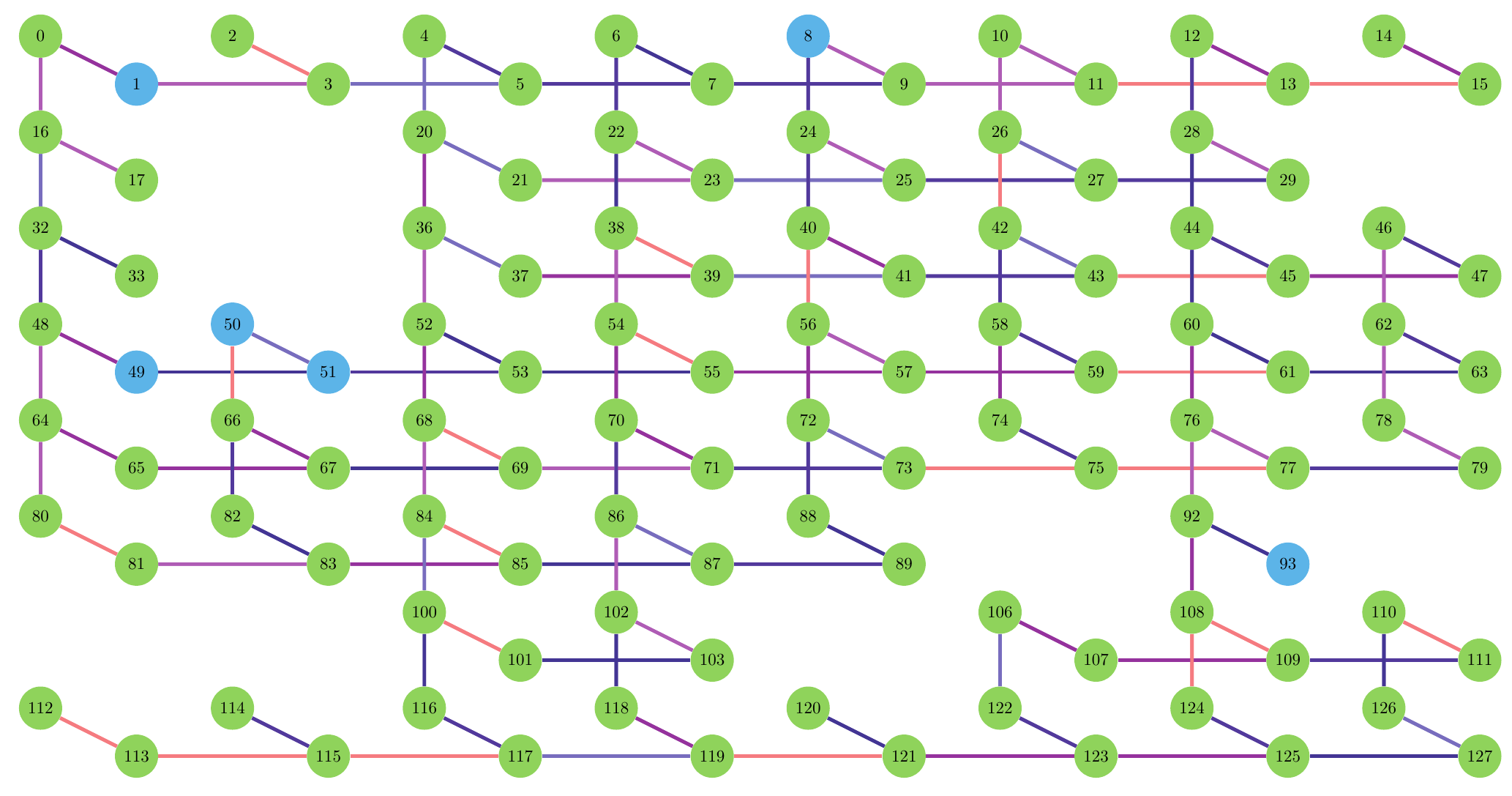}\label{fig:logical_decodable}}
\subfigure[\ ]{\includegraphics[width=0.48\textwidth]{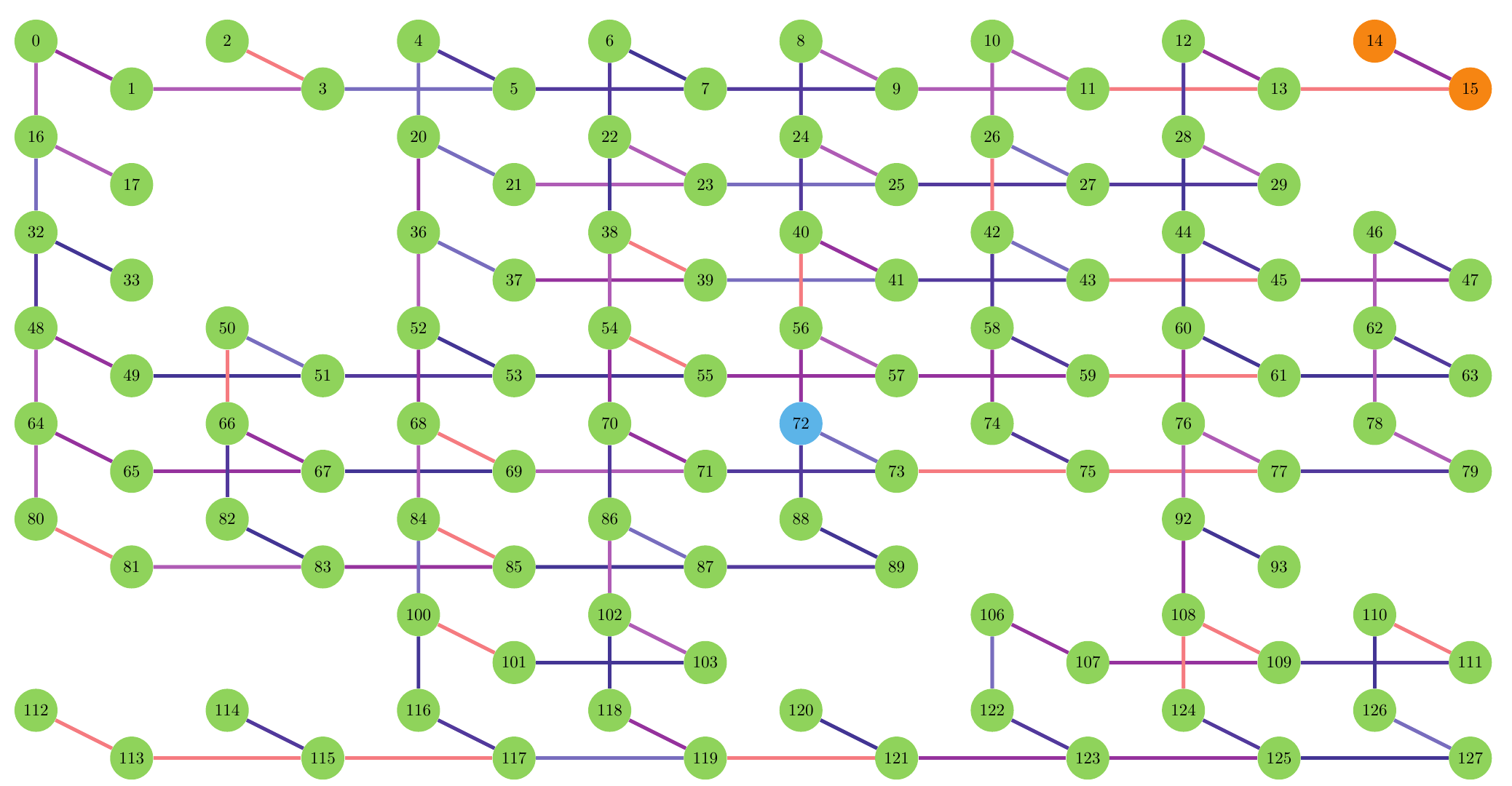}\label{fig:problem_decodable}}
\subfigure[\ ]{\includegraphics[width=0.48\textwidth]{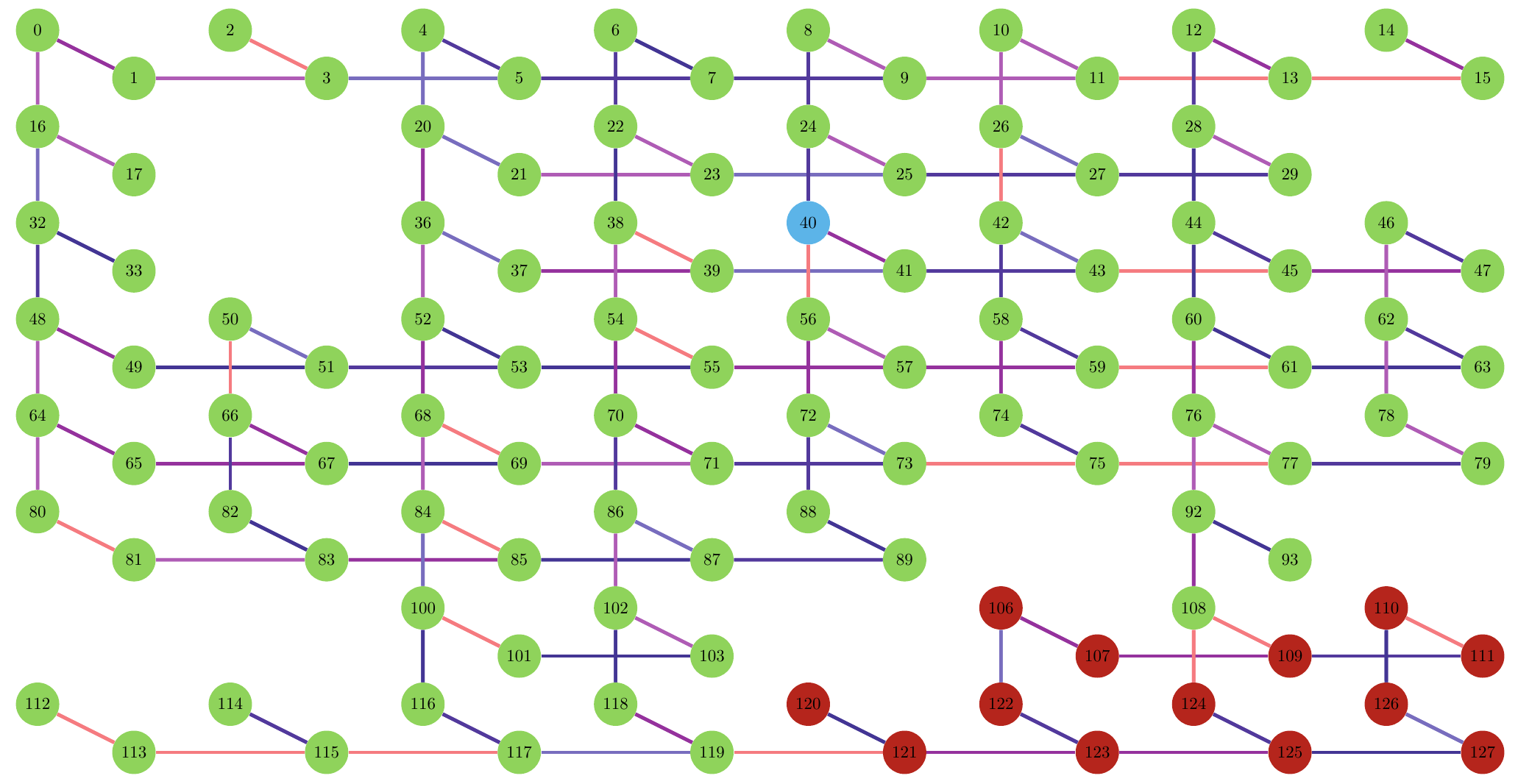}\label{fig:undecodable}}
\caption{(Color online) {Examples of error types and decodability.} All states shown are experimentally observed examples. Green circles are correct logical qubits, i.e., with all their physical qubits aligned with the ground state; blue, orange and red circles are logical qubits with, respectively, one, two, and three bit flip errors. The magnitude (though not the sign) of the problem couplings is color-coded: pink lines indicate $\lvert J_{ij} \rvert =\frac{1}{6}$, and the shade of the line darkens through gradations to indigo for $\lvert J_{ij}\rvert =1$. (a) A state decodable by both logical group decoding and problem group decoding. In this typical example we observe two logical qubits with a single bit flip error each (blue circles) among otherwise correct logical qubits (green circles). The state is logical group decodable because single bit flips are majority-vote correctable. It is also problem group decodable because there are only two bit flips, which is not enough to ruin all three copies of the problem embedded within the QAC scheme. (b) A logical group decodable state. The $6$ single bit flips are all decodable via majority vote so the state is logical group decodable, but the state is not problem group decodable because each problem group is corrupted by at least one bit flip. (c) A problem group decodable state. This state has two logical qubits in the upper right hand corner (orange circles) that are loosely coupled to the rest of the problem (pink line, $\lvert J_{ij} \rvert =\frac{1}{6}$) and which each have two physical problem qubits flipped from the ground state values. The problem qubits that flipped are correlated between the two logical qubits; they belong to the same copy of the problem because the problem coupling is strong between counterpart problem qubits (purple line). This leaves one copy of the problem fully intact, and the state can be decoded using problem group decoding. (d) An undecodable state. There is a cluster of logical qubit flips (red circles) in the lower right hand corner. That region is loosely coupled to the rest of the problem; all links going out of it are weak (pink) couplings. This means that the state with these logical qubits flipped together is a low-lying final excited state of the logical Ising problem, which has been suppressed via the repetition energy scale enhancement portion of QAC but is still observable in the problem instance's output statistics. This state belongs to the cluster observable near Hamming weight $20$ in Fig.~\ref{fig:E_vs_hamming} in the main text.}
\label{fig:LGD-PGD}
\end{figure*}

\begin{figure*}
\subfigure[]{\includegraphics[width=.48\textwidth]{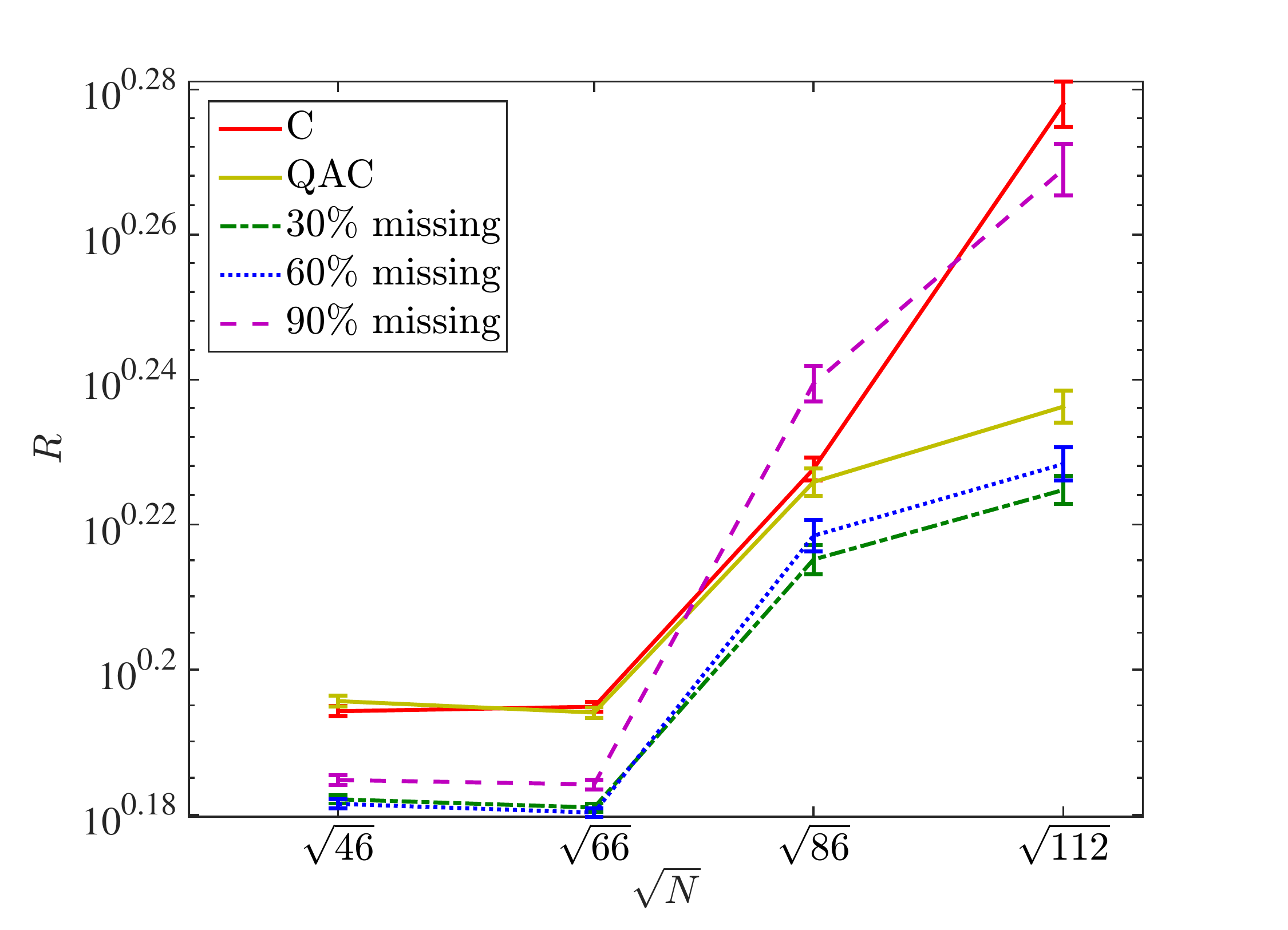}}
\subfigure[]{\includegraphics[width=.48\textwidth]{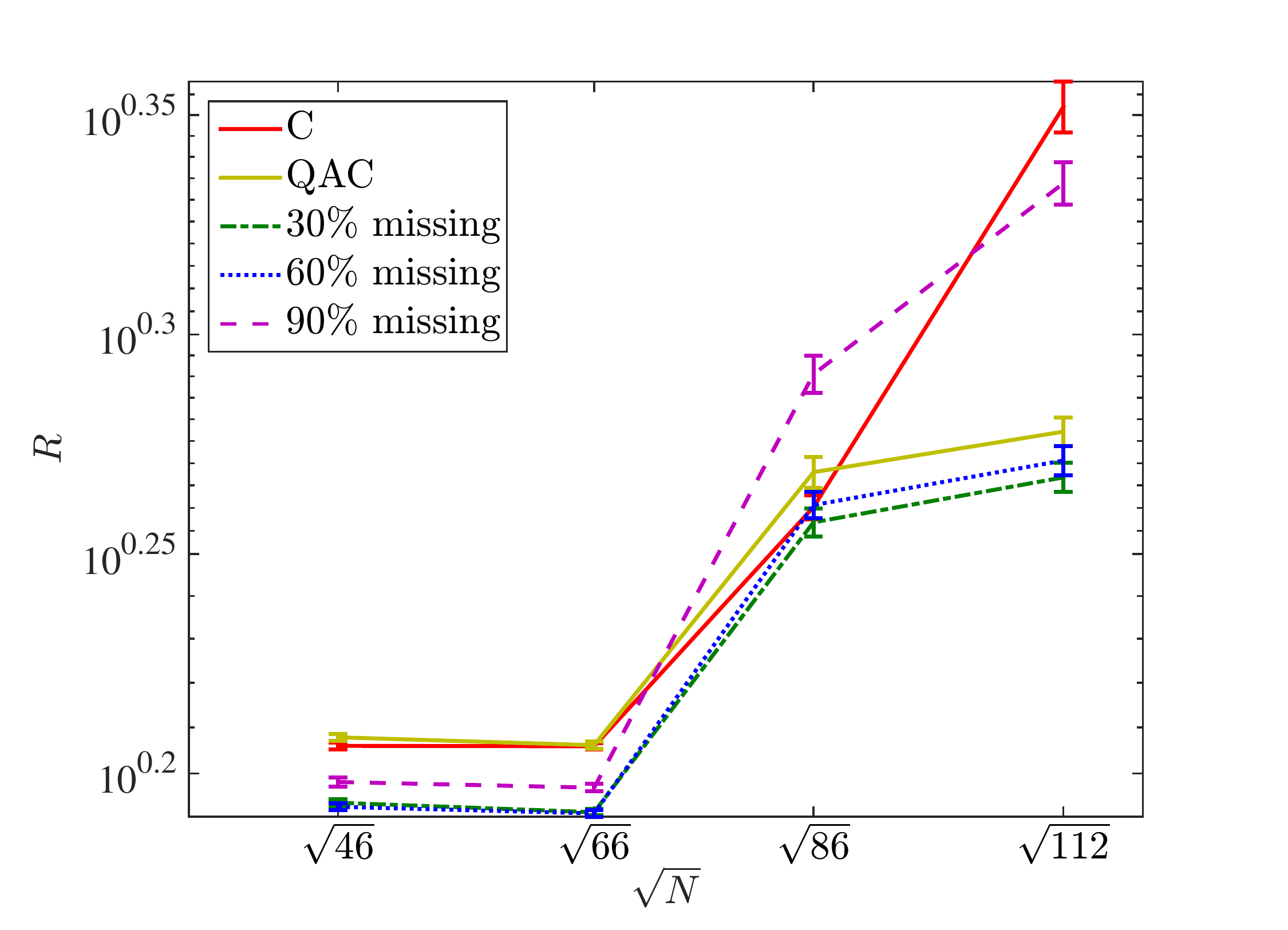}}
\caption{(Color online) {Additional percentiles supplementing Fig.~\ref{fig:robustness} of the main text showing the effect of missing penalty qubits.} Effect of missing penalty qubits for $\alpha=1$ on the $50$th and $75$th percentiles for $\beta$ optimized for the number of missing qubits. The C strategy (solid red) and QAC strategy (solid yellow) lines are the same data that was displayed in Fig.~4(a). The dash-dotted green, dotted blue, and dashed purple series show the effects of randomly removing $30\%$, $60\%$, and $90\%$ of the penalty qubits, respectively. The performance at $30\%$ and $60\%$ loss tracks the original QAC closely, suggesting the code is highly resilient to penalty qubit loss if the penalty magnitude is adjusted accordingly. }
\label{fig:robustness2}
\end{figure*}

\begin{figure*}[t]
\subfigure[\ U]{\includegraphics[width=0.32\textwidth]{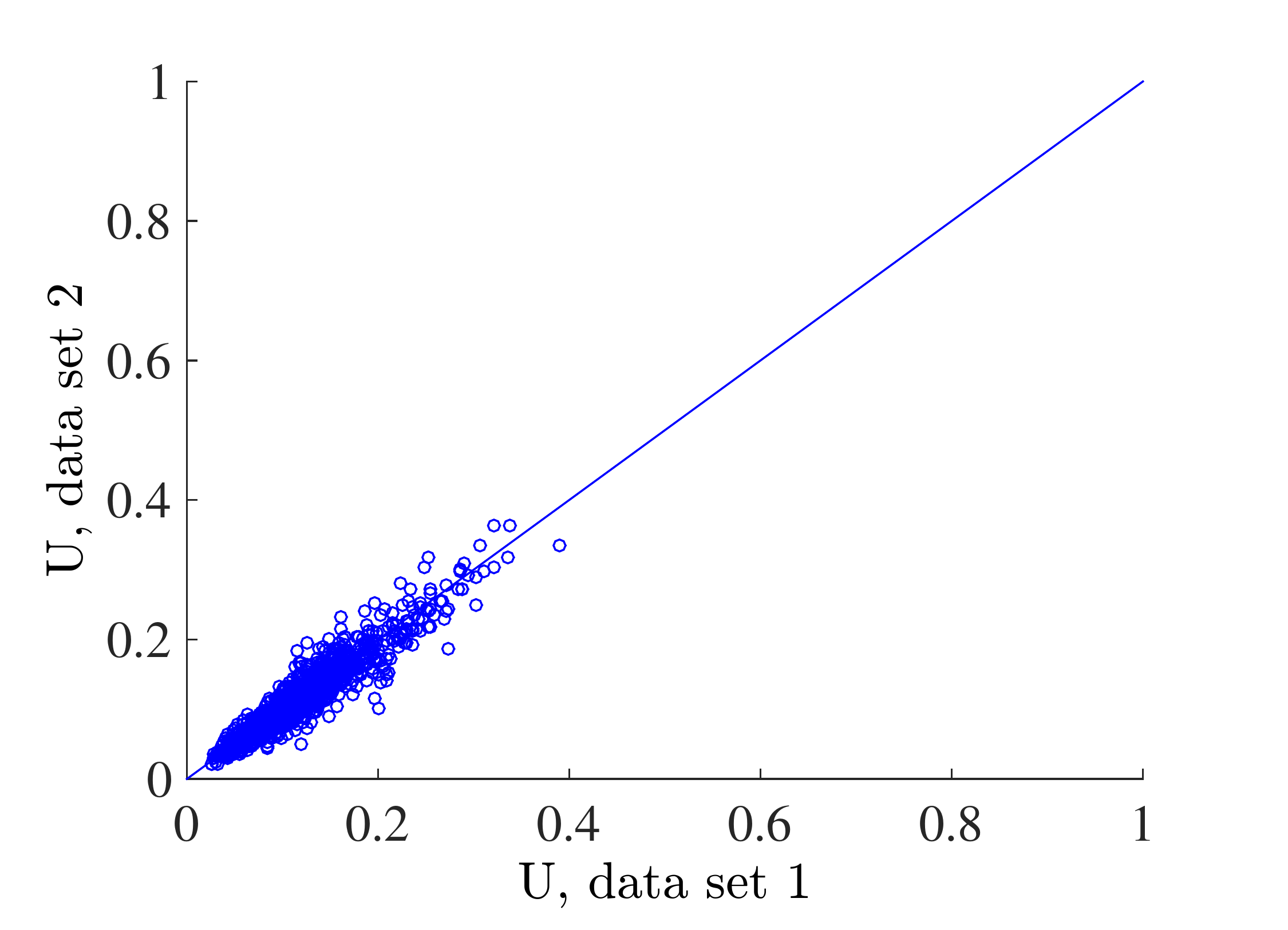} \label{fig:correlation_U_86q_a05}}
\subfigure[\ C]{\includegraphics[width=0.32\textwidth]{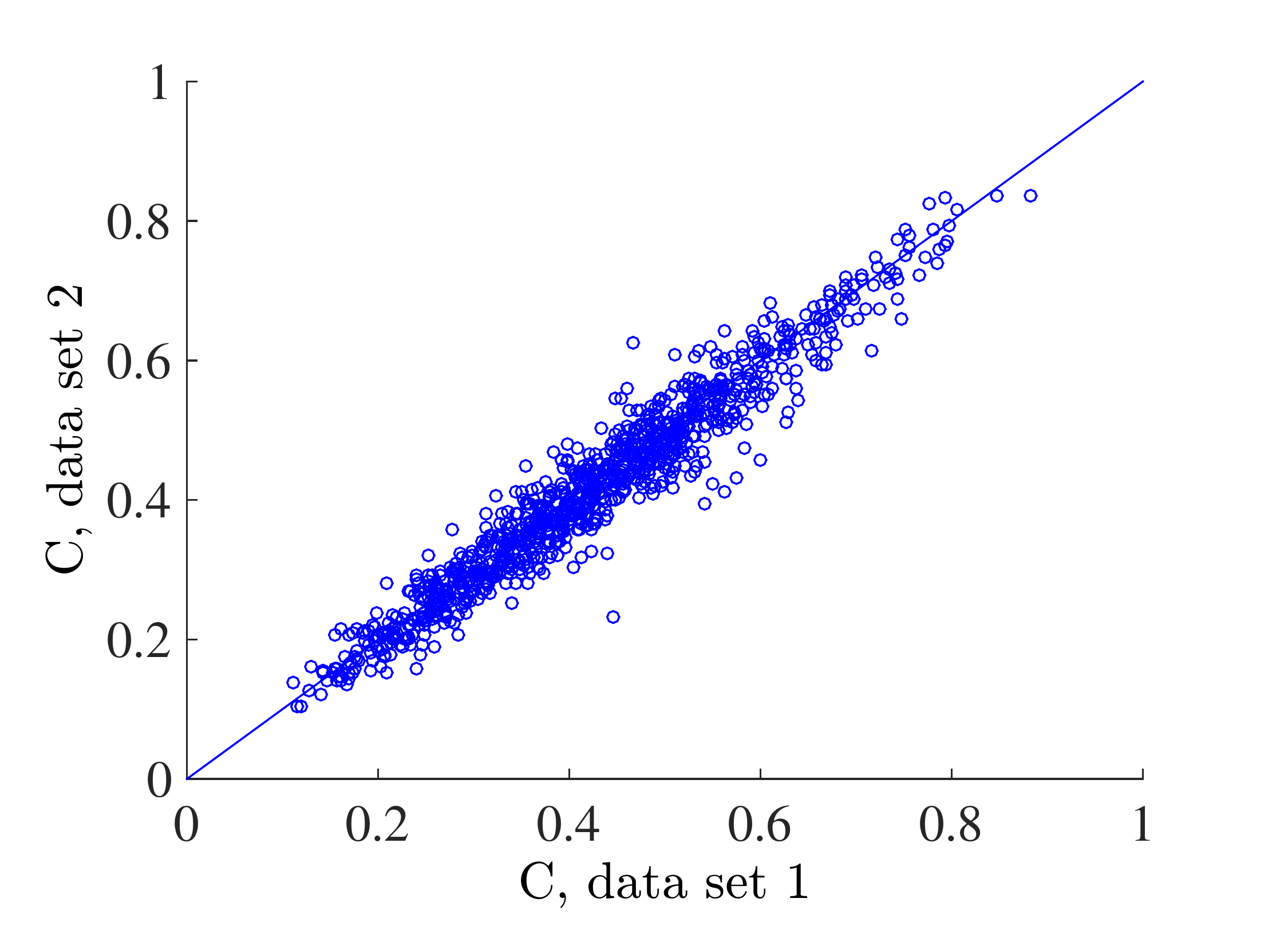} \label{fig:correlation_C_86q_a05}}
\subfigure[\ QAC]{\includegraphics[width=0.32\textwidth]{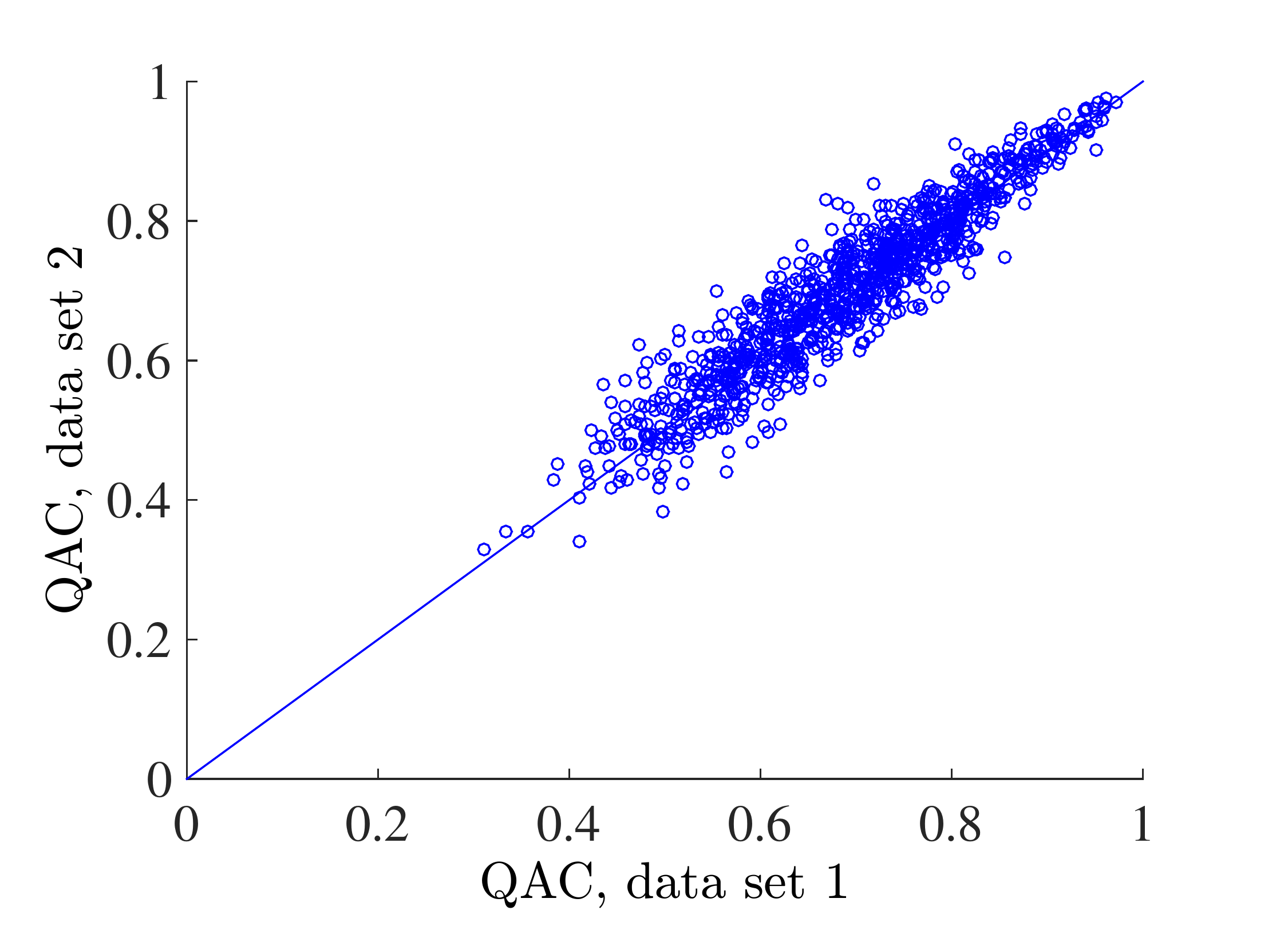} \label{fig:correlation_QAC_86q_a05}}
\caption{{Correlation of different $\alpha = 0.5$, $\overline{N}=86$ data sets.} (a) The U case, i.e., a single copy of the problem without any error correction. (b) The C strategy. (c) The QAC strategy. In all cases the correlation is excellent, with the corresponding Pearson correlation coefficients being $\rho = 0.942$ for U, $\rho = 0.969$ for C, and $\rho = 0.941$ for QAC.}
\label{fig:cor_a05}
\end{figure*}

\begin{figure}[t]
\includegraphics[width=0.5\textwidth]{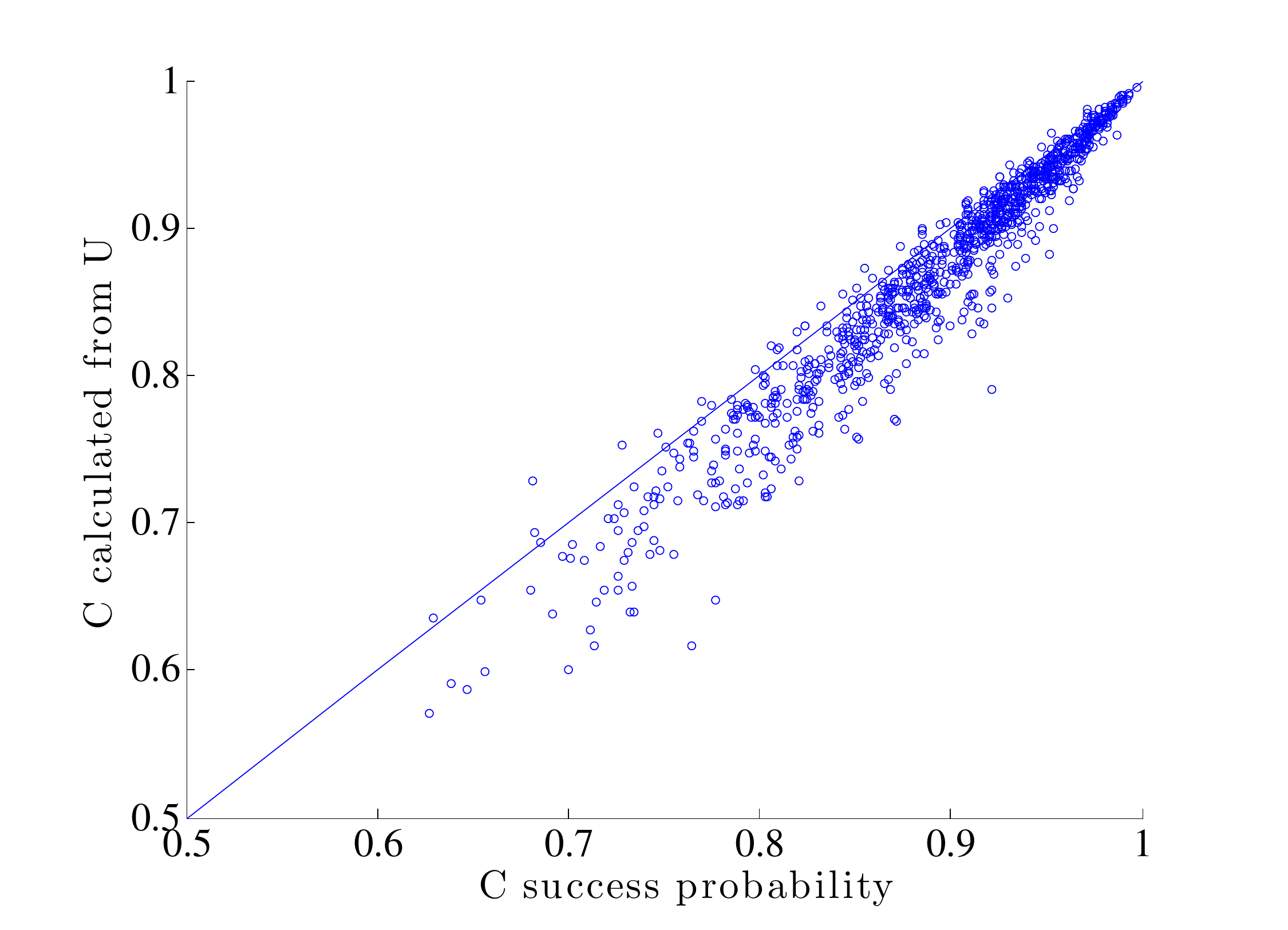}
\caption{{Correlation between actual C and theoretical C.} The plot shows the probabilities for $1000$ random Ising instances with $\overline{N}=112$ perfect logical qubits, computed using the C strategy and from a single unprotected copy (U), i.e., $p_{\textrm{C},i}$ \textit{vs} $p_{\textrm{U},i}$ for each instance $i$. A strong correlation is observed, with Pearson correlation coefficient of $0.968$.}
\label{fig:independence_112q}
\end{figure}

\subsection{Explanation for cases in which C outperforms QAC} 
As shown in Fig.~\ref{fig:scatter112}, there is a small subset of instances in which the C strategy outperforms the QAC strategy.
Of this subset, only $5$ instances have $\beta_{\textrm{opt}} = 0$.  For these instances C outperforms QAC simply because it actually contains four copies of the encoded problem \textit{vs} three copies for QAC. For the remainder of the instances in this subset, having $\beta_{\textrm{opt}}>0$, the QAC strategy improves the success probability relative to simply using three copies of the encoded problem.  

Why then does QAC not provide a sufficient improvement to outperform C in these cases?  As seen in Fig.~\ref{fig:graphs}(c), there are holes in the logical connectivity graph due to the fact that we used only perfect logical qubits in our experiments.  Qubits around these holes have fewer couplings and therefore the energy penalty is lower for violated couplings than for qubits in the intact areas of the graph. This is exacerbated when these couplings are weak, causing the energy cost for a logical error to be small especially if the physical qubits are tied together by a finite $\beta_{\textrm{opt}}$, as in the case of the QAC strategy.  This leads to a pronounced increase in undecodable, low-energy eigenstates, as depicted in Fig.~\ref{fig:E_vs_hamming2}.  It turns out that this is typical, i.e., instances in which C outperforms QAC are dominated by undecodable, low-energy excited states, arising from the flipping of large clusters of logical qubits that are weakly tied to the rest.This can be compared to what happens in the more common case, in which QAC outperforms C, as shown in Fig.~\ref{fig:E_vs_hamming}.

\section{Error types and decodability} \label{app:error_types}

In order to more closely examine the effects of the QAC encoding and the capabilities of the problem group and logical group decoding methods, it is instructive to consider a representative problem instance. 
Figure~\ref{fig:E_vs_hamming} of the main text shows all of the states that were observed in $1000$ annealing cycles of an $\overline{N}=112$ problem instance that lies near the $95$th percentile in terms of time to solution, colored by the decoding method to which they yielded. Looking deeper, Fig.~\ref{fig:LGD-PGD} shows sample states from the decodability categories of Fig.~\ref{fig:E_vs_hamming} of the main text, illuminating the error mechanisms for which each type of decoding is particularly suited.


Overall, we see that both decoding methods succeed for low Hamming weight thermal errors. While logical group decoding performs well for excited states consisting of a large number of random errors, problem group decoding is better equipped to address errors that are correlated through the problem coupling terms present in the QAC Hamiltonian. 
Although logical (rather than physical) errors are not expected to yield to any trivial-complexity postprocessing decoding method such as the ones examined here, it is possible that a strategy involving some kind of local classical optimization may allow even these states to be recovered.

\section{Robustness to physical qubit loss}

In the main text we presented results demonstrating the robustness of QAC to physical qubit loss. Specifically, we investigated the case where the missing physical qubits are few enough to be embedded as the penalty qubits within logical qubits, so that all problem couplings would remain intact. Toward that end, we performed quantum annealing on the $1000$ random problem instances of each size again, but this time we removed $30\%$, $60\%$, and then finally $90\%$ of the penalty qubits from each instance at random.

Figure~\ref{fig:robustness2} shows additional percentiles (median and $75$th) supplementing Fig.~\ref{fig:robustness} of the main text ($95$th). We observe that the separation between the C and QAC strategies at $\overline{N}=112$ persists even when up to $60\%$ of the penalty qubits are removed.

\section{Correlation tests} \label{sec:CorrelationTest}

Since our data collection lasted several weeks, we checked the stability of our results by performing two separate sets of experiments for $\alpha = 0.5$ and $\overline{N}=86$, separated by $25$ days. We computed the Pearson correlation coefficient,
$\rho_{XY} = \frac{\textrm{cov}(X,Y)}{\sigma_X \sigma_Y}$ (the covariance of the two variables $X$ and $Y$ divided by the product of their standard deviations) for $X$ and $Y$ being the first and second data set, respectively. The results are shown in Fig.~\ref{fig:cor_a05}, confirming that our data collection procedure was stable over time.

In a separate test we checked the correlation between the C success probability obtained by actually running $4$ copies of perfect logical qubits in parallel and the theoretical C success probability obtained by running a single copy, and using it in the binomial expression for the success probability of $4$ independent copies, i.e., $1-(1-p_{\textrm{U},i})^4$, where $p_{\textrm{U},i}$ is the single copy success probability for instance $i$. In more detail, to account for gauge averaging the procedure we used is the following procedure. Let $p_{\textrm{U},g,i}$ denote the experimentally observed success probability within the U case for a given instance $i$ and gauge $g$. Then $p_{\textrm{U},i} = \frac{1}{G}\sum_{g=1}^{G}1-(1-p_{\textrm{U},g,i})^4$ is the theoretical value of C from the U case, where we gauge-average the success probabilities. The result is shown in Fig.~\ref{fig:independence_112q} and the correlation is high, confirming that the $4$ copies in the actual C strategy are to a large degree independent. Deviations are likely due to residual cross-talk effects \cite{q-sig2}. In the main text we used $p_{\textrm{C},i}$ in our comparisons of the C and QAC strategy, rather than $p_{\textrm{U},i}$. Since as can be seen in Fig.~\ref{fig:independence_112q} for most instances $p_{\textrm{C},i} > p_{\textrm{U},i}$, this is a more stringent test of the QAC strategy. The procedure for computing $p_{\textrm{C},i}$ is the same as outlined in the Appendix \ref{app:Methods} for the gauge-averaged number of repetitions $R_i$.

\section{Energy gap enhancement in a solvable transverse field Ising model on a ring} \label{app:Analytics}
Here we provide details supporting the discussion of the Ising Hamiltonian on a ring with local fields presented in the main text.  We first briefly review the model.

Let $\mathcal{P}_d$ be the Pauli group over $d$ qubits, and consider an Abelian subgroup $\mathcal{S} \in \mathcal{P}_d$ with $d-1$ generators $g_i$ given by:
\beq \label{eq:generators}
g_i = \sigma_i^z \sigma_{i+1}^z \ , i \in \{1, \dots, d-1\} \ .
\eeq
This defines a distance $d$ repetition stabilizer code that detects all bit flips $\sigma_i^x$ since they anti-commute with at least one of the stabilizer generators in Eq.~\eqref{eq:generators}. The code comprises a single logical qubit with codespace spanned by $\ket{\overline{0}} = \ket{0 \dots 0}$ and $\ket{\overline{1}} = \ket{1 \dots 1}$. We can choose the logical operators as $\overline{X} = \otimes_{i=1}^d \sigma_i^x$ and $\overline{Z} = \frac{1}{d} \sum_{i=1}^d \sigma_i^z$.  

Consider a single qubit Hamiltonian $H_{\mathrm{Ising}} = -h \sigma^z$ (with the ground state denoted by $\ket{0}$ and the excited state given by $\ket{1}$) and encode it using the above repetition code.  Rather than replacing $\sigma^z$ by $\overline{Z}$, we use $\overline{\sigma}^z \equiv d \overline{Z}$:
\beq
\overline{H}_{\mathrm{Ising}} = -h  \sum_{i=1}^d \sigma_i^z \ .
\eeq
We can add to this Hamiltonian a penalty that is the sum over the generators $g_i$ and their product $\prod_{i=1}^{d-1} g_i = \sigma_1^z \sigma_d^z$. With the convention $\sigma_{d+1}^z = \sigma_1^z$ the penalty term is then
\beq
H_P = - \beta \sum_{i=1}^d \sigma_i^z \sigma_{i+1}^z \ ,
\eeq 
which describes a one-dimensional Ising chain with periodic boundary conditions, and penalizes all single bit flip errors. We use this as the final Hamiltonian of our quantum annealing algorithm.  The total time-dependent Hamiltonian is that of a transverse field Ising model on a ring with a local field:
\beq  
\label{eq:HQA}
H(s) = -A(s) \sum_{i=1}^d \sigma_i^x + B(s) \left(  \overline{H}_{\mathrm{Ising}} + H_P\right) .
\eeq
For convenience we use the dimensionless time variable $s=t/t_f$. Note that $A(s)$ and $B(s)$ have dimensions of energy while $h$ and $\beta$ are both dimensionless. Also note that in order to encode the entire quantum annealing algorithm \cite{jordan2006error} we would have had to replace the transverse field term by $\overline{X}$. However, the $\overline{X}$ operator is a $d$-weight operator, which is not physically available.

We proceed to analyze $H(s)$ by treating the penalty term as a perturbation that is switched on at the end of the annealing evolution, with time going backward. Thus both $\beta$ and $s$ are considered small parameters.

\subsection*{Unperturbed Hamiltonian}
Consider first the $d=1$ case (single qubit Hamiltonian)
\beq
H(s) = - A_0 \left( 1 - s \right) \sigma^x - A_0 s h \sigma^z \ ,
\eeq
where for simplicity we have additionally set $A(s) = A_0 (1-s)$ and $B(s) = A_0 s$.  This can be easily diagonalized to yield the (dimensionless) eigenenergies
\beq
\eps_\pm/A_0 = \pm  \sqrt{(1-s)^2 + h^2 s^2} \equiv \pm  \lambda(s) \ ,
\eeq
with respective orthonormal eigenstates
\bes
\begin{align}
\ket{\eps_+(s)} &= \frac{1}{c_+(s)} \left[ \frac{h s-\lambda(s)}{1 -s }  \ket{0} + \ket{1} \right] \ , \\ 
\ket{\eps_-(s)}  &= \frac{1}{c_-(s)} \left[ \left(h s + \lambda(s) \right) \ket{0} + \left( 1- s \right) \ket{1} \right]\ ,
\end{align}
\ees
where $c_\pm(s)$ are normalization constants.  When we make $d$ copies of this system
\beq
H(s) = - A_0 \left( 1 - s \right) \sum_{i=1}^d \sigma_i^x - A_0 s h \sum_{i=1}^d \sigma_i^z \ ,
\eeq
the ground state is given by $\ket{\eps_\GS (s)} = \otimes_{i=1}^d \ket{\eps_-(s)}$ with energy $\eps^{(0)}_\GS(s) = -A_0 d \lambda(s)$, and the $d$-fold degenerate first excited states are given by $\ket{\eps_k(s)} = \otimes_{i=1}^{k-1} \ket{\eps_-(s)}_i \otimes \ket{\eps_+(s)} \otimes_{i=k+1}^d \ket{\eps_-(s)}_i $, i.e., $k$ labels which qubit is in the excited state. These first excited states have energy $-(d-2)A_0\lambda(s)$, so the unperturbed gap is $\Delta^{(0)}(s) = 2A_0\lambda(s)$. This gap is minimized at 
\beq
s_{\min}^{(0)} = \frac{1}{1+h^2} \ ,
\eeq
where 
\beq
\Delta_{\min}^{(0)} \equiv \Delta^{(0)}(s_{\min}^{(0)}) = 2A_0 \frac{|h|}{\sqrt{1 + h^2}}\ .
\eeq
%
\subsection*{Perturbation} 
\label{sec:perturbation}
%
We now introduce the penalty term $H_P = - \beta s V$  as a perturbation in $\beta s$, with $V = \sum_{i=1}^d \sigma_{i}^z \sigma_{i+1}^z$. From first order perturbation theory, the corrected dimensionless ground state energy is
\begin{eqnarray}
\eps_\GS(s)/A_0 &=& \eps^{(0)}_\GS(s)/A_0 - \beta s  \bra{\eps_\GS(s)} V \ket{\eps_\GS(s)} \nonumber \\
&=& -d \lambda(s) - d \beta s \bra{\eps_-(s)} \sigma^z \ket{\eps_-(s)}^2 \ .
\end{eqnarray}
For the first excited states, we employ first order degenerate perturbation theory, whereby we need to calculate the projected perturbation $PVP$, where $P = \sum_k \ket{\eps_{k}}$ projects on the ground state, i.e., the matrix elements $ \sum_i  \bra{\eps_k }\sigma_i^z \sigma_{i+1}^z \ket{\eps_{k'}}$. The only non-zero matrix elements are
\bes
\begin{align}
&  \sum_i  \bra{\eps_k }\sigma_i^z \sigma_{i+1}^z \ket{\eps_k}  \nonumber \\
&= (d-2)  \bra{\eps_-} \sigma^z \ket{\eps_-} ^2 + 2 \bra{\eps_+} \sigma^z \ket{\eps_+} \bra{\eps_-} \sigma^z \ket{\eps_-} \equiv a \ ,    \\
& \sum_i  \bra{\eps_k }\sigma_i^z \sigma_{i+1}^z \ket{\eps_{k+1}}  = \sum_i  \bra{\eps_{k-1} }\sigma_i^z \sigma_{i+1}^z \ket{\eps_{k}}  \nonumber \\
& =   \bra{\eps_-} \sigma^z \ket{\eps_+}^2 \equiv b \ .
  \end{align}
\ees
This defines a translationally invariant $d \times d$ tridiagonal matrix (with corner terms):
\beq
\left(
\begin{array}{cccccc}
a & b & 0 & 0 &\dots & b \\
b & a & b & 0 & \dots & 0 \\
0 & b & a & b & \dots & 0 \\
\vdots &  & \ddots & \ddots & \ddots\\
b & 0 & \dots & 0 &b & a
\end{array}
\right) \ .
\eeq
The eigenvalues of this matrix can be found analytically and are given by \cite{NLA:NLA1811}:
\beq
\lambda_n = a + 2 |b| \cos \left( \frac{2 \pi n}{d} \right) \ , \quad n = 0, \dots, d-1 .
\eeq
Thus the corrected dimensionless first excited state energy is $\eps_1(s)/A_0 = -(d-2)\lambda(s) -\beta s \max_n \lambda_n = -(d-2)\lambda(s) -\beta s(a+2|b|)$. Therefore, to first order in perturbation theory the dimensionless gap  $\Delta(s)/{A_0} = [\eps_1(s)-\eps_\GS(s)]/A_0$ to the first excited state is given, after some algebra, by:
\beq
\label{eq:gap}
\Delta(s)/{A_0} = 2\lambda(s) - 2\beta s \frac{(1-s)^2 - 2 (h s)^2}{\lambda(s)^2} \ .
\eeq
To find the minimum of this function we expand $s_{\min}$ (the location of the minimum) to first order in $\beta$, keeping in mind that we are interested in the end of the evolution (near $s=0$) and the limit where $\beta \ll h$. Thus, writing $s_{\min} = s_0 + s_1 \beta + O(\beta^2)$ we find that $d\Delta(s)/ds=0$ to first order in $\beta$ at
\bes
\begin{align}
s_{\min} &= s_{\min}^{(0)} - \beta s_1 + O(\beta^2)\ , \\
s_1 &= \frac{|h| (8 - h^2)}{(1+h^2)^{5/2}} \ ,
\end{align}
\ees
where $s_{\min}^{(0)}$ is the unperturbed value. At this $s_{\min}$ we have
\bes
\begin{align}
\Delta_{\min} &\equiv \Delta(s_{\min}) = \Delta_{\min}^{(0)} + \Delta_{\min}^{(1)} + O(\beta^2) \ , \\
\Delta_{\min}^{(1)} &= A_0 \beta \frac{2 (2- h^2)}{(1 + h^2)^2} \ ,
\end{align}
\ees
where likewise $\Delta_{\min}^{(0)}$ is the unperturbed value. Since $\Delta_{\min}^{(1)}>0$ if $h< \sqrt{2}$, and likewise then $s_1 > 0$, we find that for sufficiently small $h$ a perturbative $\beta$ has the effect of both increasing the minimum gap as well as shifting it to earlier in the evolution.

\end{document}